\newif\ifmycheck
\mychecktrue 
\mycheckfalse

\documentclass[twocolumn,tighten]{aastex61}

\providecommand\tabletype{deluxetable*}
\providecommand\bibsty{apj}
\providecommand\tablesize{\scriptsize}

\usepackage{graphicx}
\graphicspath{{./figures/}}

\makeatletter
\providecommand*{\input@path}{}
\g@addto@macro\input@path{{./tables/}}% append
\makeatother

\usepackage[]{xcolor} 
\usepackage{natbib} 
\usepackage{hyperref}
\usepackage{amsmath}
\usepackage{multirow}
\usepackage{accents}
\hypersetup{    
  colorlinks      = true,
  linkcolor       = {black},
  linkbordercolor = {white},
  citecolor       = {black},
  citebordercolor = {white},
  urlcolor        = {blue},
  urlbordercolor  = {white},
}
\usepackage[all]{hypcap}
\usepackage{import}

\usepackage{afterpage}
\usepackage{lineno}
\usepackage{soul} 
\usepackage{amsmath}
\usepackage{xspace}
\usepackage{xifthen}
\usepackage{eso-pic}

\newcommand{\ie}{i.e.\xspace}
\newcommand{\eg}{e.g.\xspace}
\newcommand{\etc}{etc.\xspace}

\newcommand{\vs}{vs.\xspace}

\newcommand{\NOOP}[1]{}

\newcommand{\CHECK}[1]{#1}

\ifmycheck {
  
  \renewcommand{\CHECK}[1]{{\bf \textcolor{orange}{#1}}}
  
} \fi

\newcommand{\breakcell}[2][c]{%
  \begin{tabular}[#1]{@{}c@{}}#2\end{tabular}}

\newcommand{\reportnum}[2]{
  \AddToShipoutPictureBG*{%
    \AtPageUpperLeft{%
      \hspace{0.75\paperwidth}%
      \raisebox{#1\baselineskip}{%
        \makebox[0pt][l]{\textnormal{#2}}
  }}}%
}

\mathchardef\mhyphen="2D

\newcommand{\roughly}{\ensuremath{ {\sim}\,} }

\newlength{\dhatheight}

\newcommand{\code}[1]{\texttt{#1}\xspace}

\newcommand{\var}[1]{\ensuremath{\texttt{\MakeUppercase{#1}}}\xspace}

\newcommand{\unit}[1]{\ensuremath{\mathrm{\,#1}}\xspace}

\newcommand{\erg}{\unit{erg}}
\newcommand{\degree}{\ensuremath{{}^{\circ}}\xspace}
\newcommand{\mas}{\unit{mas}}
\newcommand{\amin}{\unit{arcmin}}

\newcommand{\nm}{\unit{nm}}

\newcommand{\cm}{\unit{cm}}

\newcommand{\second}{\unit{s}}

\newcommand{\magn}{\unit{mag}}
\newcommand{\mmag}{\unit{mmag}}
\providecommand{\deg}{}
\renewcommand{\deg}{\unit{deg}}

\newcommand{\secref}[1]{Section~\ref{sec:#1}}
\newcommand{\appref}[1]{Appendix~\ref{app:#1}}
\newcommand{\tabref}[1]{Table~\ref{tab:#1}}

\newcommand{\figref}[1]{Figure~\ref{fig:#1}}

\newcommand{\bandvar}[2][]{%
  \ifthenelse{\isempty{#1}}{\var{#2}}{\var{#2\_#1}}%
}

\newcommand{\spreaderrmodel}[1][]{\bandvar[#1]{spreaderr\_model}}
\newcommand{\wavgspreadmodel}[1][]{\bandvar[#1]{wavg\_spread\_model}}

\newcommand{\magauto}[1][]{\bandvar[#1]{mag\_auto}}
\newcommand{\magmodel}[1][]{\bandvar[#1]{mag\_model}}

\newcommand{\magerrauto}[1][]{\bandvar[#1]{magerr\_auto}}

\newcommand{\flags}[1][]{\bandvar[#1]{flags}}
\newcommand{\nepochs}[1][]{\bandvar[#1]{nepochs}}

\newcommand{\LCDM}{\ensuremath{\rm \Lambda CDM}\xspace}

\newcommand{\ra}{{\ensuremath{\alpha_{2000}}}\xspace}
\newcommand{\dec}{{\ensuremath{\delta_{2000}}}\xspace}

\newcommand{\maglimg}{23.4\xspace}
\newcommand{\maglimr}{23.2\xspace}
\newcommand{\maglimi}{22.5\xspace}
\newcommand{\maglimz}{21.8\xspace}
\newcommand{\maglimY}{20.1\xspace}
\newcommand{\moflimg}{23.7\xspace}
\newcommand{\moflimr}{23.5\xspace}
\newcommand{\moflimi}{22.9\xspace}
\newcommand{\moflimz}{22.2\xspace}

\newcommand{\photoz}{photo-$z$\xspace}

\newcommand{\Photoz}{Photo-$z$\xspace}
\newcommand{\devauc}{De Vaucouleurs'\xspace}
\newcommand{\uberseg}{{\"u}berseg\xspace}

\newcommand{\mof}{MOF\xspace}
\newcommand{\MOF}{\mof}

\newcommand{\SExtractor}{\code{SExtractor}}
\newcommand{\sextractor}{\SExtractor}
\newcommand{\PSFEx}{\code{PSFEx}}
\newcommand{\psfex}{\PSFEx}
\newcommand{\SWARP}{\code{SWarp}}
\newcommand{\swarp}{\SWARP}
\newcommand{\SCAMP}{\code{SCAMP}}
\newcommand{\scamp}{\SCAMP}
\newcommand{\ngmix}{\code{ngmix}}

\newcommand{\HEALPix}{\code{HEALPix}}
\newcommand{\healpix}{\HEALPix}
\newcommand{\healpy}{\code{healpy}}
\newcommand{\nside}{\code{nside}}

\newcommand{\mangle}{\code{mangle}}

\newcommand{\modest}{\code{MODEST\_CLASS}}
\newcommand{\teff}{\ensuremath{t_{\rm eff}}\xspace}
\newcommand{\Teff}{\ensuremath{T_{\rm eff}}\xspace}

\newcommand{\sdss}{\ensuremath{\mathrm{sdss}}\xspace}
\newcommand{\ukidss}{\ensuremath{\mathrm{ukidss}}\xspace}
\newcommand{\des}{\ensuremath{\mathrm{des}}\xspace}
\newcommand{\apass}{\ensuremath{\mathrm{apass}}\xspace}
\newcommand{\twomass}{\ensuremath{\mathrm{2mass}}\xspace}
\newcommand{\cfht}{\ensuremath{\mathrm{CFHT}}\xspace}

\newcommand{\gold}{{Y1A1 GOLD}\xspace}
\newcommand{\coadd}{{Y1A1 COADD}\xspace}
\newcommand{\finalcut}{{Y1A1 FINALCUT}\xspace}

\newcommand{\PSM}{{PSM}\xspace}
\newcommand{\GCM}{{GCM}\xspace}
\newcommand{\FGCM}{{FGCM}\xspace}

\newcommand{\OR}{\ensuremath{{\rm \,OR\,}}}
\newcommand{\AND}{\ensuremath{{\rm \,AND\,}}}

\reportnum{-10.0}{DES-2015-0118}
\reportnum{-11.0}{FERMILAB-PUB-17-180-AE}

\begin{document}

%\title{The Dark Energy Survey First-Year Cosmology Data Set} 
\title{Dark Energy Survey Year 1 Results:
Photometric Data Set for Cosmology} 

% Author list file generated with: mkauthlist 1.2.2 
% mkauthlist -j aastex61 -a data/author_order.csv data/DES-2015-0118_author_list_v2.csv authors.tex 

\author{A.~Drlica-Wagner}
\affiliation{Fermi National Accelerator Laboratory, P. O. Box 500, Batavia, IL 60510, USA}
\author{I.~Sevilla-Noarbe}
\affiliation{Centro de Investigaciones Energ\'eticas, Medioambientales y Tecnol\'ogicas (CIEMAT), Madrid, Spain}
\author{E.~S.~Rykoff}
\affiliation{Kavli Institute for Particle Astrophysics \& Cosmology, P. O. Box 2450, Stanford University, Stanford, CA 94305, USA}
\affiliation{SLAC National Accelerator Laboratory, Menlo Park, CA 94025, USA}
\author{R.~A.~Gruendl}
\affiliation{Department of Astronomy, University of Illinois, 1002 W. Green Street, Urbana, IL 61801, USA}
\affiliation{National Center for Supercomputing Applications, 1205 West Clark St., Urbana, IL 61801, USA}
\author{B.~Yanny}
\affiliation{Fermi National Accelerator Laboratory, P. O. Box 500, Batavia, IL 60510, USA}
\author{D.~L.~Tucker}
\affiliation{Fermi National Accelerator Laboratory, P. O. Box 500, Batavia, IL 60510, USA}
\author{B.~Hoyle}
\affiliation{Universit\"ats-Sternwarte, Fakult\"at f\"ur Physik, Ludwig-Maximilians Universit\"at M\"unchen, Scheinerstr. 1, 81679 M\"unchen, Germany}
\author{A. Carnero Rosell}
\affiliation{Laborat\'orio Interinstitucional de e-Astronomia - LIneA, Rua Gal. Jos\'e Cristino 77, Rio de Janeiro, RJ - 20921-400, Brazil}
\affiliation{Observat\'orio Nacional, Rua Gal. Jos\'e Cristino 77, Rio de Janeiro, RJ - 20921-400, Brazil}
\author{G.~M.~Bernstein}
\affiliation{Department of Physics and Astronomy, University of Pennsylvania, Philadelphia, PA 19104, USA}
\author{K.~Bechtol}
\affiliation{LSST, 933 North Cherry Avenue, Tucson, AZ 85721, USA}
\author{M.~R.~Becker}
\affiliation{Department of Physics, Stanford University, 382 Via Pueblo Mall, Stanford, CA 94305, USA}
\affiliation{Kavli Institute for Particle Astrophysics \& Cosmology, P. O. Box 2450, Stanford University, Stanford, CA 94305, USA}
\author{A.~Benoit-L{\'e}vy}
\affiliation{CNRS, UMR 7095, Institut d'Astrophysique de Paris, F-75014, Paris, France}
\affiliation{Department of Physics \& Astronomy, University College London, Gower Street, London, WC1E 6BT, UK}
\affiliation{Sorbonne Universit\'es, UPMC Univ Paris 06, UMR 7095, Institut d'Astrophysique de Paris, F-75014, Paris, France}
\author{E.~Bertin}
\affiliation{CNRS, UMR 7095, Institut d'Astrophysique de Paris, F-75014, Paris, France}
\affiliation{Sorbonne Universit\'es, UPMC Univ Paris 06, UMR 7095, Institut d'Astrophysique de Paris, F-75014, Paris, France}
\author{M.~Carrasco~Kind}
\affiliation{Department of Astronomy, University of Illinois, 1002 W. Green Street, Urbana, IL 61801, USA}
\affiliation{National Center for Supercomputing Applications, 1205 West Clark St., Urbana, IL 61801, USA}
\author{C.~Davis}
\affiliation{Kavli Institute for Particle Astrophysics \& Cosmology, P. O. Box 2450, Stanford University, Stanford, CA 94305, USA}
\author{J.~de Vicente}
\affiliation{Centro de Investigaciones Energ\'eticas, Medioambientales y Tecnol\'ogicas (CIEMAT), Madrid, Spain}
\author{H.~T.~Diehl}
\affiliation{Fermi National Accelerator Laboratory, P. O. Box 500, Batavia, IL 60510, USA}
\author{D.~Gruen}
\affiliation{Kavli Institute for Particle Astrophysics \& Cosmology, P. O. Box 2450, Stanford University, Stanford, CA 94305, USA}
\affiliation{SLAC National Accelerator Laboratory, Menlo Park, CA 94025, USA}
\author{W.~G.~Hartley}
\affiliation{Department of Physics \& Astronomy, University College London, Gower Street, London, WC1E 6BT, UK}
\affiliation{Department of Physics, ETH Zurich, Wolfgang-Pauli-Strasse 16, CH-8093 Zurich, Switzerland}
\author{B.~Leistedt}
\affiliation{New York University, CCPP,  New York, NY 10003, USA}
\author{T.~S.~Li}
\affiliation{Fermi National Accelerator Laboratory, P. O. Box 500, Batavia, IL 60510, USA}
\author{J.~L.~Marshall}
\affiliation{George P. and Cynthia Woods Mitchell Institute for Fundamental Physics and Astronomy, and Department of Physics and Astronomy, Texas A\&M University, College Station, TX 77843,  USA}
\author{E.~Neilsen}
\affiliation{Fermi National Accelerator Laboratory, P. O. Box 500, Batavia, IL 60510, USA}
\author{M.~M.~Rau}
\affiliation{Faculty of Physics, Ludwig-Maximilians-Universit\"at, Scheinerstr. 1, 81679 Munich, Germany}
\affiliation{Universit\"ats-Sternwarte, Fakult\"at f\"ur Physik, Ludwig-Maximilians Universit\"at M\"unchen, Scheinerstr. 1, 81679 M\"unchen, Germany}
\author{E.~Sheldon}
\affiliation{Brookhaven National Laboratory, Bldg 510, Upton, NY 11973, USA}
\author{J.~A.~Smith}
\affiliation{Austin Peay State University, Dept. Physics-Astronomy, P.O. Box 4608 Clarksville, TN 37044, USA}
\author{M.~A.~Troxel}
\affiliation{Center for Cosmology and Astro-Particle Physics, The Ohio State University, Columbus, OH 43210, USA}
\affiliation{Department of Physics, The Ohio State University, Columbus, OH 43210, USA}
\author{S.~Wyatt}
\affiliation{Austin Peay State University, Dept. Physics-Astronomy, P.O. Box 4608 Clarksville, TN 37044, USA}
\affiliation{Fermi National Accelerator Laboratory, P. O. Box 500, Batavia, IL 60510, USA}
\author{Y.~Zhang}
\affiliation{Fermi National Accelerator Laboratory, P. O. Box 500, Batavia, IL 60510, USA}
\author{T. M. C.~Abbott}
\affiliation{Cerro Tololo Inter-American Observatory, National Optical Astronomy Observatory, Casilla 603, La Serena, Chile}
\author{F.~B.~Abdalla}
\affiliation{Department of Physics \& Astronomy, University College London, Gower Street, London, WC1E 6BT, UK}
\affiliation{Department of Physics and Electronics, Rhodes University, PO Box 94, Grahamstown, 6140, South Africa}
\author{S.~Allam}
\affiliation{Fermi National Accelerator Laboratory, P. O. Box 500, Batavia, IL 60510, USA}
\author{M.~Banerji}
\affiliation{Institute of Astronomy, University of Cambridge, Madingley Road, Cambridge CB3 0HA, UK}
\affiliation{Kavli Institute for Cosmology, University of Cambridge, Madingley Road, Cambridge CB3 0HA, UK}
\author{D.~Brooks}
\affiliation{Department of Physics \& Astronomy, University College London, Gower Street, London, WC1E 6BT, UK}
\author{E.~Buckley-Geer}
\affiliation{Fermi National Accelerator Laboratory, P. O. Box 500, Batavia, IL 60510, USA}
\author{D.~L.~Burke}
\affiliation{Kavli Institute for Particle Astrophysics \& Cosmology, P. O. Box 2450, Stanford University, Stanford, CA 94305, USA}
\affiliation{SLAC National Accelerator Laboratory, Menlo Park, CA 94025, USA}
\author{D.~Capozzi}
\affiliation{Institute of Cosmology \& Gravitation, University of Portsmouth, Portsmouth, PO1 3FX, UK}
\author{J.~Carretero}
\affiliation{Institut de F\'{\i}sica d'Altes Energies (IFAE), The Barcelona Institute of Science and Technology, Campus UAB, 08193 Bellaterra (Barcelona) Spain}
\author{C.~E.~Cunha}
\affiliation{Kavli Institute for Particle Astrophysics \& Cosmology, P. O. Box 2450, Stanford University, Stanford, CA 94305, USA}
\author{C.~B.~D'Andrea}
\affiliation{Department of Physics and Astronomy, University of Pennsylvania, Philadelphia, PA 19104, USA}
\author{L.~N.~da Costa}
\affiliation{Laborat\'orio Interinstitucional de e-Astronomia - LIneA, Rua Gal. Jos\'e Cristino 77, Rio de Janeiro, RJ - 20921-400, Brazil}
\affiliation{Observat\'orio Nacional, Rua Gal. Jos\'e Cristino 77, Rio de Janeiro, RJ - 20921-400, Brazil}
\author{D.~L.~DePoy}
\affiliation{George P. and Cynthia Woods Mitchell Institute for Fundamental Physics and Astronomy, and Department of Physics and Astronomy, Texas A\&M University, College Station, TX 77843,  USA}
\author{S.~Desai}
\affiliation{Department of Physics, IIT Hyderabad, Kandi, Telangana 502285, India}
\author{J.~P.~Dietrich}
\affiliation{Excellence Cluster Universe, Boltzmannstr.\ 2, 85748 Garching, Germany}
\affiliation{Faculty of Physics, Ludwig-Maximilians-Universit\"at, Scheinerstr. 1, 81679 Munich, Germany}
\author{P.~Doel}
\affiliation{Department of Physics \& Astronomy, University College London, Gower Street, London, WC1E 6BT, UK}
\author{A.~E.~Evrard}
\affiliation{Department of Astronomy, University of Michigan, Ann Arbor, MI 48109, USA}
\affiliation{Department of Physics, University of Michigan, Ann Arbor, MI 48109, USA}
\author{A.~Fausti Neto}
\affiliation{Laborat\'orio Interinstitucional de e-Astronomia - LIneA, Rua Gal. Jos\'e Cristino 77, Rio de Janeiro, RJ - 20921-400, Brazil}
\author{B.~Flaugher}
\affiliation{Fermi National Accelerator Laboratory, P. O. Box 500, Batavia, IL 60510, USA}
\author{P.~Fosalba}
\affiliation{Institut de Ci\`encies de l'Espai, IEEC-CSIC, Campus UAB, Carrer de Can Magrans, s/n,  08193 Bellaterra, Barcelona, Spain}
\author{J.~Frieman}
\affiliation{Fermi National Accelerator Laboratory, P. O. Box 500, Batavia, IL 60510, USA}
\affiliation{Kavli Institute for Cosmological Physics, University of Chicago, Chicago, IL 60637, USA}
\author{J.~Garc\'ia-Bellido}
\affiliation{Instituto de Fisica Teorica UAM/CSIC, Universidad Autonoma de Madrid, 28049 Madrid, Spain}
\author{D.~W.~Gerdes}
\affiliation{Department of Astronomy, University of Michigan, Ann Arbor, MI 48109, USA}
\affiliation{Department of Physics, University of Michigan, Ann Arbor, MI 48109, USA}
\author{T.~Giannantonio}
\affiliation{Institute of Astronomy, University of Cambridge, Madingley Road, Cambridge CB3 0HA, UK}
\affiliation{Kavli Institute for Cosmology, University of Cambridge, Madingley Road, Cambridge CB3 0HA, UK}
\affiliation{Universit\"ats-Sternwarte, Fakult\"at f\"ur Physik, Ludwig-Maximilians Universit\"at M\"unchen, Scheinerstr. 1, 81679 M\"unchen, Germany}
\author{J.~Gschwend}
\affiliation{Laborat\'orio Interinstitucional de e-Astronomia - LIneA, Rua Gal. Jos\'e Cristino 77, Rio de Janeiro, RJ - 20921-400, Brazil}
\affiliation{Observat\'orio Nacional, Rua Gal. Jos\'e Cristino 77, Rio de Janeiro, RJ - 20921-400, Brazil}
\author{G.~Gutierrez}
\affiliation{Fermi National Accelerator Laboratory, P. O. Box 500, Batavia, IL 60510, USA}
\author{K.~Honscheid}
\affiliation{Center for Cosmology and Astro-Particle Physics, The Ohio State University, Columbus, OH 43210, USA}
\affiliation{Department of Physics, The Ohio State University, Columbus, OH 43210, USA}
\author{D.~J.~James}
\affiliation{Astronomy Department, University of Washington, Box 351580, Seattle, WA 98195, USA}
\affiliation{Cerro Tololo Inter-American Observatory, National Optical Astronomy Observatory, Casilla 603, La Serena, Chile}
\author{T.~Jeltema}
\affiliation{Santa Cruz Institute for Particle Physics, Santa Cruz, CA 95064, USA}
\author{K.~Kuehn}
\affiliation{Australian Astronomical Observatory, North Ryde, NSW 2113, Australia}
\author{S.~Kuhlmann}
\affiliation{Argonne National Laboratory, 9700 South Cass Avenue, Lemont, IL 60439, USA}
\author{N.~Kuropatkin}
\affiliation{Fermi National Accelerator Laboratory, P. O. Box 500, Batavia, IL 60510, USA}
\author{O.~Lahav}
\affiliation{Department of Physics \& Astronomy, University College London, Gower Street, London, WC1E 6BT, UK}
\author{M.~Lima}
\affiliation{Departamento de F\'isica Matem\'atica, Instituto de F\'isica, Universidade de S\~ao Paulo, CP 66318, S\~ao Paulo, SP, 05314-970, Brazil}
\affiliation{Laborat\'orio Interinstitucional de e-Astronomia - LIneA, Rua Gal. Jos\'e Cristino 77, Rio de Janeiro, RJ - 20921-400, Brazil}
\author{H.~Lin}
\affiliation{Fermi National Accelerator Laboratory, P. O. Box 500, Batavia, IL 60510, USA}
\author{M.~A.~G.~Maia}
\affiliation{Laborat\'orio Interinstitucional de e-Astronomia - LIneA, Rua Gal. Jos\'e Cristino 77, Rio de Janeiro, RJ - 20921-400, Brazil}
\affiliation{Observat\'orio Nacional, Rua Gal. Jos\'e Cristino 77, Rio de Janeiro, RJ - 20921-400, Brazil}
\author{P.~Martini}
\affiliation{Center for Cosmology and Astro-Particle Physics, The Ohio State University, Columbus, OH 43210, USA}
\affiliation{Department of Astronomy, The Ohio State University, Columbus, OH 43210, USA}
\author{R.~G.~McMahon}
\affiliation{Institute of Astronomy, University of Cambridge, Madingley Road, Cambridge CB3 0HA, UK}
\affiliation{Kavli Institute for Cosmology, University of Cambridge, Madingley Road, Cambridge CB3 0HA, UK}
\author{P.~Melchior}
\affiliation{Department of Astrophysical Sciences, Princeton University, Peyton Hall, Princeton, NJ 08544, USA}
\author{F.~Menanteau}
\affiliation{Department of Astronomy, University of Illinois, 1002 W. Green Street, Urbana, IL 61801, USA}
\affiliation{National Center for Supercomputing Applications, 1205 West Clark St., Urbana, IL 61801, USA}
\author{R.~Miquel}
\affiliation{Instituci\'o Catalana de Recerca i Estudis Avan\c{c}ats, E-08010 Barcelona, Spain}
\affiliation{Institut de F\'{\i}sica d'Altes Energies (IFAE), The Barcelona Institute of Science and Technology, Campus UAB, 08193 Bellaterra (Barcelona) Spain}
\author{R.~C.~Nichol}
\affiliation{Institute of Cosmology \& Gravitation, University of Portsmouth, Portsmouth, PO1 3FX, UK}
\author{R.~L.~C.~Ogando}
\affiliation{Laborat\'orio Interinstitucional de e-Astronomia - LIneA, Rua Gal. Jos\'e Cristino 77, Rio de Janeiro, RJ - 20921-400, Brazil}
\affiliation{Observat\'orio Nacional, Rua Gal. Jos\'e Cristino 77, Rio de Janeiro, RJ - 20921-400, Brazil}
\author{A.~A.~Plazas}
\affiliation{Jet Propulsion Laboratory, California Institute of Technology, 4800 Oak Grove Dr., Pasadena, CA 91109, USA}
\author{A.~K.~Romer}
\affiliation{Department of Physics and Astronomy, Pevensey Building, University of Sussex, Brighton, BN1 9QH, UK}
\author{A.~Roodman}
\affiliation{Kavli Institute for Particle Astrophysics \& Cosmology, P. O. Box 2450, Stanford University, Stanford, CA 94305, USA}
\affiliation{SLAC National Accelerator Laboratory, Menlo Park, CA 94025, USA}
\author{E.~Sanchez}
\affiliation{Centro de Investigaciones Energ\'eticas, Medioambientales y Tecnol\'ogicas (CIEMAT), Madrid, Spain}
\author{V.~Scarpine}
\affiliation{Fermi National Accelerator Laboratory, P. O. Box 500, Batavia, IL 60510, USA}
\author{R.~Schindler}
\affiliation{SLAC National Accelerator Laboratory, Menlo Park, CA 94025, USA}
\author{M.~Schubnell}
\affiliation{Department of Physics, University of Michigan, Ann Arbor, MI 48109, USA}
\author{M.~Smith}
\affiliation{School of Physics and Astronomy, University of Southampton,  Southampton, SO17 1BJ, UK}
\author{R.~C.~Smith}
\affiliation{Cerro Tololo Inter-American Observatory, National Optical Astronomy Observatory, Casilla 603, La Serena, Chile}
\author{M.~Soares-Santos}
\affiliation{Fermi National Accelerator Laboratory, P. O. Box 500, Batavia, IL 60510, USA}
\author{F.~Sobreira}
\affiliation{Instituto de F\'isica Gleb Wataghin, Universidade Estadual de Campinas, 13083-859, Campinas, SP, Brazil}
\affiliation{Laborat\'orio Interinstitucional de e-Astronomia - LIneA, Rua Gal. Jos\'e Cristino 77, Rio de Janeiro, RJ - 20921-400, Brazil}
\author{E.~Suchyta}
\affiliation{Computer Science and Mathematics Division, Oak Ridge National Laboratory, Oak Ridge, TN 37831}
\author{G.~Tarle}
\affiliation{Department of Physics, University of Michigan, Ann Arbor, MI 48109, USA}
\author{V.~Vikram}
\affiliation{Argonne National Laboratory, 9700 South Cass Avenue, Lemont, IL 60439, USA}
\author{A.~R.~Walker}
\affiliation{Cerro Tololo Inter-American Observatory, National Optical Astronomy Observatory, Casilla 603, La Serena, Chile}
\author{R.~H.~Wechsler}
\affiliation{Department of Physics, Stanford University, 382 Via Pueblo Mall, Stanford, CA 94305, USA}
\affiliation{Kavli Institute for Particle Astrophysics \& Cosmology, P. O. Box 2450, Stanford University, Stanford, CA 94305, USA}
\affiliation{SLAC National Accelerator Laboratory, Menlo Park, CA 94025, USA}
\author{J.~Zuntz}
\affiliation{Institute for Astronomy, University of Edinburgh, Edinburgh EH9 3HJ, UK}

\collaboration{(DES Collaboration)}

\correspondingauthor{Alex Drlica-Wagner}
\email{kadrlica@fnal.gov}

%TC:break Abstract
\begin{abstract}

We describe the creation, content, and validation of the Dark Energy Survey (DES) internal year-one cosmology data set, \gold, in support of upcoming cosmological analyses.
The \gold data set is assembled from multiple epochs of DES imaging and consists of calibrated photometric zeropoints, object catalogs, and ancillary data products---\eg, maps of survey depth and observing conditions, star-galaxy classification, and photometric redshift estimates---that are necessary for accurate cosmological analyses.
The \gold wide-area object catalog consists of $\roughly 137$ million objects detected in coadded images covering $\roughly 1800 \deg^2$ in the DES $grizY$ filters.
The $10\sigma$ limiting magnitude for galaxies is $g = \maglimg$, $r = \maglimr$, $i = \maglimi$, $z = \maglimz$, and $Y = \maglimY$.
Photometric calibration of \gold was performed by combining nightly zeropoint solutions with stellar locus regression, and the absolute calibration accuracy is better than 2\% over the survey area.
DES \gold is the largest photometric data set at the achieved depth to date, enabling precise measurements of cosmic acceleration at $z \lesssim 1$. 
\end{abstract}
%TC:break _main_


\section{Introduction}
% (Drlica-Wagner, Rykoff, Sevilla)
\label{sec:intro}

The Dark Energy Survey \citep[DES;][]{DES:2005,DES:2016ktf} is a photometric survey utilizing the Dark Energy Camera \citep[DECam;][]{Flaugher:2015} on the Blanco 4m telescope at Cerro Tololo Inter-American Observatory (CTIO) in Chile to observe $\roughly 5000 \deg^2$ of the southern sky in five broadband filters, $g,r,i,z,Y$, ranging from $\roughly400\nm$ to $\roughly1060\nm$ \citep{Li:2016,Y3FGCM}.\footnote{The DECam filter throughput is publicly available at \url{http://www.ctio.noao.edu/noao/node/1033}.}
The primary goal of DES is to study the origin of cosmic acceleration and the nature of dark energy through four key probes: weak lensing, large-scale structure, galaxy clusters, and Type Ia supernovae.
More generally, DES provides a rich scientific data set and has already had a significant impact beyond cosmology \citep[\eg,][]{DES:2016ktf}.

Precision measurements of dark energy with DES rely on an unprecedented survey data set and a comprehensive understanding of the survey performance.
It is necessary to identify, characterize, and mitigate the influences of variable observing conditions, data processing artifacts, photometric calibration nonuniformity, and astrophysical foregrounds.
For example, photometric calibration must be accurate and uniform to avoid introducing noise and bias into photometric redshift estimates.
Studies of galaxy clustering depend on a detailed knowledge of survey coverage,  galaxy detection efficiency, and the accuracy of recovered galaxy properties.
Furthermore, detailed modeling of the point-spread function (PSF) and instrument response is required to perform galaxy shape measurements on objects that are fainter than the the detection limit of a single DES image.
The scale and complexity of assembling, characterizing, and validating the DES data motivate a collaborative effort that draws upon and enables a wide range of scientific analyses.

\begin{figure*}[t]
\center
\includegraphics[width=0.8\textwidth]{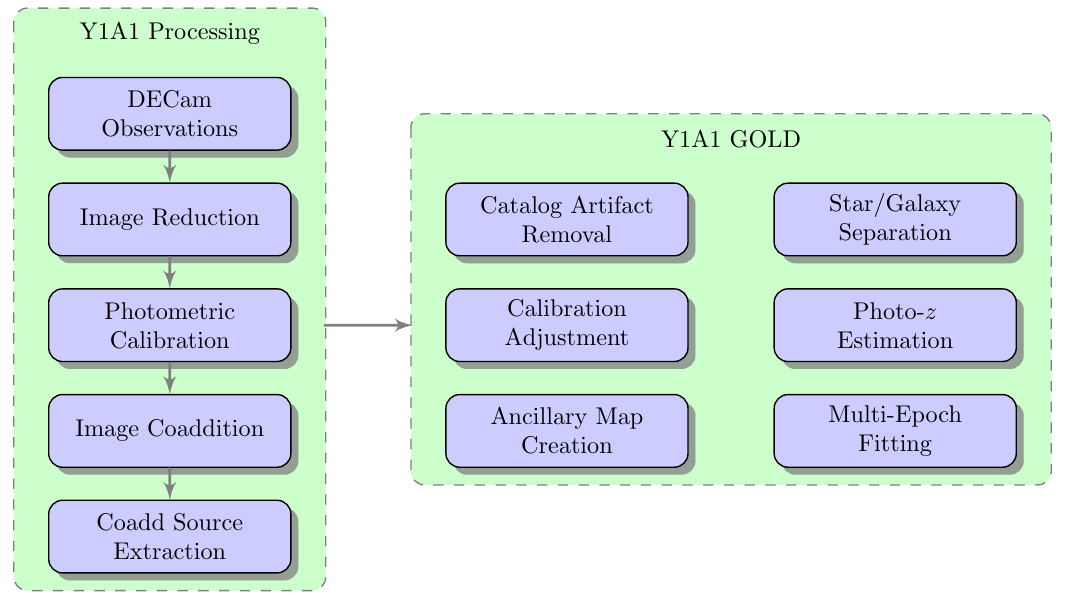}
\caption{\label{fig:flowchart} 
Schematic of the constituents of the Y1A1 processing (left) and the additional \gold products (right).
} 
\end{figure*}

Here we describe the creation, composition, and validation of the DES first-year (Y1) data set in support of cosmological analyses (shown schematically in \figref{flowchart}).
While this data set is currently proprietary to the DES Collaboration, this document is intended to serve as a reference for these data products when they become publicly available.\footnote{Note that the DES Y1 cosmology data set described here is distinct from the forthcoming DES public data release, which will include data from the first 3 years of DES.}
Observing for DES Y1 spanned from 2013 August to 2014 February and covered $\roughly 40\%$ of the DES footprint, averaging three to four visits per band.
The resulting images were processed through the DES data management (DESDM) system \citep{Ngeow:2006,Mohr:2008,Sevilla:2011,Mohr:2012,Desai:2012,Morganson:2017} an assembled into the DES year-one annual data set (Y1A1). 
Y1A1 consists of reduced single-epoch images and object catalogs (known as ``\finalcut''), along with multi-epoch coadded images and associated multi-band catalogs (known as ``\coadd'').
Photometric calibration of Y1A1 was performed globally on a CCD-to-CCD basis, and maps of the survey coverage and depth were assembled with the \mangle software suite \citep{Hamilton:2004,Swanson:2008}.
The Y1A1 data set covers $\roughly 2000 \deg^2$ in any single filter with inhomogeneous coverage and depth.
In total, $\roughly 1800 \deg^2$ of the Y1A1 footprint has simultaneous coverage in all five DES filters.

The desired precision of DES cosmological analyses motivates further refinement of Y1A1.
The resulting data set, referred to as \gold, is accompanied by extensive validation and ancillary data products to facilitate cosmological analyses.
The primary components of \gold are (\figref{flowchart}): 
(1) a multi-band photometric object catalog subselected from the \coadd object catalog; 
(2) an adjusted photometric calibration to improve uniformity over the survey footprint; 
(3) shape and photometry information from a simultaneous multi-epoch, multi-object fit;
(4) a set of ancillary maps quantifying survey characteristics using the HEALPix rasterization scheme \citep{Gorski:2005}; and
(5) several value-added quantities for high-level analyses (\ie, a star-galaxy classifier and \photoz estimates).
When creating the \gold object catalog, several classes of non-physical, spurious, or otherwise problematic objects were identified and either flagged or removed.
The calibrated magnitudes of objects were also corrected for interstellar extinction using a stellar locus regression (SLR) technique.
The ancillary data products associated with \gold accurately quantify the characteristics of the survey, further mitigating the impact of systematic uncertainties.
A high-level summary of the performance of \gold is tabulated in \tabref{summary}.
 
Our purpose here is to document the production and performance of the \gold data set in support of upcoming DES cosmology analyses.
We start by describing the DES Y1 observations in \secref{observations} and briefly reviewing the image reduction pipeline applied to produce the Y1A1 data set in \secref{desdm}. 
In \secref{calibration} we describe the photometric calibration of the Y1A1 data, and in \secref{coadd} we describe the image coaddition process.
In \secref{catalog} we discuss the creation of unique object catalogs, while in Sections \ref{sec:maps} and \ref{sec:vac} we describe the ancillary maps and value-added quantities produced to complement the \gold catalog. 
We briefly conclude in \secref{discussion}.

% Summary table
\begin{\tabletype}{l c c c c c c}
\tablewidth{0pt}
\tabletypesize{\tablesize}
\tablecaption{\gold Data Quality Summary\label{tab:summary}}
\tablehead{
Parameter & \multicolumn{5}{c}{Band} & Reference \\
 & $g$ & $r$ & $i$ & $z$ & $Y$ & 
}
\startdata
Median PSF FWHM & 1.25\arcsec & 1.07\arcsec & 0.97\arcsec & 0.89\arcsec & 1.07\arcsec & \secref{borismaps} \\
Sky Coverage (in all bands) & \multicolumn{4}{c}{$1786 \deg^2$} & $1773 \deg^2$ & \secref{footprint} \\ 
Astrometric Accuracy & \multicolumn{5}{c}{$25 \mas$ (relative); ${<}\,300 \mas$ (external)} & \secref{astrometry} \\ 
Absolute Photometric Uncertainty (mmag) & 14 & 4 & 2 & 15 & 32 & \secref{cal_acc} \\ 
Relative Photometric Uniformity (mmag) & 19 & 22 & 20 & 20 & 18 & \secref{cal_acc} \\ 
Completeness Limit (95\%) & $23.6$ & $23.4$ & $22.9$ & $22.4$ &  & \secref{goldcat} \\ 
Coadd Galaxy Magnitude Limit ($10\sigma$)\tablenotemark{a} & $\maglimg^{+0.14}_{-0.40}$ & $\maglimr^{+0.13}_{-0.37}$ & $\maglimi^{+0.14}_{-0.34}$ & $\maglimz^{+0.12}_{-0.37}$ & $\maglimY^{+0.18}_{-0.33}$ & \secref{depth} \\ 
Multi-Epoch Galaxy Magnitude Limit ($10\sigma$)\tablenotemark{a} & $\moflimg^{+0.07}_{-0.40}$ & $\moflimr^{+0.16}_{-0.29}$ & $\moflimi^{+0.14}_{-0.30}$ & $\moflimz^{+0.14}_{-0.32}$ & \ldots & \secref{depth} \\ 
Galaxy Selection ($i \leq 22$) & \multicolumn{5}{c}{Efficiency $>98\%$; Contamination $<3\%$} & \secref{sgsep} \\  
Stellar Selection ($i \leq 22$) & \multicolumn{5}{c}{Efficiency $>86\%$; Contamination $<6\%$} & \secref{sgsep} \\ 
%Photo-z Accuracy & \multicolumn{5}{c}{ \CHECK{$\Delta z_{50} = -0.027 \pm 0.008$} } & \secref{photoz} \\
\enddata
%\tablecomments{Values for the astrometry and photometric uniformity represent the mode of the distribution calculated over the \gold footprint.}
\tablenotetext{a}{The quoted values correspond to the mode, 16\textsuperscript{th} percentile, and 84\textsuperscript{th} percentiles of the magnitude limit distribution. Using the median instead of the mode reduces the magnitude limit by $\roughly 0.05$ mag.}
\end{\tabletype}

\section{Data Collection}
% (Drlica-Wagner, Diehl, Neilsen)
\label{sec:observations}

% ADW: Pre-survey started on Aug 15; main survey started on Aug 31.
DES has been allocated 105 nights per year on the Blanco telescope starting in 2013.
The first year of DES observing spanned from 2013 August 31 to 2014 February 9 and consisted of both full and half nights.\footnote{Several exposures taken during engineering time earlier in 2013 August were also included in the the Y1A1 data set.}
Details on DES operation and data collection are provided by \citet{Diehl:2014}; here we briefly summarize some of the key details relevant to the creation of \gold. 
 
\begin{figure*}[t]
\center
\includegraphics[width=0.8\textwidth]{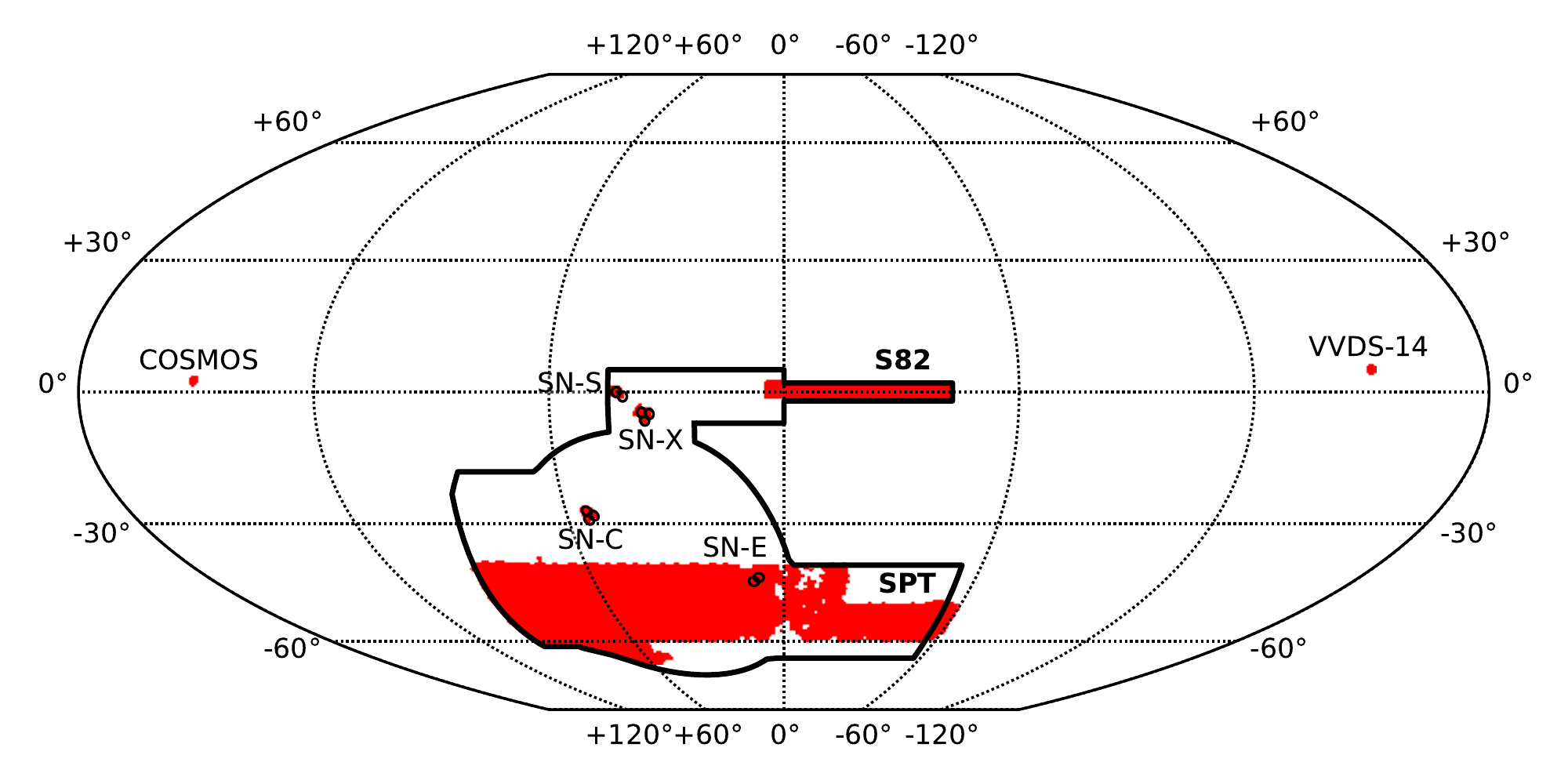}
\caption{\label{fig:footprint} 
DES \gold sky coverage in celestial coordinates (red) plotted in McBryde-Thomas flat polar quartic projection. Specific regions of \gold footprint, including the SN and auxiliary fields, are explicitly labeled (see \secref{observations}). The nominal DES five-year footprint is outlined in black.
} 
\end{figure*}

DES consists of two observing programs: a shallower wide-area survey and a deeper time-domain (``supernova'' or ``SN'') survey (\figref{footprint}). 
The DES wide-area survey footprint covers $\roughly 5000 \deg^2$ with 90s exposures in $griz$ and 45s exposures in $Y$.  
A single imaging pass over this footprint, called a ``tiling'', collects science data over $\roughly 83\%$ of the survey footprint owing to inefficiencies in the pointing layout and camera footprint (\eg, area not covered owing to gaps between CCDs, nonfunctioning CCDs, and problematic area near the edges of the CCDs). 
The DECam pointings for each tiling are shifted relative to each other by a large fraction of the camera field of view in a dither pattern designed to maximize uniformity and distribute repeated detections of a given object over the focal plane. 
During Y1, DES observed $\roughly 2000\deg^2$ of the wide-area survey footprint with three to four dithered tilings per filter. 
The Y1 footprint consisted of two areas: one near the celestial equator including Stripe 82 \citep[S82;][]{Annis:2011}, and a much larger area that was also observed by the South Pole Telescope \citep[SPT;][]{Carlstrom:2011}. 
During Y1, DES collected 17,671 wide-area survey exposures in a variety of observing conditions \citep{Diehl:2014}.
 
The SN survey observes 10 fields in four filters ($griz$) on a regular cadence to detect and characterize supernova through difference imaging \citep{Kessler:2015}. 
Longer exposure times ($\geq 150\second$) and frequent repeated visits result in a significantly deeper survey in the SN fields.
All 10 SN fields reside within the DES wide-area footprint, but only two were covered by wide-area imaging in Y1 (\figref{footprint}). 
Over the course of Y1, DES collected a total of 2699 time-domain survey exposures. 
 
In addition to the wide-area and SN survey fields, two auxiliary fields outside the DES footprint were observed to aid in the training of photometric redshifts and star-galaxy classification. 
Fields overlapping with COSMOS \citep{Scoville:2007} and VVDS-14h \citep{LeFevre:2005} were observed during the DES Science Verification (SV) period.%
\footnote{Data from the DECam Science Verification period is available at: \url{https://des.ncsa.illinois.edu/releases/sva1}.}
These observations are deeper than most of the Y1 wide-area survey.
 
During DES operation, sets of biases and flat-field calibration exposures were taken in each filter before each night of observing. 
Standard-star fields were observed at three different airmasses at the beginning and end of each night unless conditions were obviously nonphotometric.%
\footnote{On DES half nights only two standard-star fields were observed at the midpoint of the night.} 
Cloud cover was monitored continuously during observing by the RASICAM all-sky infrared camera \citep{Lewis:2010,Reil:2014}.
 
DES images the sky whenever weather allows the Blanco dome to be open, resulting in some exposures being taken in very poor conditions. 
Thus, data quality monitoring is essential to select exposures that meet the scientific requirements of the survey.
The quality of exposures is evaluated based on the PSF, sky brightness, and sky transparency.
For each exposure, we define \teff to be the ratio between the actual exposure time and the exposure time necessary to achieve the same signal-to-noise ratio (S/N) for point sources observed in nominal conditions \citep{Neilsen:2015}.\footnote{The effective exposure time is defined as $\Teff = \teff T_{\rm exp}$, where $T_{\rm exp}$ is the shutter-open exposure time.}
To pass preliminary data quality cuts, wide-area survey exposures must have $\teff > 0.3$ in $r$, $i$, and $z$ band and $\teff > 0.2$ in $g$ and $Y$ band. 
The median measured \teff for Y1 was $\teff = 0.75$ in the $r$, $i$ and $z$ band and $\teff = 0.49$ in the $g$ and $Y$ band.
In contrast, the preliminary data quality cuts for SN exposures require ${\rm FWHM} < 2''$ and that a 20th-magnitude simulated source have signal-to-noise ratio $>$20 ($>$80) for the shallow (deep) exposures \citep{Kessler:2015}.
Of the exposures taken during Y1, 82\% of the wide-area exposures and 95\% of the SN exposures passed their respective data quality criteria (\ie, did not require re-observation).
A number of additional exposures were removed from the Y1 data set due to instrumental artifacts (scattered light and internal reflections from bright stars, contaminating light from airplanes, poor telescope tracking, shutter malfunctions, dome occultations, \etc). 
In total, the \finalcut processing consists of 16,857 DECam exposures, including wide field, SN, auxiliary fields, and standard stars.
 
\section{Image Processing}
\label{sec:desdm}
% (Drlica-Wagner, Yanny, Gruendl, Bernstein)
% See also: http://data.darkenergysurvey.org/aux/releasenotes/DESDMrelease.html

The DESDM system is responsible for reducing, cataloging, and distributing DES data.
Earlier iterations of the DESDM image processing pipeline are outlined in \citet{Sevilla:2011} and \citet{Mohr:2012}, and a more detailed summary with updates for the forthcoming DES three-year processing is available in \citet{Bernstein:2017} and \citet{Morganson:2017}. 
Here we briefly summarize the single-epoch image processing steps applied during the DES \finalcut campaign.
The \finalcut campaign resulted in $\roughly 20$ TB of processed images and a catalog of $\roughly 740$ million detected objects.

\begin{figure}[th]
\center
\includegraphics[width=0.9\columnwidth]{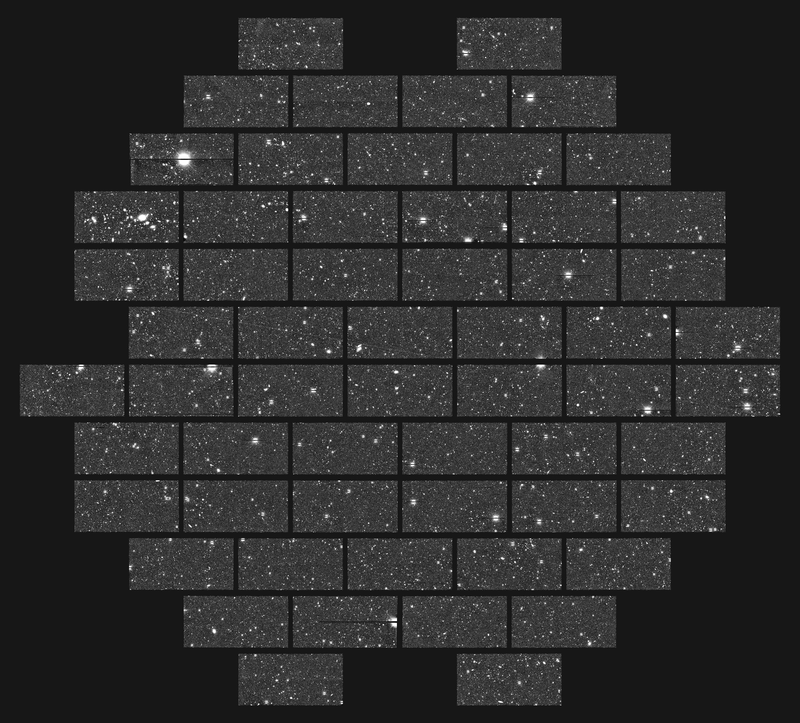}
\includegraphics[width=0.9\columnwidth]{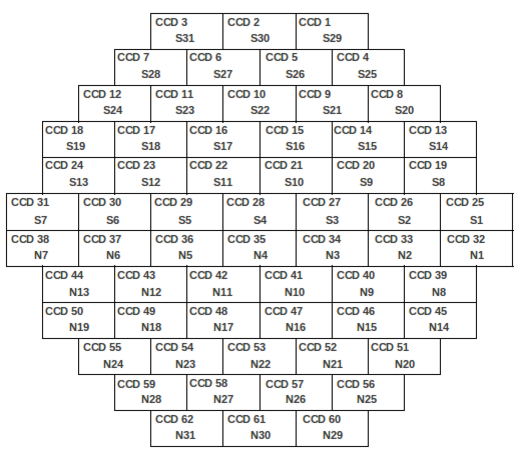}
\caption{\label{fig:decam} 
Processed DECam image from Y1A1 (top) and CCD layout (bottom). 
The three empty slots in the DECam image correspond to CCD2, CCD31, and CCD61. 
CCD61 failed during SV, while CCD2 failed part way through Y1. 
One amplifier of CCD31 has time-variable low-light-level nonlinearity and this CCD was not processed for Y1A1.}
\end{figure}
 
\begin{enumerate}
\item {\it Overscan and Crosstalk:} 
Each DECam CCD has two amplifiers for converting photo-carrier counts to analog-to-digital units (ADU).
For each amplifier, the average in the overscan region was calculated and subtracted on a row-by-row basis.  
Crosstalk is manifested as low-level leakage of electronic signals between different readout amplifiers and is observed at the level of $\roughly10^{-3}$ for pairs of amplifiers on the same CCD and $\roughly 10^{-4} - 10^{-6}$ for pairs of amplifiers on different CCDs on the same electronic back plane. 
Crosstalk was corrected by applying a matrix operation to the simultaneous readout of 140 amplifiers (including the amplifiers for the eight focus and alignment CCDs). 
The elements of the crosstalk correction matrix were derived from the median amplifier output for each ``victim'' channel as a function of the ``source'' amplifier signal for a large number of sky images.
Crosstalk between the DECam CCDs is found to be nonlinear when the signal on the source amplifier exceeds its saturation level -- i.e., the level at which the amplifier response becomes nonlinear \citep[Figure 2 of][]{Bernstein:2017}.
This crosstalk nonlinearity was incorporated into crosstalk correction.
There is no evidence for temporal variation in the crosstalk between CCDs on year timescales, and a single crosstalk matrix was used for the Y1A1 processing.
 
\item {\it Bias Correction:}
A master bias frame was constructed from the average of $\roughly 100$ zero-second exposures taken during the pre-night calibration sequences over the course of the Y1 observing season.  
This master bias was subtracted from each CCD to remove any residual fixed pattern noise not incorporated by the overscan correction.
 
\item {\it Bad-Pixel Masking:}
Bad pixel masks were created for each CCD by identifying outliers in sets of biases and $g$-band flat-field calibration exposures.
These bad pixels were masked and interpolated based on values in adjacent columns. 
The Y1A1 processing campaign used a single static bad pixel mask.
Two CCDs have failed since commissioning and were removed from Y1 processing \citep[\figref{decam};][]{Diehl:2014}.  
CCD61 failed on 2012 November 7 and data from this CCD were not used in Y1. 
CCD2 failed on 2013 November 30 and data from this CCD were only included for the early months of Y1.%
\footnote{CCD2 subsequently recovered on 2016 December 29.}

\item {\it Nonlinearity Correction:}
Several ($\roughly 10$) CCD amplifiers have a nonlinear response at low light levels (generally below 300 ADU/pixel). 
For DES, this affects the sky level in short ($\roughly 15\second$) standard-star observations and wide-survey dark-sky $g$-band observations ($90\second$). 
For most other filters/exposure times, the night sky alone is enough to give a sufficient number of counts per pixel to make the nonlinearity correction negligible. 
The nonlinearity effect can be several percent at very low light levels.
 At very high light levels ($> 2\times 10^{4}$ ADU/pixel), there is also a small nonlinear behavior ($\lesssim 2\%$). 

We corrected for nonlinearity at both low and high light levels using a fixed look-up table derived from calibration exposures obtained during the SV period. 
One amplifier on CCD31 has a time-variable nonlinear gain at the 20\% level, and this CCD was excluded from Y1 processing.%
\footnote{The other amplifier on CCD31 is stable and has been recovered in more recent processing.}
The Y1A1 data processing did not correct for charge-induced pixel shifts (\ie, the ``brighter-fatter'' effect; \citealt{Antilogus:2014}; \citealt{Gruen:2015}), although corrections have been incorporated into more recent reductions of the DES data \citep{Bernstein:2017}.
 
\item {\it Pupil Correction:}
An additive correction was applied for pupil ghosting in each exposure. 
As part of this process, ``star flats'' were created in each filter and CCD by taking multiple dithered exposures of a dense stellar field and fitting a cubic polynomial to variations in the observed brightnesses of stars.
The pupil ghost correction was constructed on a CCD by CCD basis for each exposure from the star flat and the level of sky background (including scattered light and the night-sky pupil image).
The pupil correction was scaled and subtracted from each CCD individually. 
This technique can leave gradients of several percent in the sky background level (worst in $z$ and $Y$ band), which propagate into the reduced science images and are corrected during photometric calibration (\appref{calibration}).%  
\footnote{More recent implementations of the data processing pipeline fit the additive correction over the full focal plane rather than CCD by CCD \citep{Morganson:2017}.}
 
\item {\it Flat Fielding:} 
The response of DECam to the night sky is more stable than nightly variations in the illumination of the flat-field screen taken during pre-night calibrations.
Therefore, in Y1A1 we created a single average flat-field frame for each filter from $\roughly 100$ individual flat-field exposures. 
The science exposures were divided by the average flat-field frames normalized on a CCD by CCD basis.
The pupil and flat-field corrections used for Y1A1 processing remove small-scale fluctuations in the background due to pixel-size variations, \ie, tree rings, edge brightening, and tape bumps \citep{Plazas:2014}.  
However, this correction is approximate and results in photometric measurement residuals at the level of $\roughly 0.5\%$.
A more rigorous correction has been applied in subsequent DES data reductions \citep{Bernstein:2017,Morganson:2017}.
 
\item {\it Weight Plane Creation:}
A weight plane image was created containing the inverse variance of the flat-fielded image value in each pixel.  
The variance estimate summed the expected Poisson noise and read noise. 
Saturated pixels were flagged and set to zero in the weight plane.
The weight plane is used to assign relative weights to images during the coaddition process.
% CTS = [e-]; GAIN = [e-/ADU]; IMG = CTS/GAIN = [ADU]; RDN = [e-]
% WGT = 1 / ( Var(CTS) + (RDN)**2 ) This is Y3A1
%    -> 1 / ( Var(CTS)/GAIN**2 + (RDN / GAIN)**2 )
%    -> 1 / ( IMG/GAIN + (RDN/GAIN)**2 ) This is Y1A1
 
\item {\it Fringe Frame Correction:}
Fringing is visible in $z$- and $Y$-band exposures.
The fringing pattern is nearly identical in these bands but has a larger amplitude in the $Y$ band.
A set of templates was constructed from a stack of $\roughly 120$ $z$- and $Y$-band exposures from DES SV.
These template images were median filtered and averaged on a pixel-by-pixel level to construct a fringe frame. 
In the reduction pipeline, each CCD of the $z$- and $Y$-band exposures had its median sky level measured, and this sky level was used to scale the fringe frame, which was then subtracted on a CCD-by-CCD basis. 
The scaling method was identical to that used to scale and subtract the pupil pattern.
The vast majority of exposures have a fringe residual that is $< 0.1\%$ of the sky background level. 
Exposures taken under the brightest conditions accepted for $Y$-band observing can have a fringe residual that is $\sim 0.4\%$ of the sky background level.
 
\item {\it Illumination Correction:}
Light reflected from the flat-field screen fills the telescope pupil differently than the focused light of distant stars.
To account for pixel-level differences in the throughput of the flat-field images, we applied a multiplicative correction to the DECam response based on the star flats.
After dividing by the star flats the residual difference in response between CCDs is typically $< 2\%$ peak to peak (\appref{psm}).

\item {\it Preliminary Astrometric Solution:}
A world coordinate system (WCS) was installed in the image header at the time of observation using a fixed distortion map derived from the star flats and an optical axis read from the telescope encoders. 
The pointing of each image was updated matching the centers of bright stars measured with \SExtractor \citep{Bertin:1996,Bertin:2002} to the UCAC-4 catalog using \scamp \citep{Bertin:2006}.
This WCS was replaced by a superior one during the coaddition step (\secref{coadd}), and the astrometric accuracy of the \gold catalog is described in \secref{astrometry}.

\item {\it Artifact Removal:}
Bright stars ($\lesssim 16 \magn$) saturate the $90\second$ DES exposures in $griz$.
Saturated pixels are set to zero in the image weight map plane.
Brighter stars can produce charge overflow into pixels in the CCD readout direction. 
These overflow pixels are flagged in the mask plane, zeroed in the weight plane, and interpolated in the image plane. 
In addition, corresponding pixels on the victim amplifier of the CCD are masked owing to large nonlinear crosstalk.
Extremely bright oversaturated stars can leave a secondary charge overflow in the readout register of the amplifier, conventionally called ``edge bleeds'' (see Fig.~5 in \citealt{Bernstein:2017}).
Edge bleeds can be located some distance from the bright star and are strongest in the rows near the readout register.  
These rows are identified and masked.

Energy deposited from cosmic-ray interactions with the CCDs were detected on single images using the \code{findCosmicRays} algorithm adopted from the LSST software stack.\footnote{\url{https://lsst-web.ncsa.illinois.edu/doxygen/x_masterDoxyDoc/namespacelsst_1_1meas_1_1algorithms.html}}
The cosmic-ray pixels were flagged in the mask plane, zeroed in the weight plane, and interpolated in the image plane.
 
Long streaks produced by rapidly moving objects (\ie, meteors and Earth-orbiting satellites) were detected using a Hough transform algorithm and were also masked \citep{Melchior:2016}. 
 
\item {\it Single-Epoch Catalog Creation:}
Object catalogs were produced for each CCD using the \code{AstrOmatic} package \citep{Bertin:2006}. 
Photometric fluxes were derived using \psfex and \sextractor for fixed apertures, the PSF model, and a galaxy model. 
The local sky background on each CCD was estimated by \sextractor.
The single-epoch \finalcut catalogs served as an input into the photometric calibration described in \secref{calibration}.
 
\end{enumerate}
 
\section{Photometric Calibration}
% (Tucker, Drlica-Wagner, Rykoff, ...)
\label{sec:calibration}
 
The photometric calibration of Y1A1 was a multistep process largely following the procedure of \citet{Tucker:2007}.
Photometric calibration was performed on the single-epoch \finalcut images first on a nightly and then on a global basis.  
An additional calibration adjustment was derived from the stellar locus and applied at catalog level.
Below we briefly describe the steps in the photometric calibration of Y1A1; a more detailed discussion of the Y1A1 photometric calibration can be found in \appref{calibration}.

\subsection{Nightly Photometric Calibration}
\label{sec:psm}
 
A preliminary photometric calibration of the Y1A1 data was performed on a nightly basis.
Standard-star fields were observed at various airmasses at the beginning and end of each night.
These images were reduced and the centroids of stars were matched to a set of {\em primary} standard stars from Sload Digital Sky Survey (SDSS) DR9 \citep{Smith:2002}.
The DES {\em secondary} standards were then transformed to an initial DES AB photometric system via a set of transformation equations derived from SDSS DR9 and supplemented by UKIDSS DR6 (\appref{transform}).
This tied the DES flux calibration of the secondary standards to SDSS and to the AB magnitude system \citep[\ie,][]{Padmanabhan:2008}.

The transformed nightly standards were used to fit a set of nightly photometric equations to model the spatial and temporal dependence of the DECam instrument throughput (Equations (\ref{eqn:psm_g})--(\ref{eqn:psm_Y})).
These equations track the accumulation of dust on the Blanco primary mirror, the relative throughput of the atmosphere at CTIO, and variations in the throughput and shape of the filter response at the location of each CCD.
The nightly photometric equations produce an initial photometric calibration for all exposures taken on photometric nights.
The relative calibration scatter for the nightly solution on a typical photometric night is $\roughly 0.02 \magn$ rms.
This nightly photometric calibration was used to anchor the relative global calibration of non-photometric exposures described in the next section.
A more detailed description can be found in \appref{psm}.

\subsection{Global Calibration}
\label{sec:gcm}

We implemented a global calibration module (\GCM) to derive calibrated zeropoints for all exposures, including those taken under non-photometric conditions, and to improve on the relative calibration accuracy achieved by the nightly photometric solution.
The \GCM procedure follows that of \citet{Glazebrook:1994} and is described in more detail in \appref{gcm}.
Briefly, the Y1A1 data were split into regions of contiguous, overlapping images where at least one image had been previously calibrated. 
The calibrated images served as a reference against which other images in the grouping were calibrated.
To be calibrated by the GCM, an image needed to either overlap a calibrated image or have an unbroken path of overlapping images to a calibrated image.
 
\begin{figure*}[t]
\center
\includegraphics[width=0.85\textwidth]{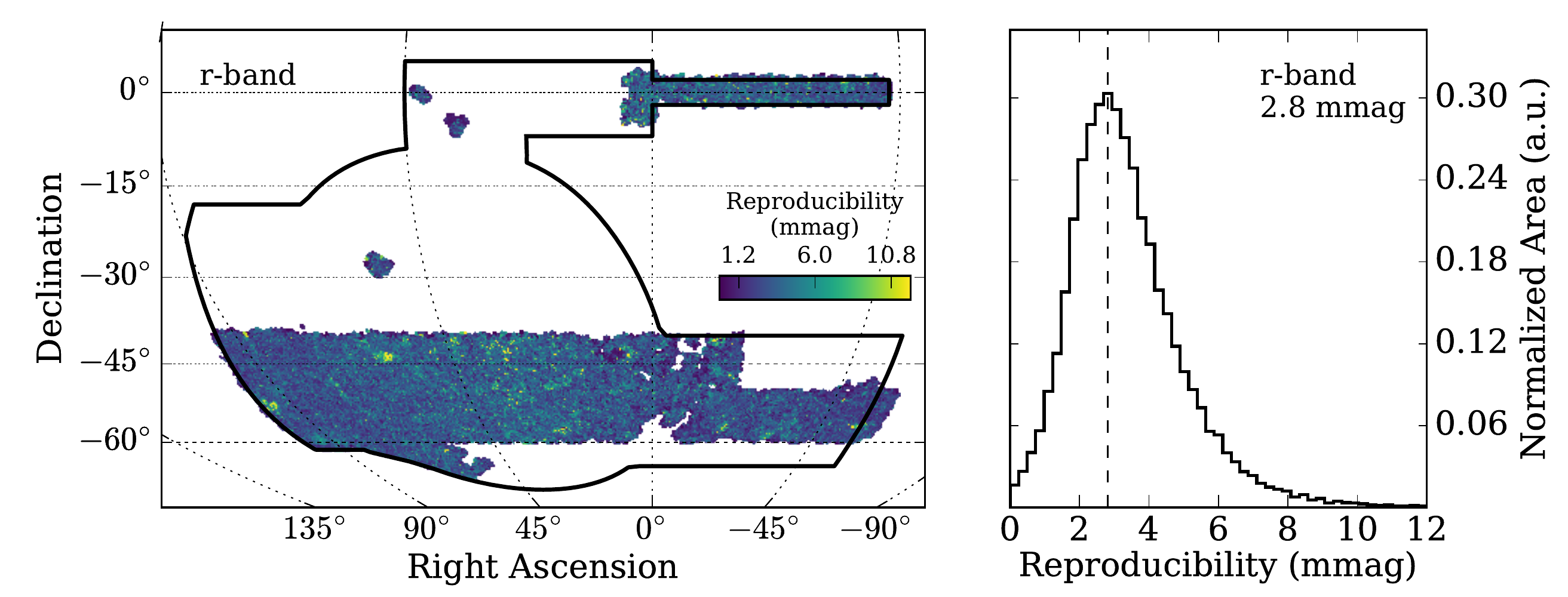}
\caption{\label{fig:gcm_zprms_r} Internal reproducibility uncertainty for the Y1A1 $r$-band photometric zeropoints calculated by comparing the rms calibrated magnitudes of stars in overlapping CCDs.
The mode of the rms internal calibration uncertainty is $2.8~\mmag$.
Similar figures for other bands are shown in \appref{gcm}.
}
\end{figure*}

Following the prescription of \citet{Glazebrook:1994}, we estimate the rms magnitude residual for each CCD image from overlap with other CCD images.
The rms distribution over all CCD images is a measure of the internal reproducibility uncertainty on small scales (the scales of overlapping CCD images) and is a measure of the precision of the overall \GCM solution.  
We find the rms to be $\roughly3\mmag$ (\figref{gcm_zprms_r}).
This uncertainty is relevant when analyzing light curves of variable objects but does not represent the internal consistency/uniformity of the relative calibrations on large scales.

While the \GCM method is very precise, small systematic gradients in the flat fields of individual images can cause low-amplitude gradients over large scales.  
To ``anchor'' the fit against large-scale gradients, we used the set of nightly secondary standard stars and a sparse grid work of {\it tertiary} standard stars observed under photometric conditions and calibrated by the nightly photometric equations.
The tertiary standards were chosen such that they would anchor the global solution on scales $>10$--$15\deg$, but on smaller scales the calibration would be dominated by the solution from overlapping uncalibrated exposures.
We further examine the uniformity and absolute calibration accuracy (relative to the AB system) in \secref{slr} and \secref{cal_acc}.
 
\subsection{Photometric Calibration Adjustment}
\label{sec:slr}

The global calibration is found to be uniform at the $\roughly 2\%$ level in each band over the majority of the Y1A1 survey footprint (discussed in \secref{cal_acc}).
However, non-uniformity in the colors of objects can severely impact DES science by introducing a spatial dependence on object selection and \photoz estimation. 
The SLR technique uses the distinct shape of the stellar locus in color-color space to provide a relative calibration of exposures in different bands \citep[\eg,][]{Ivezic:2004a, MacDonald:2004a,High:2009a,Gilbank:2011a,Desai:2012,Coupon:2012a,Kelly:2014a}.
To correct for residual spatial non-uniformity in the calibration and account for Galactic reddening (including uncertainties in the amplitude of reddening and possible variations in the effective Milky Way dust law), we have applied a secondary adjustment to the calibration of the coadd object catalogs derived from the stellar locus.
Gradients in stellar population are subdominant to other calibration uncertainties in Y1A1 given the DES filter bandpasses and high Galactic latitude of the survey \citep[e.g.,][]{High:2009a,Kelly:2014a}.
We followed the procedure of \citet{Drlica-Wagner:2015} and applied a modified version of the \code{BigMACS} SLR code \citep{Kelly:2014a}\footnote{\url{https://code.google.com/p/big-macs-calibrate/}} coupled with an empirical stellar locus to derive zeropoint adjustments to improve the color uniformity of stars across the Y1A1 footprint.
The SLR adjustment was tied to the $i$-band magnitude derived from the GCM, dereddened using the \citet[SFD;][]{Schlegel:1998} dust map with a reddening law from \citet{O'Donnell:1994}. 
The SLR zeropoint adjustments were interpolated to the positions of each object in the catalog and were applied directly to the magnitudes of objects derived from the coadded images.
In this way, the calibrated magnitudes of the \gold catalog are {\it already corrected for interstellar extinction}.
After the SLR adjustment, the color of stars was found to be uniform at the $\roughly 1\%$ level across the footprint, which was verified using the red sequence of galaxies.
More detail on the SLR calibration adjustment can be found in \appref{slr}.

\subsection{Photometric Calibration Accuracy}
\label{sec:cal_acc}
 
\begin{figure*}[t]
\center
\includegraphics[width=0.85\textwidth]{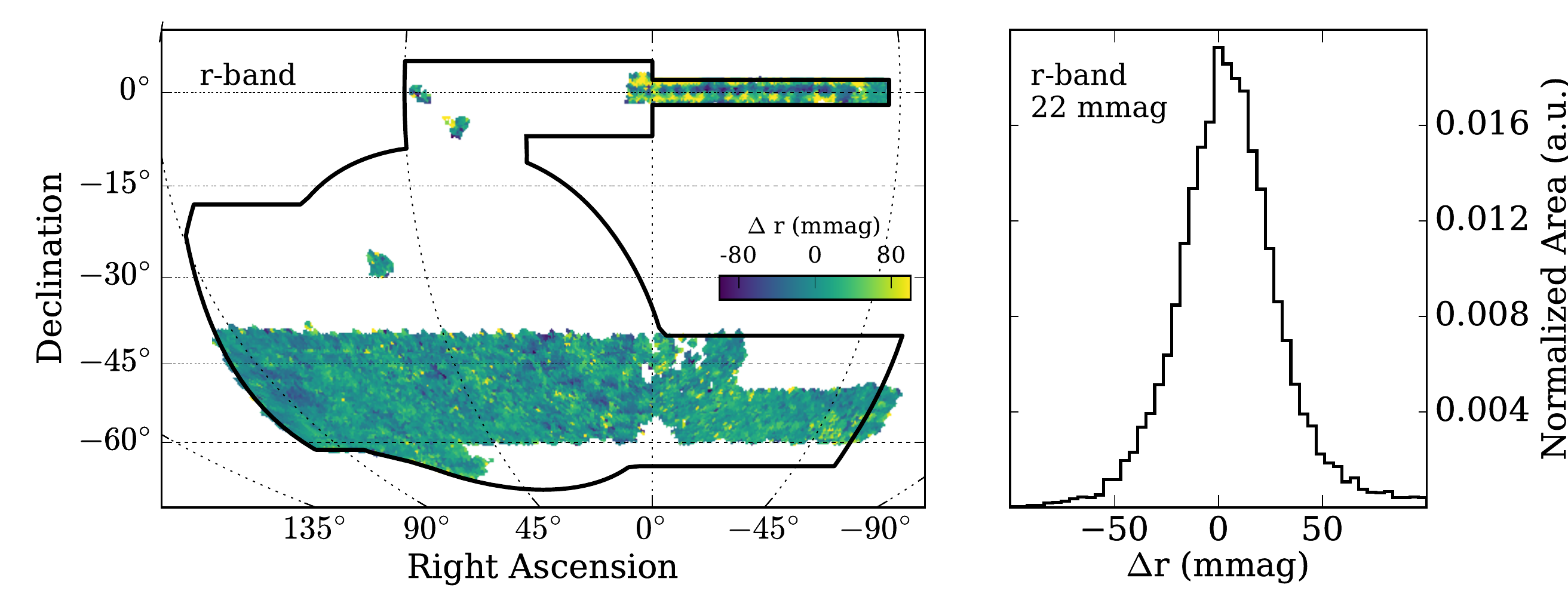}
\caption{\label{fig:apass2mass} Comparison of stellar magnitudes from the DES Y1A1 \GCM and those estimated from APASS/2MASS transformed into the DES filter system (\appref{transform}). 
The sky plot (left) shows the median magnitude offset for stars binned into $\roughly 0.2 \deg^2$ \healpix pixels.
The \GCM calibrated magnitudes are consistent with the transformed values from APASS/2MASS with a half-width of $\sigma_{68} = 22 \mmag$ (calculated between the 16th and 84th percentiles). 
Similar figures for other bands are shown in \appref{calibration}.
}
\end{figure*}
 
To quantify the accuracy of photometric calibration, we would like to characterize the statistical distribution of $\Delta m = m_{\rm meas} - m_{\rm true}$, where $m_{\rm meas}$ and $m_{\rm true}$ are the measured and true magnitude of catalog objects, respectively.
The characterization of the $\Delta m$ distribution can be split into two components: (1) an ``absolute'' calibration accuracy that represents a linear shift of the $\Delta m$ distribution (\eg, the mean of the distribution), and (2) a ``relative'' calibration accuracy that represents the spread of the $\Delta m$ distribution (\eg, standard deviation of the distribution).
In reality, values of $m_{\rm true}$ are not available, and we must make use of the calibrated magnitudes from other surveys or synthetic models, which have their own associated uncertainties.
We describe several calibration validation studies below and summarize the results in \tabref{calib_summary}.
 
The absolute calibration of the \gold is tied to SDSS through the DES secondary standard stars. 
As an independent cross-check on the absolute photometric calibration, we examined the CALSPEC standard star, C26202 \citep{Bohlin:2014}. 
We calculated synthetic magnitudes for C26206 by convolving {\it Hubble Space Telescope} (HST) spectra (\code{stisnic\_006})\footnote{\url{http://www.stsci.edu/hst/observatory/crds/calspec.html}} with the focal-plane-averaged DECam filter throughput including atmospheric attenuation at an airmass of 1.3 \citep{MODTRAN:1999}. 
% g=16.6947, r=16.3403, i= 16.2571, z = 16.2453, and Y= 16.2683.
The predicted magnitude of C26202 in each of the DES $grizY$ bands is $g=16.695$, $r=16.340$, $i= 16.257$, $z = 16.245$, and $Y=  16.268$. 
These predicted magnitudes were then compared against the pre-SLR corrected magnitudes measured by the GCM to give a ``top-of-the-atmosphere'' estimate of the absolute calibration uncertainty. 
We derive an absolute offset (in mag) of $\delta g=0.014$, $\delta r=0.004$, $\delta i=0.002$, $\delta z=0.015$, and $\delta Y=0.032$, which we quote as the absolute photometric calibration uncertainty in \tabref{summary}.
 
Our primary technique for quantifying the relative photometric accuracy of \gold is by comparing the calibrated magnitudes of stars against those derived from a combination of APASS \citep{Henden:2014} and 2MASS \citep{Skrutskie:2006} (\figref{apass2mass}).
We perform a ``top-of-the-atmosphere'' comparison by calculating the difference between the GCM calibrated magnitude and the APASS/2MASS magnitude transformed to the DES system (\appref{transform}).
We  derive the relative calibration uncertainty as the half-width between the 16th and 84th percentiles of the difference in magnitude over the footprint: $\sigma_{68}(g)=0.019$, $\sigma_{68}(r)=0.022$, $\sigma_{68}(i)=0.020$, $\sigma_{68}(z)=0.020$, and $\sigma_{68}(Y)=0.018$ (\tabref{summary}). 
These values include calibration uncertainties from both DES and APASS/2MASS, and are thus a conservative upper bound on the Y1A1 \GCM accuracy.
We further compare the SLR-adjusted \gold photometry to the transformed APASS/2MASS photometry dereddened using the SFD maps and reddening law of \citet{O'Donnell:1994}. 
We find a dispersion of $\sigma_{68}(g)=0.025$, $\sigma_{68}(r)=0.024$, $\sigma_{68}(i)=0.020$, $\sigma_{68}(z)=0.018$, and $\sigma_{68}(Y)=0.015$. 
These values include an additional contribution from differences in the reddening correction derived from the SLR and the SFD dust maps, which results in larger uncertainty in the bluer filters where interstellar reddening is more extreme.
These comparisons are shown in more detail in \appref{cal_acc}.
 
As an additional cross-check, we compared the ``top-of-the-atmosphere'' Y1A1 GCM calibration against a global calibration of the contiguous DES three-year data set (Y3A1).
The absolute calibration of the Y3A1 data set was also tied to C26202, but made use of additional observations of this object. 
From comparisons against other CALSPEC standards (LDS749B and WD0308-565), the absolute calibration of Y3 is believed to be accurate at the $\roughly 1\%$ level.
The relative calibration of Y3 was performed over the contiguous Y3A1 footprint using an independent forward global calibration method (\FGCM) and is found to be uniform at the $0.7\%$ level \citep{Y3FGCM}.
We checked the absolute calibration of \gold by matching stars against their Y3 counterparts over the \gold footprint. 
We found that the absolute offset between Y1A1 \GCM and Y3A1 \FGCM was $\delta g =0.023$, $\delta r <0.001$, $\delta i=0.004$, $\delta z=0.011$, and $\delta Y=0.05$, while the relative calibration spread was $\sigma_{68}(g)=0.014$, $\sigma_{68}(r)=0.007$, $\sigma_{68}(i)=0.008$, $\sigma_{68}(z)=0.013$, and $\sigma_{68}(Y)=0.015$. 
These numbers are in good agreement with those quoted above, and support the expectation that the relative calibration uncertainty in \tabref{summary} is a conservative estimate.

% Tables 
\begin{\tabletype}{l c c c c c}
\tabletypesize{\tablesize}
\tablewidth{1.2\columnwidth}
\tablecaption{Photometric Calibration Validation\label{tab:calib_summary}}
\tablehead{
Technique & \multicolumn{5}{c}{Band} \\ [-0.25em]
 & $g$ & $r$ & $i$ & $z$ & $Y$ \\ [-0.25em]
 & (mmag) & (mmag) & (mmag) & (mmag) & (mmag)
}
\startdata 
\multicolumn{6}{c}{\it Absolute Photometric Offset} \\
GCM \vs C26202 & 14 & 4 & 2 & 15 & 32 \\
GCM \vs Y3 FGCM & 23 & $<1$ & 4 & 11 & 50 \\ 
\tableline 
\multicolumn{6}{c}{\it Relative Photometric Uniformity} \\ 
GCM \vs APASS/2MASS & 19 & 22 & 20 & 20 & 18 \\
GCM+SLR \vs APASS/2MASS+SFD & 25 & 24 & 20 & 18 & 15  \\
GCM \vs Y3 FGCM & 14 & 7 & 8 & 13 & 15 \\ 
\enddata
\tablecomments{Summary of photometric calibration performance for the \gold data set. See \secref{cal_acc} for more details.}
\end{\tabletype}

\section{Image Coaddition}
\label{sec:coadd}
 
Image coaddition allows DES to detect fainter objects and mitigates the impact of residual transient imaging artifacts (\eg, unmasked cosmic rays, satellite streaks, \etc).
Combining multiple dithered exposures also positions objects at different points on the focal plane, mitigating systematics associated with the non-uniform response of the instrument.
 
DESDM produced image coadds from the weighted average of overlapping single-epoch images.
The pixels of the input images were remapped onto a uniform pixel grid using \swarp with the \code{LANCZOS3} kernel \citep{Bertin:2002,Bertin:2010}.  
The remapped pixel grid was defined on coadd \textit{tiles} spanning $0.73\deg \times 0.73\deg$ and comprising $10^4 \times 10^4$ remapped pixels (a pixel scale of $0\farcs263/{\rm pix}$, comparable to the physical pixel scale of DECam). 
For each tile, one coadded image was produced for each photometric band.

Before performing image coaddition, several image quality checks were run to identify and blacklist CCD images with severe imaging artifacts.
CCD images affected by strong scattered light artifacts were identified by a ray tracing algorithm using the Yale bright star catalog \citep{Hoffleit:1991}, the telescope pointing, and a detailed model of the DECam optics, filter changer, and shutter assemblies. % photometry from the Gunn-Stryker atlas
Several exposures have excess noise in one or more of the DECam CCD backplanes. 
These CCD images were identified through visual inspection and through the detection of a large number of spurious catalog objects.
In addition, CCD images that were affected by bright meteor trails and airplanes were identified through visual inspection.
Less than $1\%$ of CCD images were blacklisted and removed from the coadd process.
 
When DESDM created coadded images, the PSFs of the individual input images were not homogenized.
This decision was motivated by studies of SV data where PSF homogenization was found to produce correlated sky noise, which made it difficult to properly estimate the photometric uncertainties of galaxies.
While non-homogenized PSF coaddition yields better-behaved photometric uncertainties, it can introduce sharp PSF discontinuities on the $\roughly 0\fdg1$ scale that are difficult to model with conventional polynomial approximation techniques \citep[\ie, \PSFEx;][]{Bertin:2006}.
Some of these issues can be addressed by using quantities measured in the \finalcut catalog (\secref{catalog}); however, studies that depend sensitively on morphological characterization (\ie, weak lensing analyses) perform their own simultaneous fit of the individual single-epoch images (\secref{mof}).\footnote{Studies with PSF homogenization are ongoing, and PSF-homogenized coadds have been used for several DES science analyses using SV data \citep{Hennig:2017,Klein:2017}.}
 
In addition to the main survey, there are several regions where the DES Y1 imaging is considerably deeper than the nominal three to four tilings. 
Coadds have been created in these regions using different numbers of input images to achieve different photometric depths.
The \gold coadd catalog thus contains four different samples:
\begin{enumerate}
\item WIDE: The WIDE coadd data sample is built from exposures in the S82 and SPT regions of the Y1 wide-area survey footprint and has a depth of three to four tilings.
One of the SN fields, SN-E, resides within the SPT region; however, to maintain uniformity the WIDE data set only includes images that were taken as part of the DES wide-area survey (the SN-E exposures are included in the other data sets that follow). 
\item D04: The D04 sample is constructed by coadding images in the SN, COSMOS, and VVDS-14h fields with the goal of reaching an effective depth roughly comparable to the WIDE sample. 
Quantitatively, exposures were selected to give $\sum_{j}^{\rm exp} t_{{\rm eff},j} T_{{\rm exp},j} \simeq 4 T_{\rm wide}$, where $t_{{\rm eff},j}$ is the effective exposure time scale factor for exposure $j$ (\secref{observations}), $T_{{\rm exp},j}$ is the shutter-open time for exposure $j$, and $T_{\rm wide}$ is the wide-area exposure time in Y1 (90s in $griz$ and 45s in $Y$). 
When selecting exposures for the D04 and D10 samples, we attempted to apply data quality selections based on FWHM and $\teff$.
For the D04 sample, exposures in the $grizY$ bands were generally required to pass the wide-area survey data quality requirements (\secref{observations}) and have ${\rm FWHM}<1\farcs3$.
However, in several cases these requirements were relaxed to better approximate the desired depth.
While the D04 sample was designed to mimic the depth of the WIDE survey, the longer exposure times for the auxiliary and SN fields result in a data set that is on average $\roughly 0.2 \magn$ deeper than WIDE.
The median \var{MAG\_AUTO} $10\sigma$ limiting magnitude for galaxies (\secref{depth}) in the D04 sample is $g = 23.6$, $r=23.4$, $i=22.8$, $z=22.0$, $Y=20.3$.
The D04 data set has been used to train and test \photoz algorithms and object classification \citep[\eg,][]{Hoyle:2017}.
\item D10: The D10 sample is constructed in the SN, COSMOS, and VVDS-14h fields by coadding images to an effective depth of 10 exposures. 
The 10-exposure depth is intended to mimic the expected main survey depth at the end of DES. 
Similar to D04, general criteria requiring survey quality, ${\rm FWHM}<1\farcs3$ in $riz$ and ${\rm FWHM}<1\farcs4$ in $gY$ were applied.
The median \var{MAG\_AUTO} $10\sigma$ limiting magnitude for galaxies (\secref{depth}) in the D10 sample is $g = 24.2$, $r=24.0$, $i=23.5$, $z=22.7$, $Y=20.9$.
\item DFULL: The DFULL sample uses all high-quality images in the SN, COSMOS, and VVDS-14h fields.
The DFULL coadd applies a requirement of ${\rm FWHM}<1\farcs3$ in $riz$-band and ${\rm FWHM}<1\farcs4$ in $g$-band (no FWHM requirement is placed on $Y$-band). 
Exposures are still required to pass the survey quality cuts, but no restriction is placed on the number of exposures that go into the coadd.
The median \var{MAG\_AUTO} $10\sigma$ limiting magnitude for galaxies (\secref{depth}) in the DFULL sample is $g = 24.2$, $r=23.9$, $i=23.8$, $z=23.7$, $Y=21.2$, with $\sim 10\%$ of the area having a limiting magnitude greater than 25 in $griz$.\footnote{The median depth of the DFULL sample in $g$- and $r$-band is comparable to that of D10 owing to the fact that few additional exposures passed the survey quality and FWHM requirements outside of the deep SN fields.
The $r$-band depth is $0.05$ mag shallower in DFULL owing to a slightly larger area with more varied data quality.}
\end{enumerate}

\subsection{Astrometric Accuracy}
% (Drlica-Wagner, Yanny, Gruendl)
\label{sec:astrometry}
 
Astrometric calibration places the DES exposures onto a consistent reference frame with each other and with external catalogs.
We used \scamp \citep{Bertin:2006} to find an astrometric solution including corrections for optical distortion towards the edges of the focal plane.
During \finalcut processing, initial astrometric calibration was performed on individual exposures.
Starting with an approximate initial solution provided by the telescope control system, the \sextractor windowed image coordinates of bright stars in the DES exposures were extracted and matched against the UCAC-4 stellar catalog \citep{Zacharias:2013}.

\begin{figure*}[!t]
\center
\includegraphics[width=0.85\textwidth]{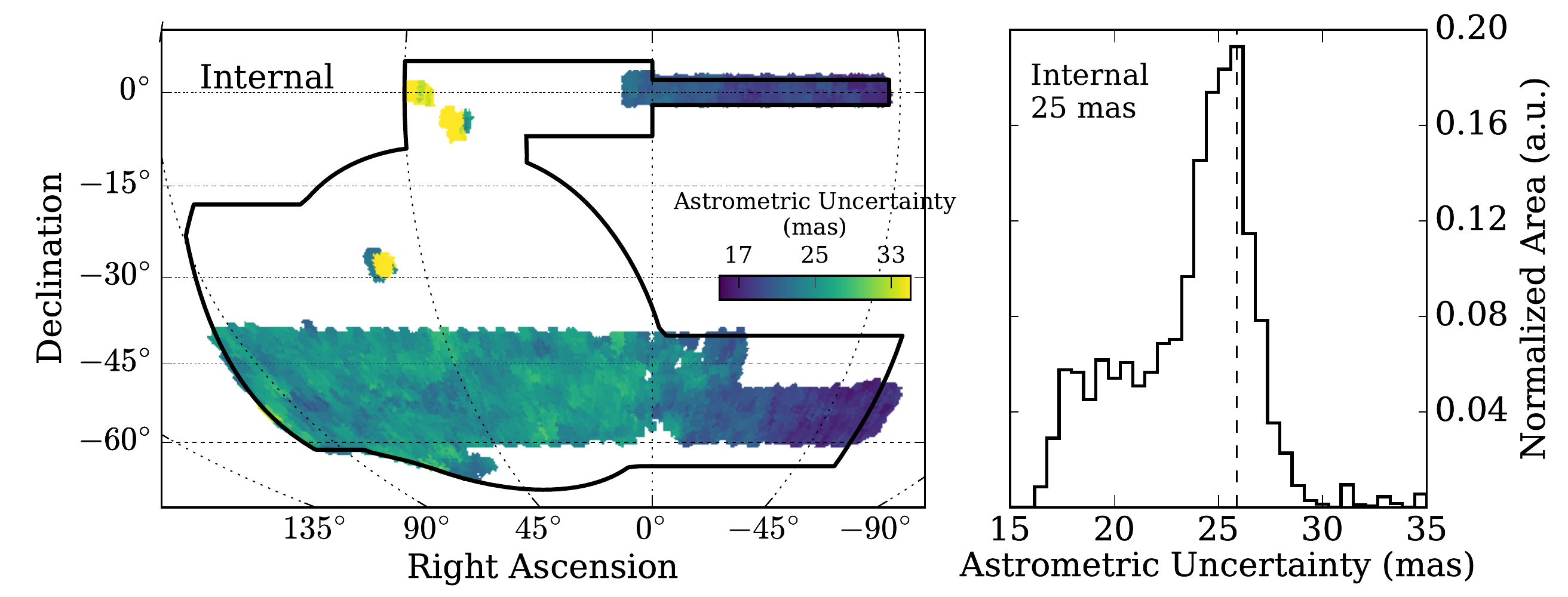}
\includegraphics[width=0.85\textwidth]{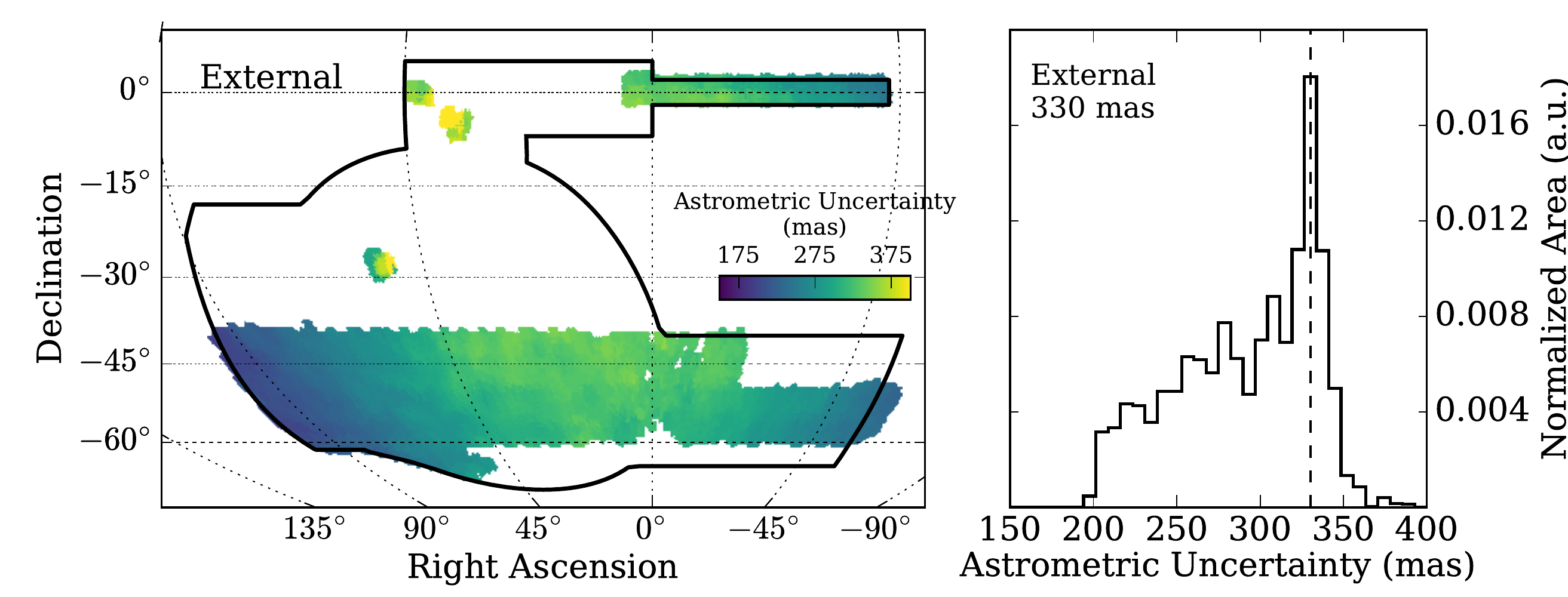}
\caption{
\label{fig:astrometry} 
(Top): Relative internal astrometric error in milliarcseconds derived by comparing the positions of stars in the individual DES exposures that go into the Y1A1 coadds. 
(Bottom): Relative external astrometric error derived by comparing the position of stars in DES and 2MASS (without correcting for proper motion). %average position?
The color scales represent the astrometric uncertainty in milliarcseconds, while the legends of the right panels report the modes of the distributions.
The SN exposure times are significantly longer than the wide-area survey exposure times leading to a fainter saturation threshold. 
This reduces the number of non-saturated bright stars and increases the astrometric uncertainty estimated by this technique (a more accurate estimate of the astrometry in the SN fields can be found in \citealt{Kessler:2015}).
} 
\end{figure*}

When building coadd tiles, an additional astrometric refinement process was performed to remap the DES input images against each other and against the 2MASS catalog \citep{Skrutskie:2006}.  
The single-epoch catalogs from all exposures overlapping a tile were input to \scamp, and a simultaneous best fit was obtained treating exposures from each filter as separate instruments.
This best-fit astrometric solution was used when combining images.
After astrometric refinement, the median internal astrometric precision of the Y1A1 wide-area coadd images is $\roughly 25 \mas$ ($3\sigma$-clipped rms dispersion around the mean for stars with $S/N > 100$).
In comparison, the median astrometric precision when compared against 2MASS is 200 -- 350$\mas$ (\figref{astrometry}).
This difference is dominated by the proper motions of high Galactic latitude stars and uncertainty in the astrometric accuracy of faint 2MASS sources.\footnote{\url{http://www.ipac.caltech.edu/2mass/releases/allsky/doc/sec2_2.html}}
This has been confirmed by comparisons between Y3 DES data and {\it Gaia} DR1 \citep{Gaia:2016} where the median astrometric uncertainty is found to be $\roughly 150 \mas$ \citep{DES:2018}.\footnote{\citet{Bernstein:2017a} show that using {\it Gaia} DR1 the astrometric solution for a single DECam exposure can be made accurate to within $3-6 \mas$.}

%While precise absolute astrometric calibration is not required for DES cosmology, we conclude that the comparison with 2MASS represents an upper limit on the absolute astrometric calibration of DES.\footnote{This has been confirmed in \citet{Bernstein:2017a} by comparing DES Y3 data against Gaia DR1 \citep{Gaia:2016}.}

\section{Object Catalogs} 
\label{sec:catalog}

\subsection{Coadd Catalog Creation}
%(Drlica-Wagner, Yanny, Sevilla)
\label{sec:coadd_catalog}

%RAG: Background subtraction was: BACK\_SIZE=256, BACK\_FILTERSIZE=3, BACKPHOTO\_TYPE=GLOBAL, BACK\_TYPE=AUTO, BACK\_VALUE=0.
%RAG: SExtractor run used: DETECT\_THRESH=1.5, DETECT\_MINAREA=6, ANALYSIS\_THRESH=1.5, DEBLEND\_MINCONT=0.001.   I believe this means the detection threshold would correspond to something like $1.5 \sqrt{6} \simeq 3.6\sigma$... (possibly modulo the convolution kernel?).

Catalogs of unique astrophysical sources were assembled from the coadded images.
The goal of the DESDM catalog production was to assemble the most inclusive catalog of sources while maintaining a low contamination fraction.
The production of catalog subsamples that are {\em complete} to a given threshold is left to subsequent science analyses.
Source detection, morphological characterization, and multi-band photometric flux measurements were performed using \SExtractor \citep{Bertin:1996,Bertin:2002}. 
Source detection used a \code{CHI-MEAN} combination of the coadded images in $r+i+z$ \citep{Szalay:1999,Bertin:2010}.
The \code{CHI-MEAN} detection image was designed to minimize discontinuities between regions with different numbers of exposures (see \appref{chimean}).
In contrast, flux and shape measurements were performed on each band individually using \sextractor in dual mode (\ie, analyzing the image for an individual band simultaneously with the detection image). The local background was estimated via $16 \times 16$ pixel boxes with $3\sigma$ clipping of bright pixels and median filtering of the boxes. The image was convolved with a $3 \times 3$ pixel structuring element of the form [[1,2,1][2,4,2][1,2,1]]. An S/N threshold of $1.5\sigma$ per pixel was applied over the convolved image to detect objects. Source localization was derived from the barycenter of the object in the $i,z,Y,r,g$ single-band coadd images (in order). 
Coadd object positions in world coordinates (J2000 epoch) were computed using the astrometric solution found during image coaddition (\secref{astrometry}).
 
The depth and PSF of the DES imaging result in overlapping isophotes for objects in crowded regions, \eg, galaxy clusters, star clusters, and dense stellar regions around the LMC. 
Incomplete deblending of overlapping objects affects the measured shapes and photometric properties of cluster galaxies, which impacts weak lensing and cluster cosmology science. 
\SExtractor attempts to deblend each detected object into sub-components using a multi-thresholding algorithm \citep{Bertin:1996}. 
An object is separated into two (or more) new objects if the intensity of the new object is greater than a fraction of the total intensity set by the \var{DEBLEND\_MINCONT} parameter, while the number of deblending thresholds is set by the \var{DEBLEND\_NTHRESH} parameter. 
The Y1A1 processing campaign adopts 0.001 and 32, respectively, for the two parameters. 
These values were optimized based on SV data to balance completeness and purity for cluster galaxies.
More aggressive deblending techniques for the DES data have been explored in \citet{Zhang:2015}.
 
\SExtractor was used to measure object photometry via several methods \citep[see][]{Sevilla:2011}.
\begin{enumerate}
\item Fixed aperture fluxes (\var{FLUX\_APER}) were measured for 12 circular apertures with different radii from $0\farcs25$ to $9\arcsec$.
\item Elliptical aperture fluxes (\var{FLUX\_AUTO}) were calculated using the second-order moments of each object to derive the elongation and orientation of the best-fit ellipse \citep{Kron:1980}. The ellipse scaling factor was derived from the first-order moment of the radial distribution.
\item PSF model fluxes (\var{FLUX\_PSF}) suitable for point-like sources were fit to the measured PSF shape. 
As mentioned in \secref{coadd}, PSF discontinuities in the Y1A1 coadd images can degrade the quality of the PSF model fluxes.
\item Exponential model fluxes (\var{FLUX\_MODEL}) suitable for galaxies were fit by convolving a one component exponential model with a local model of the PSF. These fluxes were fit both individually in each band and by fixing the model shape based on the detection image (\var{FLUX\_DETMODEL}).
\end{enumerate}

Among the morphological measurements performed by \SExtractor, two are designed to separate point-like objects (\ie, stars) from spatially extended sources (\ie, galaxies). 
The first is the \var{CLASS\_STAR} variable which uses a neural network to assess the ``stellarity'' of an object \citep{Bertin:1996}. The second variable, \var{SPREAD\_MODEL}, is derived from the Fisher's linear discriminant between a model of the PSF and an extended source model convolved with the PSF \citep{Desai:2012, Bouy:2013, Soumagnac:2015}. The application of these variables to star-galaxy separation is detailed in \secref{sgsep}.

As stated previously, catalog quantities were also derived for individual single-epoch exposures that compose the coadded images.
Objects detected on the individual exposures were associated with sources in the coadd catalog using a $1\arcsec$ matching radius.
While shallower, the single-epoch catalogs are important for probing the temporal domain.
Additionally, the photometry of the single-epoch catalogs is not subject to the PSF discontinuities present in the coadds.
For this reason, we calculated a number of photometric and morphological quantities from the average of single-epoch measurements weighted by their associated statistical uncertainties (the names of these quantities are prefixed by ``\var{WAVG}'').
In particular, the weighted-average spread-model quantity (\var{WAVG\_SPREAD\_MODEL}) has been shown to yield better star-galaxy separation \citep{Drlica-Wagner:2015} for stellar objects, and the weighted-average PSF magnitudes (\var{WAVGCALIB\_MAG\_PSF}) have been found to yield more precise stellar photometry than the corresponding coadd quantities.
In addition, uncertainties for the \var{WAVG} quantities are calculated directly from the variance in the measurements from individual exposures and thus avoid any systematics introduced in the coaddition process.

\subsection{Y1A1 GOLD Catalog Selection}
\label{sec:gold_catalog}

We assembled the \gold object catalog as a high-quality subselection of the objects extracted from the Y1A1 coadd images. 
When selecting the \gold catalog, we sought to remove spurious, non-physical objects while minimally decreasing the statistical power of any scientific investigation (\tabref{selection}).
Specifically, we required that objects be observed, but not necessarily detected, at least once in each of the $g$, $r$, $i$, and $z$ bands.
We also required that all objects have $\var{SPREADERR\_MODEL} > 0$ for the $g$, $r$, $i$, and $z$ bands to eliminate objects with unphysical \var{SPREADERR\_MODEL} values indicative of a failure in the \SExtractor photometric fit.\footnote{Objects that are not detected in a specific band have a sentinel value of $\var{SPREADERR\_MODEL} = 1$.}
We also identify several classes of objects that are extremely unusual and flag them for exclusion from most cosmological analyses (\tabref{flags}).
In addition to objects flagged by \SExtractor, we specifically identify 
(1) objects with extremely blue ($\{g-r, r-i, i-z\} < -1$) or extremely red ($\{g-r, r-i, i-z\} > 4$) colors, 
(2) bright stars that saturate some of the single-epoch inputs to the coadd image,
(3) objects that have a large ($>1\arcsec$) offset in the windowed centroid derived from the $g$ and $i$ bands.
Finally, we require that objects reside within the \gold footprint (\secref{footprint}) and flag any objects that reside in poor-quality or potentially problematic (``bad'') regions (\secref{badregions}).

% More tables
\begin{deluxetable}{l p{120pt} }
\tablewidth{0pt}
\tabletypesize{\tablesize}
\tablecaption{ \gold Catalog Selection \label{tab:selection}}
\tablehead{
Selection & Description \\
}
\startdata
$\bandvar[\{griz\}]{niter\_model} > 0$ & Select objects that were observed at least once in each of the $g,r,i,z$-bands. \\ 
$\spreaderrmodel[\{griz\}] > 0$ & $\spreaderrmodel = 0$ indicates a failure in the photometric fit.\\ 
\enddata
\end{deluxetable}

\begin{\tabletype}{c c p{120pt}}
\tablewidth{0pt}
\tabletypesize{\tablesize}
\tablecaption{ \gold Catalog Flags \label{tab:flags}}
\tablehead{
Flag Bit & Selection & Description
}
\startdata
%\multirow{2}{*}{1} & \multirow{2}{*}{$\flags[\{griz\}] > 3$} & \multirow{2}{0.4\columnwidth}{Set from the OR of the \SExtractor flags for $\{griz\}$-bands} \\
%\\ \\
%\multirow{2}{*}{2} & \multirow{2}{*}{\breakcell{$\{g-r, r-i, i-z\} < -1$ \\ $\OR \{g-r, r-i, i-z\} > 4$}} & \multirow{2}{0.4\columnwidth}{Objects with unphysical colors} \\
%\\ \\ 
%\multirow{2}{*}{4} & \multirow{2}{*}{\breakcell{$\nepochs[g] = 0 \AND \magerrauto[g] < 0.05$ \\ $\AND (\magmodel[i] - \magauto[i]) < -0.4$}} & \multirow{2}{0.4\columnwidth}{Artifacts associated with stars close to the saturation threshold.} \\
%\\ \\
%\multirow{3}{*}{8} & \multirow{3}{*}{\breakcell{($|\alphawin[g]-\alphawin[i]| > 1"$ \\ $\OR |\deltawin[g]-\deltawin[i]| > 1"$) \\ $\AND \magerrauto[g] < 0.22$}} & \multirow{3}{0.4\columnwidth}{Objects with bad astrometric colors.}\\
%\\
\multirow{2}{*}{1} & \multirow{2}{*}{$\flags[\{griz\}] > 3$} & \multirow{2}{0.4\columnwidth}{Objects flagged by \SExtractor} \\
\\ 
\multirow{2}{*}{2} & \multirow{2}{*}{\breakcell{$\{g-r, r-i, i-z\} < -1$ \\ $\OR \{g-r, r-i, i-z\} > 4$}} & \multirow{2}{0.5\columnwidth}{Objects with unphysical colors} \\
\\ [+0.25em] 
\multirow{2}{*}{4} & \multirow{2}{*}{\breakcell{$(\nepochs[g] = 0) \AND (\magerrauto[g] < 0.05)$   \\ $\AND (\magmodel[i] - \magauto[i]) < -0.4$ }} & \multirow{2}{0.5\columnwidth}{Artifacts associated with stars close to the saturation threshold} \\
\\ [+0.25em] 
\multirow{3}{*}{8} & \multirow{3}{*}{\breakcell{($|\alpha_{J2000,g}-\alpha_{J2000,i}| > 1\arcsec$ \\ $\OR |\delta_{J2000,g}-\delta_{J2000,i}| > 1\arcsec$) \\ $\AND (\magerrauto[g] < 0.05)$}} & \multirow{3}{0.5\columnwidth}{Objects with large astrometric offsets between bands}\\ 
\\ \\ 
\enddata
\end{\tabletype}

\subsection{Multi-Epoch, Multi-Object Fitting}
\label{sec:mof}

The Y1A1 coadded images provide deeper and more sensitive object detection than individual single-epoch images. 
However, the coaddition process averages across multiple images, resulting in a discontinuous PSF and correlated noise properties.
Precision measurements that rely on an accurate PSF determination, such as galaxy shape measurements for cosmic shear, require a joint fit of pixel-level data from multiple single-epoch images.

We used the \ngmix\footnote{\url{https://github.com/esheldon/ngmix}} code \citep{Sheldon:2014,SheldonHuff:2017,Jarvis:2015} to reanalyze pixel-level data from multi-epoch postage stamps of each object in the \gold coadd catalog.  
We used \PSFEx \citep{Bertin:2011} to model and interpolate the PSF at the location of each object, and then we generated an image of the PSF using the python package, \code{psfex}\footnote{\url{https://github.com/esheldon/psfex}}.
We then used the \ngmix code to fit this reconstructed PSF image to a set of three free, independent Gaussians.

We used \ngmix in ``multi-epoch'' mode to simultaneously fit a model to all available epochs and bands.  
In this mode, a model is convolved by the local PSF in each single-epoch image, and a $\chi^2$ sum is calculated over all pixels in a postage stamp. 
This is repeated for each epoch and band, and a total $\chi^2$ sum is calculated.
We then find the parameters of the model that maximize the likelihood.

\begin{figure}
\centering
\includegraphics[width=\columnwidth]{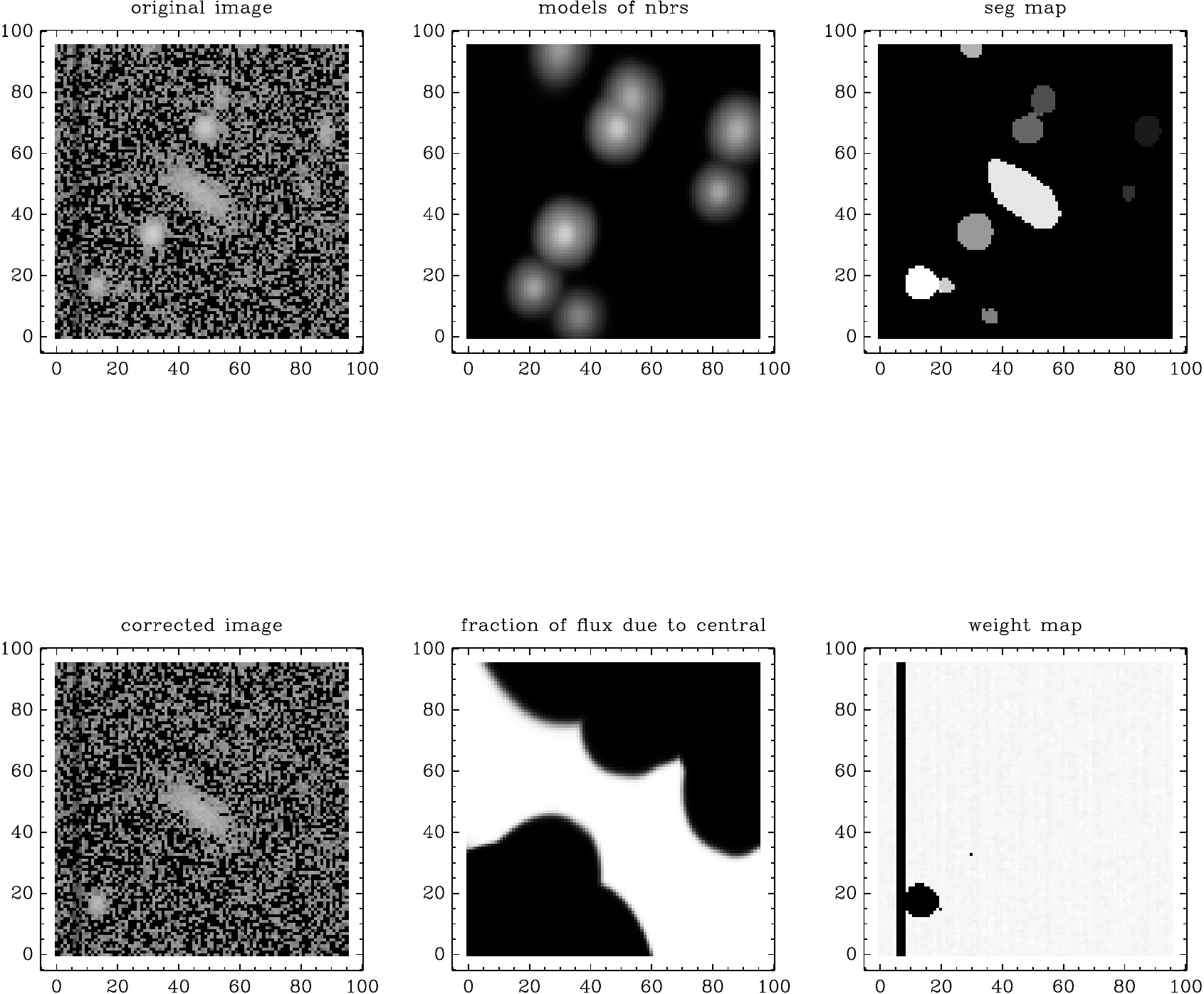}
\caption{A group of objects fit using the \MOF algorithm.
In the top row we show: (left) the sky-subtracted image, 
(center) the models for neighboring sources, and 
(right) the SExtractor segmentation map.  
In the bottom row we show: (left) the sky-subtracted image after also subtracting the light from neighbors, 
(center) the fraction of light assigned to the central object (100\% in white and 0\% in black), and 
(right) the weight map. 
Note that a bad column and a flagged object are identified in the weight map. 
The masked object was not fit, and thus its light was not subtracted.
}
\label{fig:mof}
\end{figure}

We took this procedure one step further, performing simultaneous multi-epoch,
multi-band, and multi-object fit, which we call ``\mof''.  
We first identified groups of objects using a friends-of-friends algorithm \CHECK{\citep[\eg,][]{Huchra:1982,Berlind:2006}}.
We then fit the members of the group using the following procedure:
\begin{enumerate}
  \item Perform an initial model fit to each object, masking the light from neighbors using the \uberseg algorithm \citep{Jarvis:2015}.  
  \item Fit the model to each object again, this time subtracting the light from neighbors using the models from the previous fit.
  \item Repeat the previous step until all fits converge, or a maximum of 15 iterations was reached.
\end{enumerate}
This fit was performed simultaneously in the $g,r,i,z$ bands using all available imaging epochs and assuming the same spatial model for all bands and epochs.
An example of this procedure is shown in \figref{mof}.

We found that fitting a galaxy model with fully free bulge and disk components was highly unstable, so we adopted the following approach, inspired by the ``composite'' model used in the SDSS.\footnote{\url{http://www.sdss.org/dr12/algorithms/magnitudes/\#cmodel}}
We first fit the disk and bulge models separately, represented by an exponential and \devauc profile \citep{deVaucouleurs:1948}, respectively.  
We then determined the linear combination of these models that best fit the data,
\begin{equation}
    M_{\rm tot} = f_{\rm dev}  M_{\rm dev} + (1-f_{\rm dev}) M_{\rm exp}
\end{equation}
where $M_{\rm dev}$ is the bulge model, $M_{\rm exp}$ is the disk model, and $f_{\rm dev}$ represents the fraction of light in the bulge component.
This total model is unlikely to be a good fit of the data, and we only use it as a starting point for a more refined model.  
We formed a new model that has the best $f_{\rm dev}$ determined as above, as well as the same ratio of scale lengths for the bulge and disk components.  
This new model has free parameters for the center, ellipticity, overall scale, and fluxes.
A common center, scale, and ellipticity were used for all bands, but the flux for each band was left free.

For computational efficiency, each component of this model was approximated by a sum of Gaussians \citep{HoggLang:2013}. 
This choice made convolution with the triple-Gaussian PSF model very fast. 
A fast approximation for the exponential function was also used to speed up computations \citep{Sheldon:2014}.

We imposed uninformative priors on all parameters except for the ellipticity and the fraction of light present in the bulge, $f_{\rm dev}$.  
For both of these parameters, we applied priors based on fits to deep COSMOS imaging data, provided as postage stamps with the \code{GalSim} project\footnote{\url{https://github.com/GalSim-developers/GalSim}}.
We defined convergence to be when the flux from subsequent fits to objects did not
change more than one part in a thousand, and structural parameters such as scale
and ellipticity did not change by more than a part in a million.
For incorporation into the \gold catalog, we converted \MOF fluxes to magnitudes and applied the SLR adjustment discussed in \secref{slr}.

\subsection{Catalog Completeness}
\label{sec:goldcat}

\begin{figure}[!t]
\center
\includegraphics[width=\columnwidth]{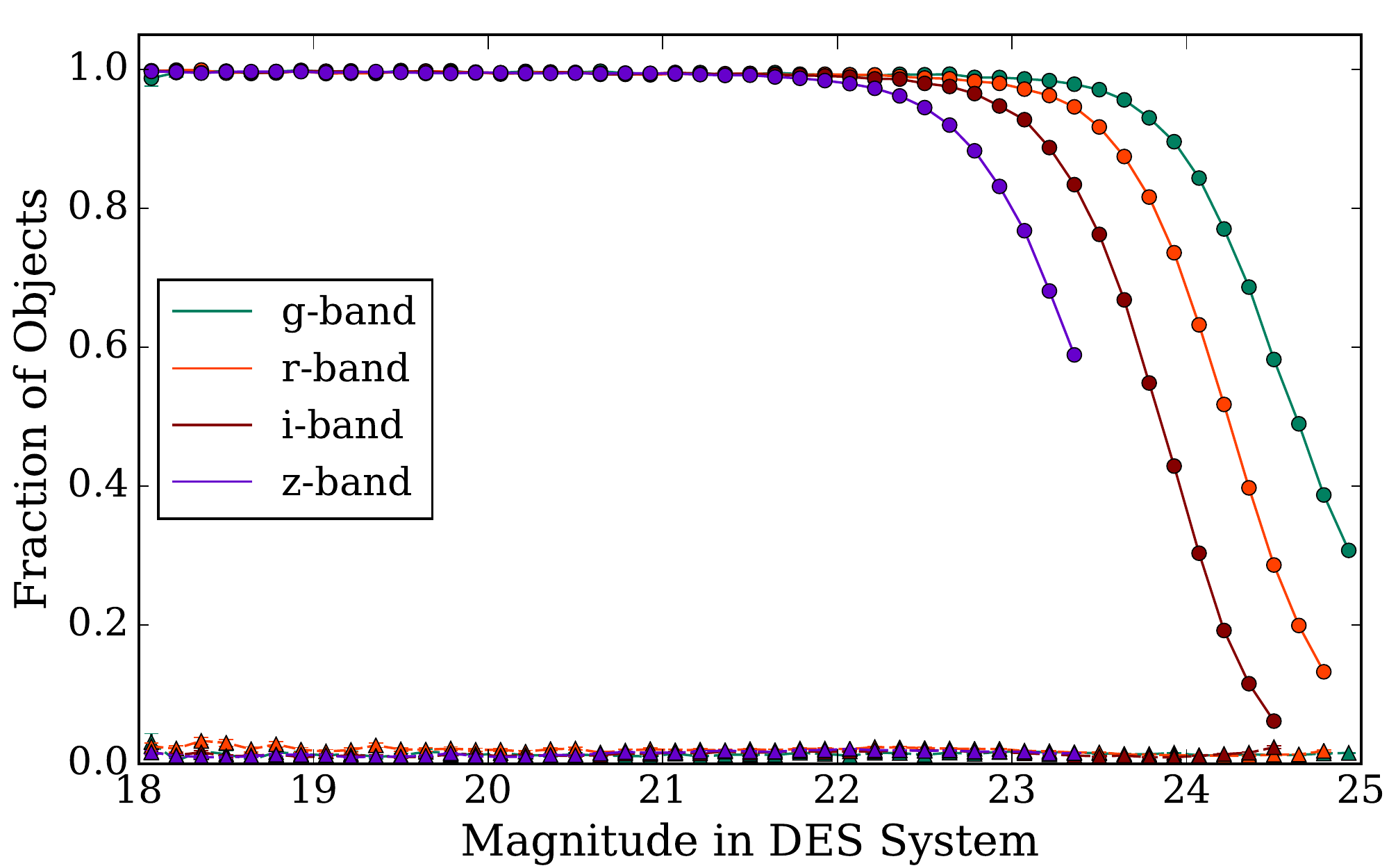}
\caption{ \label{fig:cfhtlens} 
Completeness (solid circles) and contamination (dashed triangles) of the \gold coadd object catalog determined by comparison to the CFHTLenS W4 field. 
Object matching was performed within a 1'' radius and CFHTLenS magnitudes were transformed to the DES system using the equations in \appref{transform}.
Statistics were calculated for the subset of objects that were unmasked in both surveys and have been truncated at the $5\sigma$ limiting magnitude of CFHTLenS \citep{Erben:2013}.
} 
\end{figure}

We assessed the completeness and purity of the \gold catalog by comparing it against data from the Canada-France-Hawaii Lensing Survey (CFHTLenS) W4 field \citep{Erben:2013,Hildebrandt:2012}, which overlap the S82 region of \gold.
The DES data in this overlap region has a typical $10\sigma$ limiting magnitude of $g \sim 23.1$, $r \sim 23.0$, $i \sim 22.5$, $z \sim 21.8$ (\secref{depth}).
This is comparable to the median for \gold in $i$ and $z$ bands and $\roughly 0.2 \magn$ shallower than the median in $g$ and $r$ bands (\tabref{summary}).
In this region, CFHTLenS is $\gtrsim 1 \magn$ deeper than the \gold catalog, making it a good test for object detection completeness.
We transformed the magnitude of CFHTLenS objects into the DES system (\appref{transform}) and removed objects residing in masked regions of either survey.
We associated objects between the two catalogs based on a spatial coincidence of $1\arcsec$ and required a matching magnitude within 2 mag.
We then calculated the detection completeness as the fraction of CFHTLenS objects that are matched to \gold objects as a function of the CFHTLenS magnitude transformed into the DES system.
The contamination of the \gold catalog is assessed as the fraction of \gold objects that are unmatched to CFHTLenS objects as a function of magnitude.
We find that the 95\% completeness limit of the \gold catalog is $g = 23.6$, $r = 23.4$, $i = 22.9$, and $z = 22.4$ (\figref{cfhtlens}). 
We find that for magnitudes brighter than these limits, the contamination of the \gold catalog is $\lesssim 2\%$.
The \gold catalog is $>99\%$ complete in all four bands for magnitudes brighter than 21.5.
This completeness estimate does not account for objects that are blended in both CFHTLenS and DES, which is estimated to be $\roughly 1\%$ of objects at DES depth.
We also note that Y1A1 object detection was performed on a combined $r+i+z$ detection image and no S/N threshold was applied to the measurements in individual bands when calculating completeness.

\section{Ancillary Maps}
\label{sec:maps}
 
Several ancillary maps were produced to characterize the coverage, sensitivity, observing conditions, and potentially problematic regions of \gold as a function of sky position. 
Generating ancillary maps for \gold was a multi-step process:
we created a vectorized representation of the survey coverage and limiting magnitude using \mangle \citep{Hamilton:2004,Swanson:2008}, we rasterized the \mangle maps with \HEALPix for ease of use, we estimated observing conditions over the survey footprint, and we subselected a nominal high-quality footprint. 
Finally, we flagged sky regions where the true survey performance deviates from that estimated by the ancillary data products (\ie, the regions around bright stars, astrometric failures, \etc).
Each of these steps is described in more detail below.

\subsection{Maps of Survey Coverage and Depth}
% (Benoit-Levy, Rykoff, Neilsen)
\label{sec:depth}
 
Quantifying survey coverage and limiting magnitude as a function of sky position is essential for statistically rigorous cosmological analyses.
To accurately track characteristics of the DES survey at the sub-CCD level, DESDM produces \mangle masks \citep{Hamilton:2004,Swanson:2008} as part of the \coadd pipeline.
These masks are an accurate representation of the coverage, sensitivity, and overlap of DECam exposures including dead CCDs, gaps between CCDs, masked regions around bright stars, and bright streaks from Earth-orbiting satellites.

\begin{figure*}[!ht]
\center
\includegraphics[width=.9\textwidth]{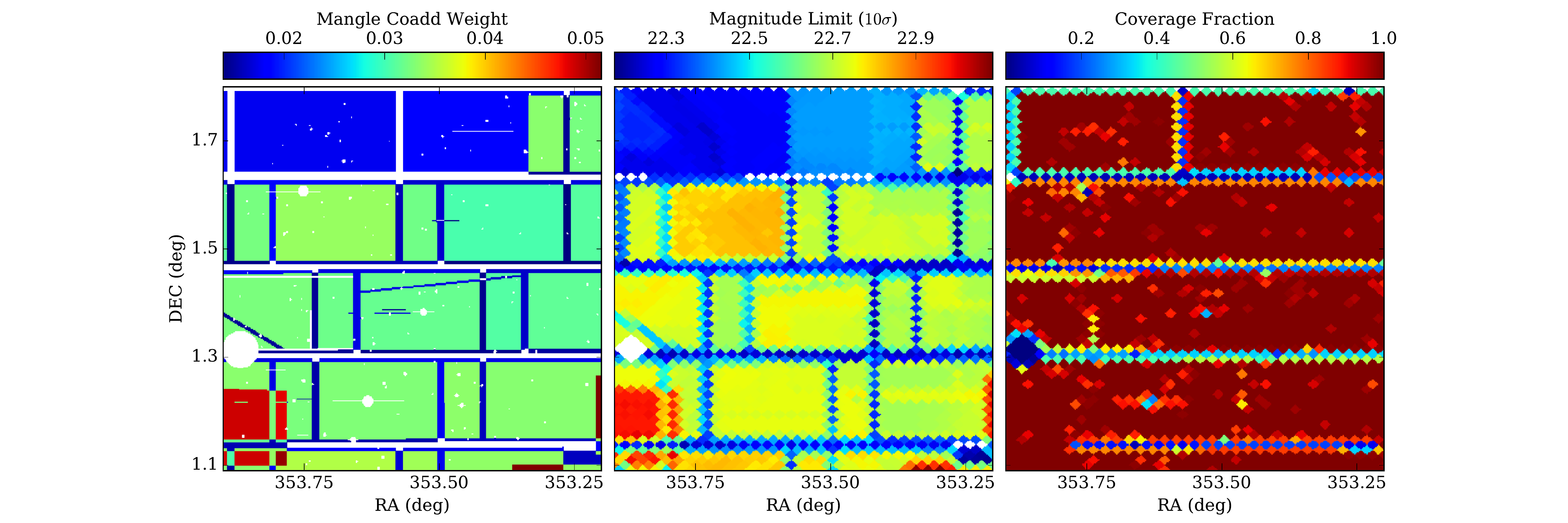}
\caption{\label{fig:mangle} 
Coverage and depth maps for a single Y1A1 coadd tile.
(Left) Vectorized \mangle weight map for an $r$-band coadd tile.
Satellite trails, star masks, and chip gaps are stored at full resolution.
(Center) Pixelized $10\sigma$ limiting magnitude map for galaxies using \healpix at $\nside = 4096$. (Right) Pixelized map of the coverage fraction at \healpix $\nside = 4096$.
This tile is located on the border of the Y1A1 footprint and has been chosen for illustrative purposes due to its variable depth and incomplete coverage. 
}
\end{figure*}

During coadd production, \mangle masks were created at the level of coadd tiles (\figref{mangle}). The steps are the following:
\begin{enumerate}
\item Polygons were created using the four corners of each input CCD image and assigned a weight equal to the median value of pixels in the CCD weight plane.
\item Satellite streaks were represented by polygons, and the area of these polygons was removed from the single-epoch CCD polygon.
\item Polygons were trimmed to fit the tile boundaries. 
\item Polygons were subdivided into disjoint regions with the \code{balkanize} command. Following the weighted-average scheme chosen for image coaddition, the total weight of a balkanized polygon is the sum of the weights of the individual polygons.
\item Regions around bright stars and bleed trails are removed from the \mangle mask. 
While the precise location of these artifacts is image dependent, it is computationally simpler to mask the stacked map with the largest shape covering a bright star or bleed trail rather than removing these regions from each single-epoch polygon. 
\item The \mangle coadd weight map was converted into a $10\sigma$ limiting magnitude map for a $2\arcsec$ diameter aperture:
\begin{equation}
m_{\rm{lim}}= m_{\rm ZP} - 2.5 \log \left( 10 \sqrt{\pi \frac{(D/2)^2}{\omega_{\rm pix}^2}\frac{1}{w_{\rm tot}}}\right),
\end{equation}
where $m_{\rm ZP}=30$, is the tile zero-point, $D=2\arcsec$, $\omega_{\rm pix}=0\farcs263$ is the pixel size, and $w_{\rm tot}$ is the total weight of the polygon.
This definition of the magnitude limit corresponds to the \var{MAG\_APER\_4} quantity measured by \sextractor.
\end{enumerate}
 
%https://opensource.ncsa.illinois.edu/confluence/display/DESDM/Healpixifing+Mangle+Masks
While the vectorized \mangle masks are a very accurate representation of the DES survey coverage, they are computationally unwieldy for many scientific analyses.
To increase the speed and ease with which survey coverage and limiting magnitude can be accessed, we generate anti-aliased \healpix maps of these quantities (\figref{mangle}). 
Pixelized maps of the survey coverage fraction were created at a resolution of $\nside=4096$ (${\rm area} = 0.73\amin^2$) by calculating the fraction of higher-resolution subpixels ($\nside=32768$, ${\rm area} = 0.01\amin^{2}$) that were contained within the \mangle mask.
Similarly, maps of the survey limiting magnitude were generated at $\nside=4096$ by calculating the mean limiting magnitude for subpixels ($\nside=32768$).
When calculating the limiting magnitude, subpixels that were not covered by the survey were excluded from the calculation, while subpixels that have been masked (\ie, bright stars, bleed trails, etc.) were assumed to have the limiting magnitude of their parent polygon.
The \healpix resolution of $\nside=4096$ was chosen as a compromise between computational accuracy and ease of use. 
This resolution was found to have a negligible effect on the correlation function of simulated galaxies on scales larger than $0\farcm5$ when combined with the survey coverage fraction maps.

\begin{figure*}[t]
\centering
\includegraphics[width=0.85\textwidth]{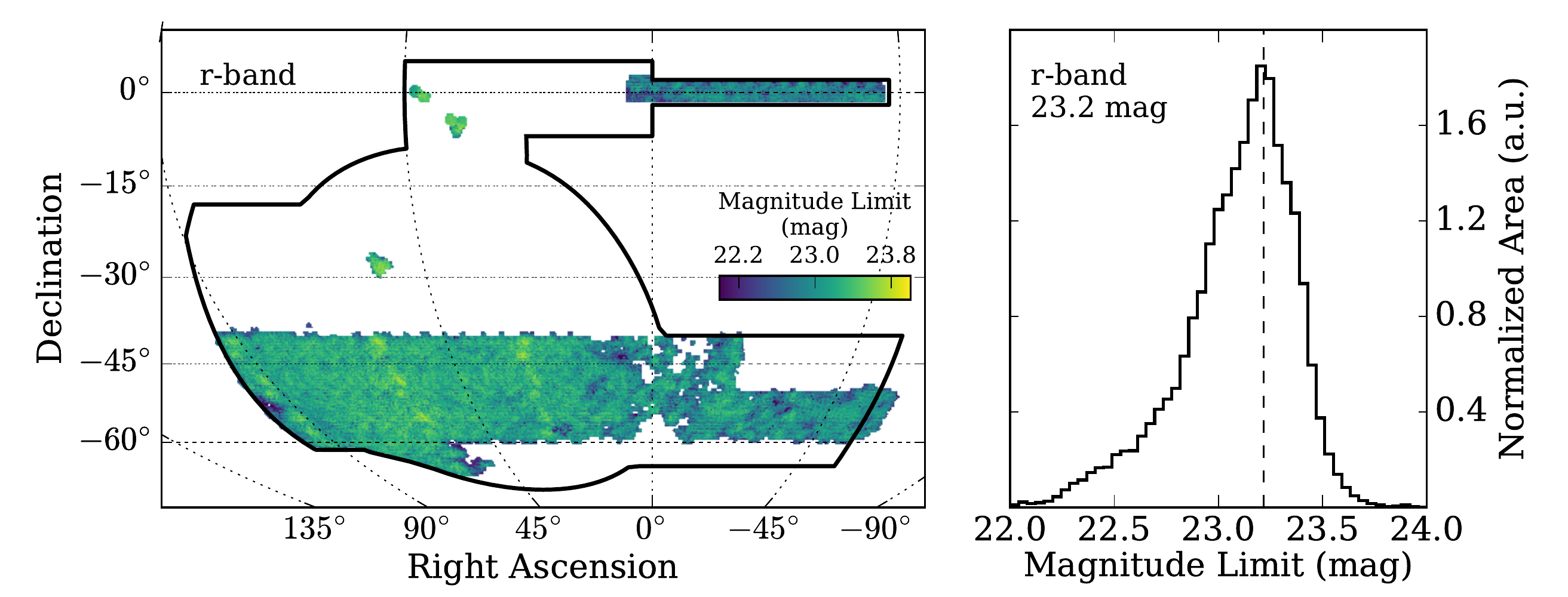}
\caption{\label{fig:depth} 
Sky map and normalized histogram for the $r$-band $10\sigma$ limiting magnitude (\var{MAG\_AUTO}) derived in HEALPix pixels over the \gold footprint. 
The mode of the limiting magnitude distribution is shown inset in the right panel.
The derivation of the limiting magnitude is described in \secref{depth}.
Similar figures for other bands are shown in \appref{depth}.
}
\end{figure*}

We followed the prescription of \citet{Rykoff:2015} to convert the \mangle coverage and depth maps into $10\sigma$ limiting magnitude maps for galaxy photometry.
We selected galaxies using the \modest star-galaxy classifier (\secref{sgsep}) and trained a random forest model to predict the $10\sigma$ limiting magnitude as a function of observing conditions.
The input vector for the random forest included the PSF FWHM, sky brightness, airmass, and exposure time for each band being fit (\secref{borismaps}).
The training was performed on coarse \healpix pixels (\nside = 1024) that contained more than 100 galaxies.
Once trained, the model was applied to the pixels at the full mask resolution of \nside = 4096.
We derived magnitude limits for both coadd \var{AUTO} magnitudes and the multi-epoch composite model magnitudes derived by the \MOF (\secref{mof}).
We applied the SLR calibration adjustment (\secref{slr}) to the resulting depth maps to correct for interstellar extinction and zeropoint non-uniformity.
The median $10\sigma$ limiting magnitudes for \var{MAG\_AUTO} are $g = \maglimg^{+0.14}_{-0.40}$,  $r = \maglimr^{+0.13}_{-0.37}$,  $i = \maglimi^{+0.14}_{-0.34}$,  $z = \maglimz^{+0.12}_{-0.37}$,  $Y = \maglimY^{+0.18}_{-0.33}$, where the uncertainties represent the 16\textsuperscript{th} and 84\textsuperscript{th} percentiles of the distribution.
In comparison, the median $10\sigma$ limiting magnitudes for the \mof \var{CM\_MAG} magnitudes are $g = \moflimg^{+0.07}_{-0.40}$, $r = \moflimr^{+0.16}_{-0.29}$, $i = \moflimi^{+0.14}_{-0.30}$, and $z = \moflimz^{+0.14}_{-0.32}$.
We find that the depth estimates are accurate at the level of 6\%-7\%, but that 3\%-4\% of this measured uncertainty is due to ``pixelization noise'' resulting from averaging over a range of depths when fitting the model on coarse pixels.
An example of the resulting depth maps for $r$ band can be found in \figref{depth}, and figures for the other bands can be found in \appref{depth}.

\subsection{Maps of Survey Characteristics}
% (Leistedt, Drlica-Wagner)
\label{sec:borismaps}
 
\begin{figure*}[ht]
\centering
\includegraphics[width=\textwidth]{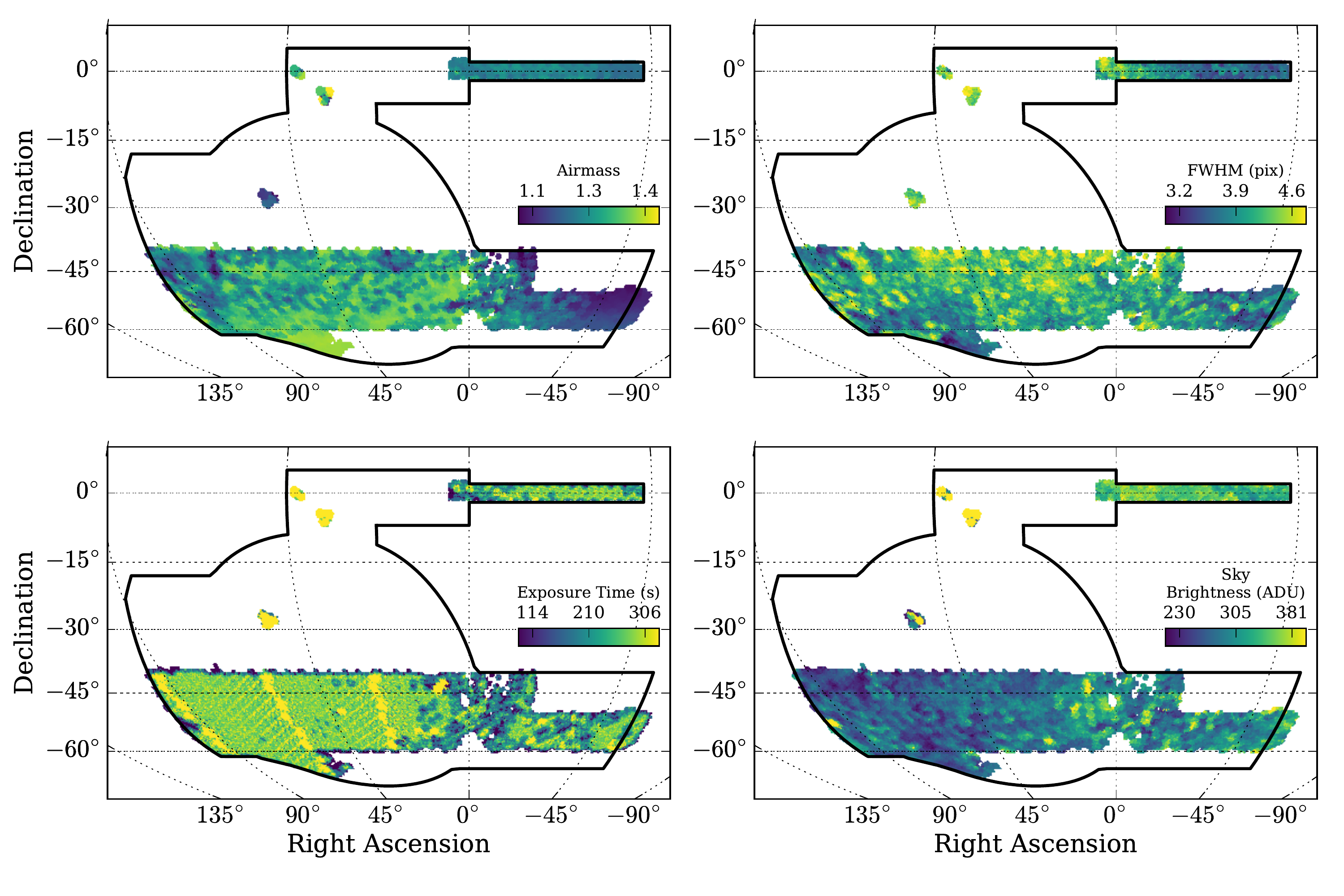}
\caption{\label{fig:systematics}
Survey characteristics of the \gold data set estimated from the inverse variance weighted stack of single-epoch images in $r$-band at each position on the sky. 
Panels correspond to mean airmass (top left), PSF FWHM in pixels (top right), exposure time in seconds (bottom left), and sky brightness in ADU (bottom right).
}
\end{figure*}

Variations in observing conditions can be a significant source of systematic uncertainty in cosmological analyses. 
In a wide-area optical survey such as DES, variable observing conditions can imprint spurious spatial correlations, noise, and depth fluctuations on the object catalogs that are used for galaxy clustering and cosmic shear analyses. 
By identifying and characterizing these systematic effects, it becomes possible to quantify and minimize their impact on scientific results. 
We followed the procedure developed by \citet{Leistedt:2016} to construct survey characteristic and coverage fraction maps for the \gold data set using \code{QuickSip}.\footnote{\url{https://github.com/ixkael/QuickSip}}
Since the nonlinear transfer function between the stack of images at any position on the sky and the final galaxy catalog is largely unknown, we created maps of many different survey observables. 
For each band, we created maps of both weighted- and unweighted-average quantities of each image. 
The main quantities expected to be used for null tests in cosmological analyses with the \gold catalog are the total exposure time, the mean PSF FWHM, the mean airmass, and the sky background.
The inverse variance weighted averages of these quantities are shown in \figref{systematics}.
Further modeling of the survey transfer function is important for DES cosmology analyses, and several approaches have already been developed \citep[\eg][]{Chang:2015,Suchyta:2016}. 
 
\subsection{Footprint Map}
%(Rykoff, Drlica-Wagner, Sevilla)
\label{sec:footprint}
 
The nominal footprint for the \gold catalog is defined using an $\nside = 4096$ \healpix map. 
For a pixel to be included in the \gold footprint, it must meet the following criteria simultaneously in the $g,r,i,z$ bands:
\begin{enumerate}
\item A \mangle coverage fraction $\geq 0.5$ implying that at least half of the pixel area has been observed or is unmasked according to \mangle (\secref{depth}).
\item A coverage fraction of $\geq 0.5$ from the survey characteristics maps (\secref{borismaps}). 
\item A minimum total exposure time of $\geq 90 \second$ (\secref{borismaps}).
\item A valid solution from the SLR calibration adjustment (\secref{slr}).
\end{enumerate}
These selection criteria reduce the total \CHECK{coadded} area of Y1A1 covered in any band, $1927 \deg^2$, to a nominal WIDE+D04 \gold footprint in $g,r,i,z$ of $1786 \deg^2$. 
Simultaneously applying the same criteria to the $Y$ band (with a minimum exposure time of $45\second$) results in a $g,r,i,z,Y$ footprint of $1773 \deg^2$.
These numbers were calculated by summing the coverage fraction of pixels in the footprint.

\subsection{Bad Region Mask}
\label{sec:badregions}
%(Rykoff, Drlica-Wagner, Sevilla)
Masks were developed to remove regions where survey artifacts make it difficult to control systematic uncertainties when doing cosmological analyses. 
Since not all science topics require the same masks (\eg, studies of galaxy evolution may not want to mask nearby galaxies), the various masks are collected into a bitmap defined in \tabref{badregions}.
Removing area associated with any of these masks results in a WIDE+D04 footprint area of $1506 \deg^2$ in $g,r,i,z$ and $1496 \deg^2$ in $g,r,i,z,Y$.

% Table
\begin{deluxetable}{c c l}
\tablewidth{0pt}
\tabletypesize{\scriptsize}
\tablecaption{ \gold Bad Region Mask \label{tab:badregions}}
\tablehead{
Flag Bit & Area & Description \\
 & ($\deg^2$)  & 
}
\startdata
1   & 30.1  & High density of astrometric discrepancies \\
2   & 119.5 & 2MASS moderate star regions ($8 < J < 12$) \\
4   & 5.4   & RC3 large galaxy region ($10 < B < 16$) \\
8   & 38.6  & 2MASS bright star regions ($5< J < 8$) \\
16  & 95.8  & Region near the LMC \\
32  & 18.4  & Yale bright star regions ($-2 < V < 5.6$) \\
64  & 1.3   & High density of unphysical colors \\ 
128 & ...   & Unused bit \\
256 & 0.7   & Milky Way globular clusters \\
512 & 7.2   & Poor COADD PSF modeling \\
\enddata
\tablecomments{
Masked regions for the Y1A1 GOLD WIDE+D04 footprint.
The masked area is calculated using the coverage fraction of the pixels that are removed from the footprint by each mask.
The criteria defining each mask can be found in \secref{badregions}.
}
\end{deluxetable}

%% 1   & 33.04 & High density of astrometric discrepancies \\
%% 2   & 133.1 & 2MASS moderate star regions ($8 < J < 12$) \\
%% 4   & 5.40  & RC3 large galaxy region ($10 < B < 16$) \\
%% 8   & 31.6  & 2MASS bright star regions ($5< J < 8$) \\
%% 16  & 98.9  & Region near the LMC \\
%% 32  & 19.75 & Yale bright star regions ($-2 < V < 5.6$) \\
%% 64  & 1.37  & High density of unphysical colors \\ 
%% 128 & ...   & Not used \\
%% 256 & 0.70  & Milky Way globular clusters \\
%%Bad regions are ordered from most least impactful (lowest bit) to the most impactful (highest bit). 

\subsubsection{Catalog Artifacts}
\label{sec:badobjects}
 
\begin{enumerate}
\item {\it Unphysical colors (bit=64):} 
This mask is designed to remove imaging artifacts that were not masked before creating coadds.
In particular, this mask removes regions where there are significant reflected light artifacts (both specular and diffuse) from bright stars, un-masked orbital satellite trails, and coadd saturation artifacts.  
This mask is pixelized at $\nside = 2048$ and pixels with $\geq 8$ objects possessing unphysical colors are masked (see \tabref{flags}).  
The threshold for flagging bad pixels was set by visual inspection of the coadd tiles.  
The resulting masked area is $1.3 \deg^2$.

\item {\it Astrometric discrepancies (bit=1):} 
We flag regions that have a high concentration of galaxies with large astrometric offsets between filters. 
We select galaxies with $i < 22$, $\var{MAGERR\_AUTO\_G} < 0.2$, and windowed positions in $g$ and $i$ band differ by more than 1\arcsec.  
This criterion has been found to select objects in regions of strongly variable background (\eg, the wings of bright stars, regions of poor sky subtraction, regions with scattered light, \etc). 
The resulting masked area in this case is $30.1 \deg^2$. 

\item {\it PSF model failures (bit=512):}
There are several regions where PSF modeling failed owing to varying depth and a discontinuous PSF. 
Coadd tiles possessing poor PSF models are identified as having a large number of stars where the coadd PSF magnitude differs from the weighted-average single epoch PSF magnitude by more than 0.2 mag. 
We flag \healpix pixels ($\nside=512$) possessing $>20$ stars with large discrepancies in PSF magnitudes.
The total region masked in this way is $7.2 \deg^2$.
 
\end{enumerate}

\subsubsection{Bright Stars}
\label{sec:starmask}

Regions around saturated stars were masked at the pixel level as part of the image processing pipeline described in \secref{desdm}.
However, catalog-level investigation revealed a residual increase in the number density of objects surrounding the brightest stars.
To avoid contamination from spurious objects in the halos of bright stars, we designed radial masks based on the brightness of the contaminating stars and the number density of surrounding objects.
These masks were developed for two bright star catalogs as described below.

\begin{enumerate}
\item {\it Yale bright star regions (bit=32):} 
 Masked regions were determined from the positions and magnitudes of stars in the Yale Bright Star Catalog \citep{Hoffleit:1991}.  
The masking radius was determined from the $V$-band magnitude of each star, following the equation:
\begin{equation}
R = 0.86835 \deg - (0.1439 \deg) \times V  .
\end{equation}
Minimum and maximum masking radii were imposed at 0.1 \deg and 0.4 \deg, respectively. The resulting masked area is $18.4 \deg^2$.
\item {\it 2MASS bright stars (bit=2,8):}
We mask regions around bright stars from the 2MASS catalog \citep{Skrutskie:2006} within a radius of
\begin{equation}
R = 0.09 \deg - (0.0073 \deg) \times J , 
\end{equation}
assuming a minimum and maximum masking radius of 0.01 \deg and 0.05 \deg, respectively.  
Many of the bright stars in 2MASS overlap with the faintest stars in the Yale Bright Star Catalog, and we find a comparable masking radius (albeit derived using different bands).
Because the fainter 2MASS stars may not be problematic for all science applications, we split the 2MASS star mask into stars with $5 < J < 8$ and stars with $8 < J < 12$.  
The masked areas are $38.6 \deg^2$ and $119.5 \deg^2$, respectively.
\end{enumerate}
 
\subsubsection{Large Foreground Objects}
\label{sec:galmask}
 
\begin{enumerate}
\item {\it The Large Magellanic Cloud (bit=16):}
The center of the Large Magellanic Cloud (LMC) is located $\roughly 5 \deg$ from the southwest edge of the DES footprint, and the stellar population of the LMC presents a number of challenges for extragalactic science. 
The high density of stars decreases the purity of galaxy samples, while the 2MASS star masks described in \secref{starmask} lead to a complex and heavily masked area.
The stellar locus of the LMC differs from that of the Milky Way making it difficult to apply the SLR calibration adjustment described in \secref{calibration}. 
For these reasons, we masked a region around the LMC with a boundary defined as $60\degree < \ra < 100\degree$ and $-70\degree < \dec < -58\degree$.
The LMC mask removed $95.8 \deg^{2}$.
 
\item {\it Bright galaxies (bit=4):} 
The Third Reference Catalog of Bright Galaxies \citep[RC3;][]{Corwin:1994} contains galaxies subtending $\gtrsim 1\arcmin$. 
Since galaxy size is highly correlated with magnitude, we continue to use a magnitude-dependent masking formulation similar to that applied to bright stars.
We masked a circular region around RC3 galaxies with $10 < B < 16$ with a magnitude-dependent selection:
\begin{equation}
R = 0.269 \deg - (0.0166 \deg) \times B .
\end{equation}
We imposed minimum and maximum masking radii such that $0.03 \deg < R < 0.1 \deg$. 
The bright galaxy mask removes $5.4 \deg^2$.

\item {\it Globular clusters (bit=256):} 
The high stellar density of Milky Way globular clusters makes them difficult regions for cosmology analyses.
We identified three globular clusters, NGC 1261, NGC 1851, and NGC 7089, and masked circular regions with radius 1.5 times the angular size reported by \citet{Sinnott:1988}.
This resulted in a total masked area of $0.7 \deg^2$.

\end{enumerate}

% Table
\begin{\tabletype}{c c l}[!t]
\tablewidth{0pt}
\tabletypesize{\footnotesize}
\tablecaption{\label{tab:modest} \gold \modest Star-Galaxy Classification}
\tablehead{
Class & Selection & Description
}
\startdata
\multirow{2}{*}{0} & \multirow{2}{*}{$\var{SPREAD\_MODEL\_I} + (5/3) \times \var{SPREADERR\_MODEL\_I} < -0.002$} & \multirow{2}{*}{Unphysical PSF fit (likely stars)} \\
\\
\multirow{2}{*}{1} & \multirow{2}{*}{\breakcell{$\var{SPREAD\_MODEL\_I} + (5/3) \times \var{SPREADERR\_MODEL\_I} > 0.005$ AND \\ NOT $(|\var{WAVG\_SPREAD\_MODEL\_I}| < 0.002$ AND $\var{MAG\_AUTO\_I}<21.5)$}} & \multirow{2}{*}{High-confidence galaxies} \\
\\ 
\multirow{2}{*}{2} & \multirow{2}{*}{$| \var{SPREAD\_MODEL\_I} + (5/3) \times \var{SPREADERR\_MODEL\_I}| < 0.002$} & \multirow{2}{*}{High-confidence stars} \\
\\ 
\multirow{2}{*}{3} & \multirow{2}{*}{$0.002 < \var{SPREAD\_MODEL\_I} + (5/3) \times \var{SPREADERR\_MODEL\_I} < 0.005$} & \multirow{2}{*}{Ambiguous classification}\\
\\ [-0.5em]
\enddata
\tablecomments{
The high-purity and high-completeness galaxy samples are defined as $\modest = 1$ and $\modest \in \{1,3\}$, respectively.
Similarly, the high-purity and high-completeness stellar samples are defined as $\modest = 2$ and $\modest \in \{0,2,3\}$, respectively.}
\end{\tabletype}

\begin{figure*}[!t]
\center
\includegraphics[width=\textwidth]{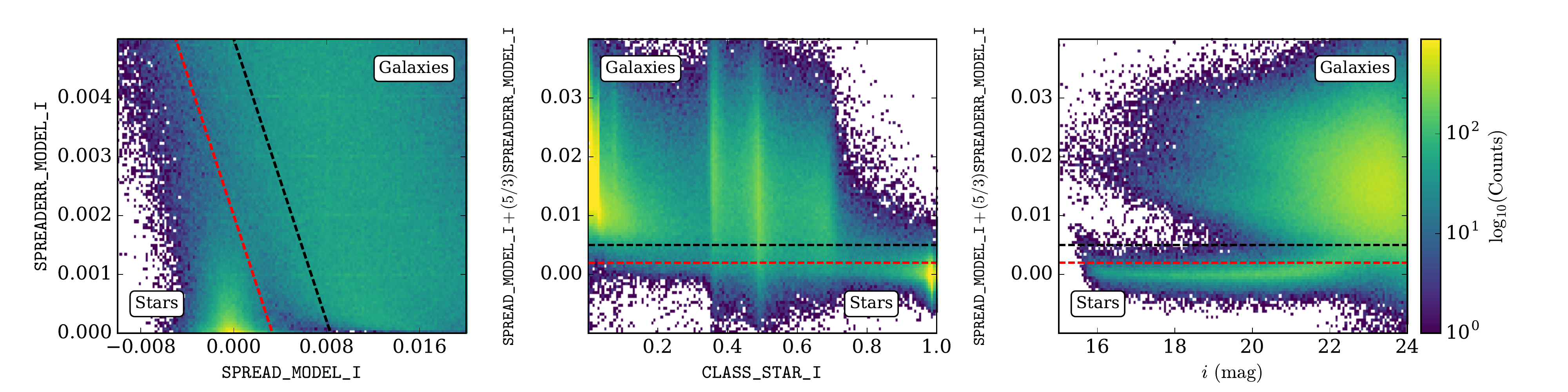}
\caption{\label{fig:modest} 
\modest star-galaxy selection for objects in a $\roughly 13 \deg^2$ region centered on $\ra,\dec = (51\degree,-45\degree)$. 
The left panel shows the distribution of \var{spread\_model\_i} and its error. 
The middle panel compares the distribution of the \SExtractor neural-network classifier, \var{class\_star}, to the \modest selection criteria.
The right panel shows a tight stellar locus in the $i$-band magnitude compared against the \modest criteria.
In all panels the black (red) lines correspond to the pure (complete) galaxy selection threshold applied on \modest.
}
\end{figure*}
 
\section{Value-Added Quantities}
\label{sec:vac}

The astrometric, photometric, and morphological parameters derived for each object are supplemented with additional information important for astrophysical and cosmological analyses.
These ``value-added quantities'' are built from the calibrated coadd object catalog and provide additional information on an object-by-object basis.
The two primary value-added quantities provided with \gold are: (1) a simple star-galaxy classifier, and (2) a set of \photoz estimates.

\subsection{Star-Galaxy Separation}
%(Sevilla, Bechtol, Drlica-Wagner)
\label{sec:sgsep}
 
As part of the \gold catalog, we produced a ``\modest'' object classification with the primary goal of selecting high-quality galaxy samples.
\modest is based on the $i$-band coadd quantity \var{SPREAD\_MODEL\_I} and its associated error, \var{SPREADERR\_MODEL\_I}.
\var{SPREAD\_MODEL} is a morphological variable defined as a normalized linear discriminant between the best-fit local PSF model and a slightly more extended model composed of a circular exponential disk convolved with the PSF \citep{Desai:2012, Soumagnac:2015}. 
The $i$ band was chosen as the reference band for object classification owing to its depth and superior PSF.
Image-level simulations of the DES data support the conclusion that $i$ band yields the best overall performance for object classification, and this result was verified using deep HST imaging on the COSMOS field.
 
We used space-based imaging of COSMOS \citep{Leauthaud:2007} and GOODS-S \citep{Giavalisco:2004} along with spectroscopic observations from VVDS \citep{LeFevre:2005} that overlapped the \gold footprint as a truth sample for developing \modest.
We defined star and galaxy samples optimized for ``high completeness'' and ``high purity'' by applying thresholds on the combination of \var{SPREAD\_MODEL\_I} and \var{SPREADERR\_MODEL\_I}.\footnote{The high-completeness and high-purity samples differ in the classification assigned to ambiguous objects.}
The object classification scheme is defined in \tabref{modest} and shown graphically in \figref{modest}.

\begin{figure}[h]
\center
\includegraphics[width=\columnwidth]{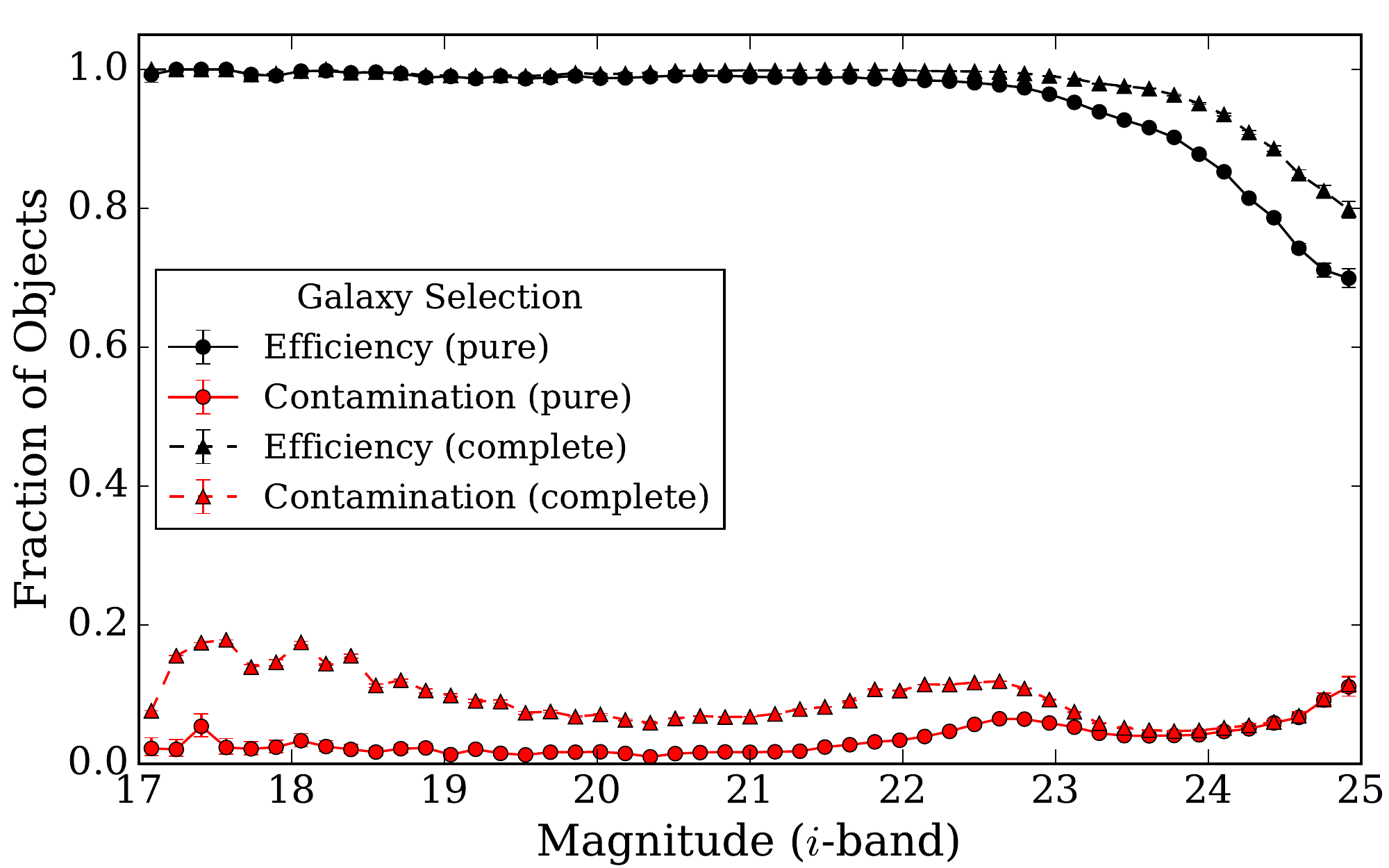}
\includegraphics[width=\columnwidth]{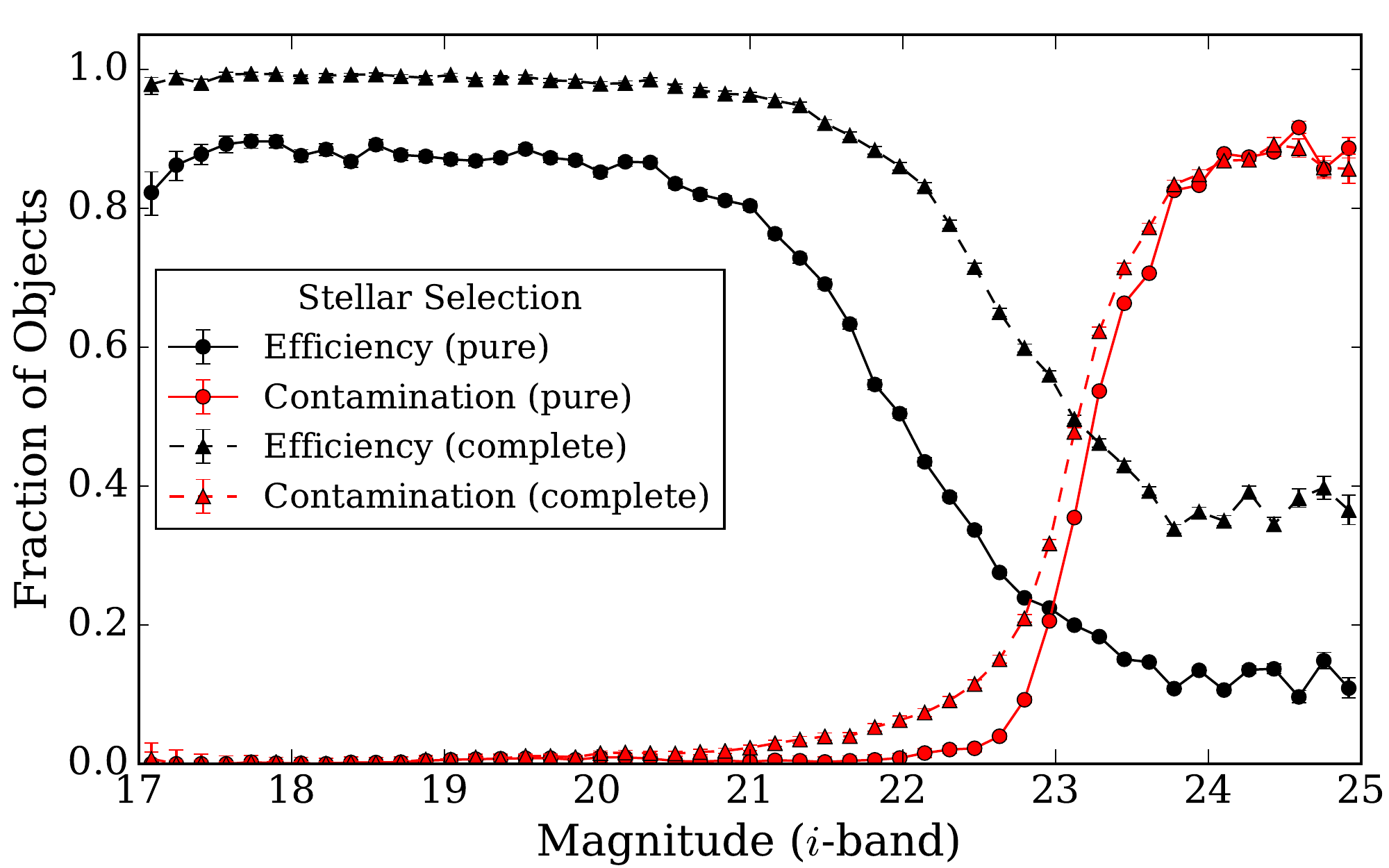}
\caption{\label{fig:modest_performance} Performance of the \modest star-galaxy classifier based on a comparison to deeper imaging from CFHTLenS. 
(Top): The measured efficiency and contamination for high-purity (solid) and high-completeness (dashed) galaxy samples. 
(Bottom): The efficiency and contamination for high-purity (solid) and high-completeness (dashed) stellar samples (note that \modest is not optimized for stellar selection).
}
\end{figure}

\begin{figure}[h]
\center
\includegraphics[width=\columnwidth]{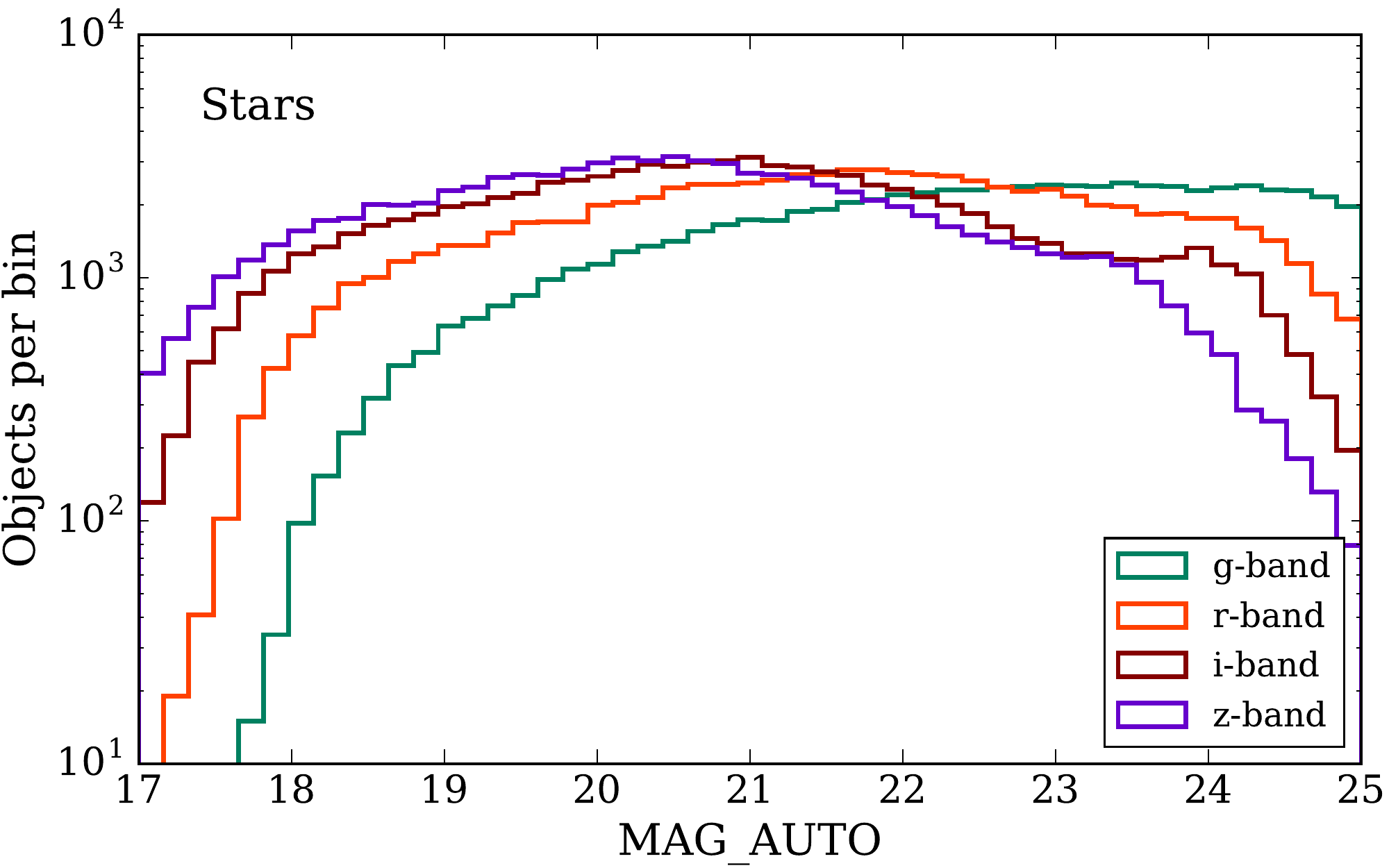}
\includegraphics[width=\columnwidth]{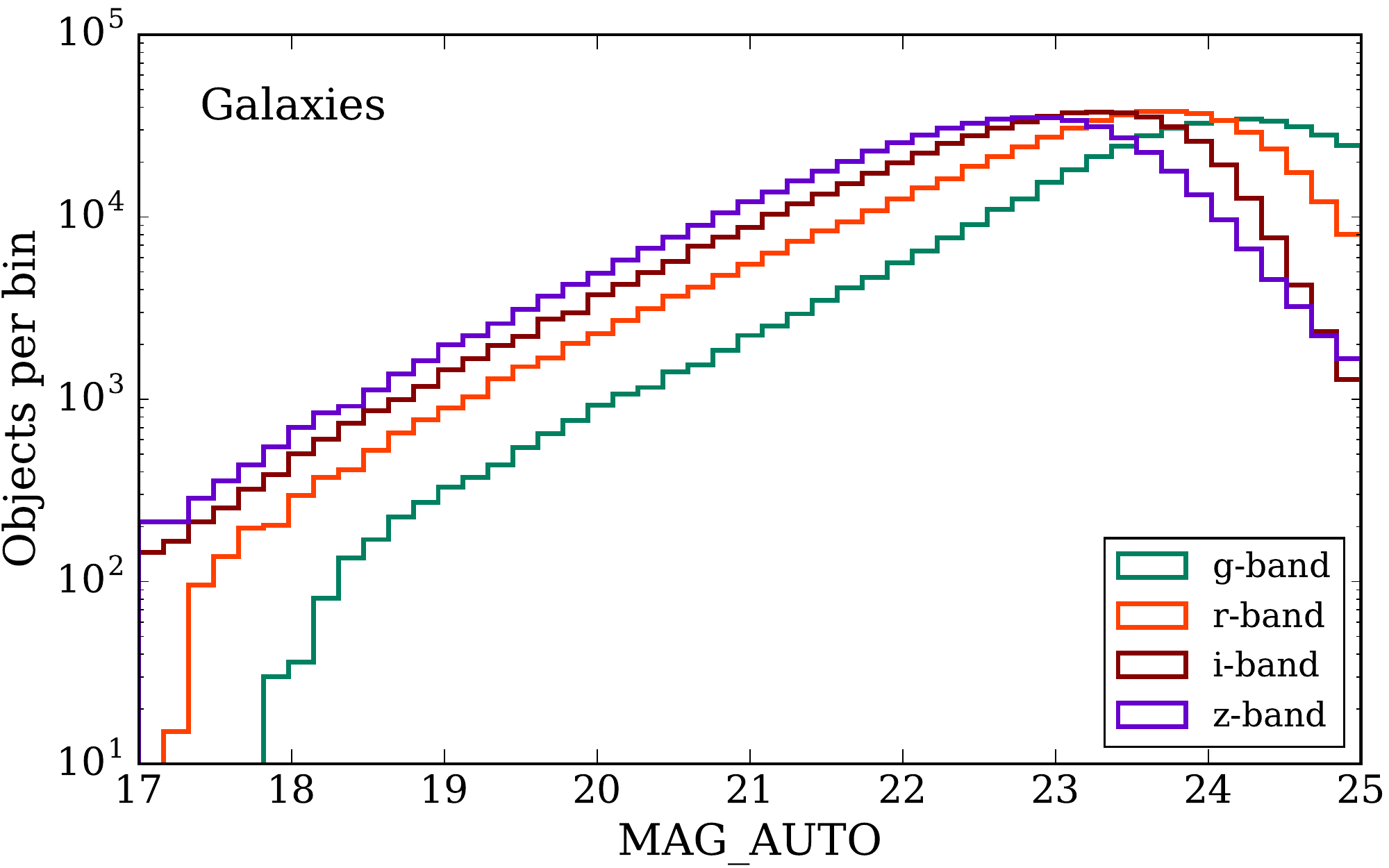}
\caption{\label{fig:modest_dndm} 
Number counts of objects passing the \var{MODEST\_CLASS} pure star selection (top) and complete galaxy selection (bottom) as a function of \var{MAG\_AUTO} magnitude in the $g$, $r$, $i$, and $z$ bands. 
The impact of galaxy contamination can be seen in the stellar number counts at magnitudes fainter than $i \gtrsim 23$.
}
\end{figure}

Following \citet{Drlica-Wagner:2015}, we validated the performance  of the \modest star-galaxy classifier on data from CFHTLenS \citep{Erben:2013,Hildebrandt:2012}.
We matched CFHTLenS catalog objects to the \gold data (\secref{catalog}) and selected high-quality samples of stars and galaxies using the \var{class\_star} and \var{fitclass} measurements by CFHTLenS \citep{Heymans:2012}.
Specifically, our CFHTLenS stellar selection was $(\var{fitclass} = 1) \OR (\var{class\_star} > 0.98)$ and our galaxy selection was $(\var{fitclass} = 0) \OR (\var{class\_star} < 0.2)$.
Note that $\roughly 7\%$ of matched CFHTLenS objects are unclassified according to this prescription, and these objects are not used for assessing the performance of \modest.

We define the ``efficiency'' of a galaxy sample as the number of true galaxies that are also classified as galaxies divided by the total number of true galaxies in the sample (i.e., the true positive rate). 
Conversely, the ``contamination'' of a galaxy sample is defined as the number of galaxies that are misclassified divided by the total number of objects classified as galaxies (i.e., the false discovery rate).
Similar definitions apply to the stellar selections, and the performance of the \modest galaxy and star selections are shown in \figref{modest_performance}.
We find that a high-purity galaxy selection has an efficiency $\gtrsim 98\%$ and a contamination rate $\lesssim 3\%$ for $i < 22$.
In contrast, the high-completeness stellar selection has an efficiency of $\gtrsim 86\%$ with a contamination  of $\lesssim 6\%$ for $i < 22$.
We estimate similar performance for \modest through a comparison against the DEEP2-3 field in the first public data release of Hyper Suprime Camera \citep{Aihara:2017}.

\begin{figure*}[!th]
\center
\includegraphics[width=\columnwidth]{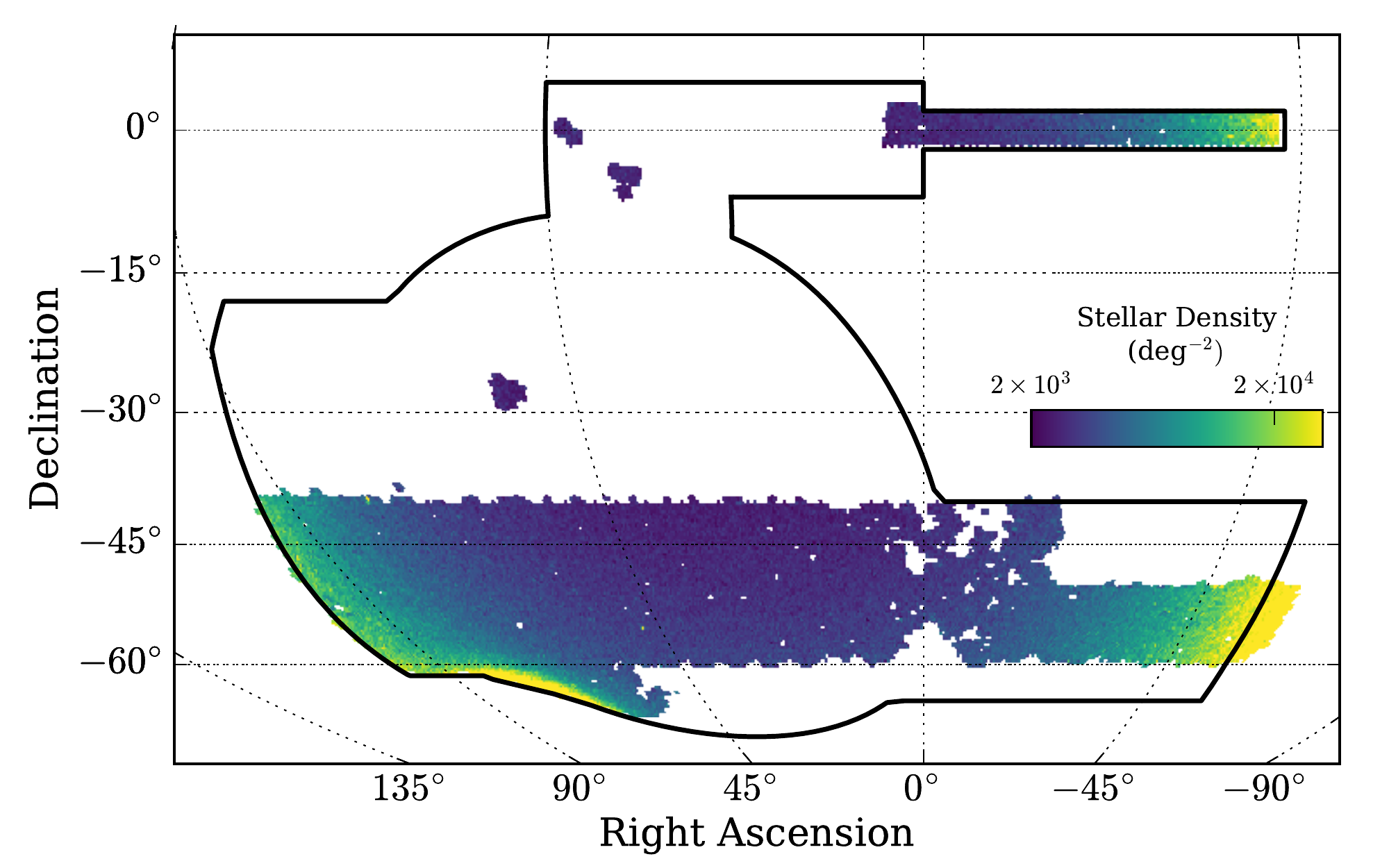}
\includegraphics[width=\columnwidth]{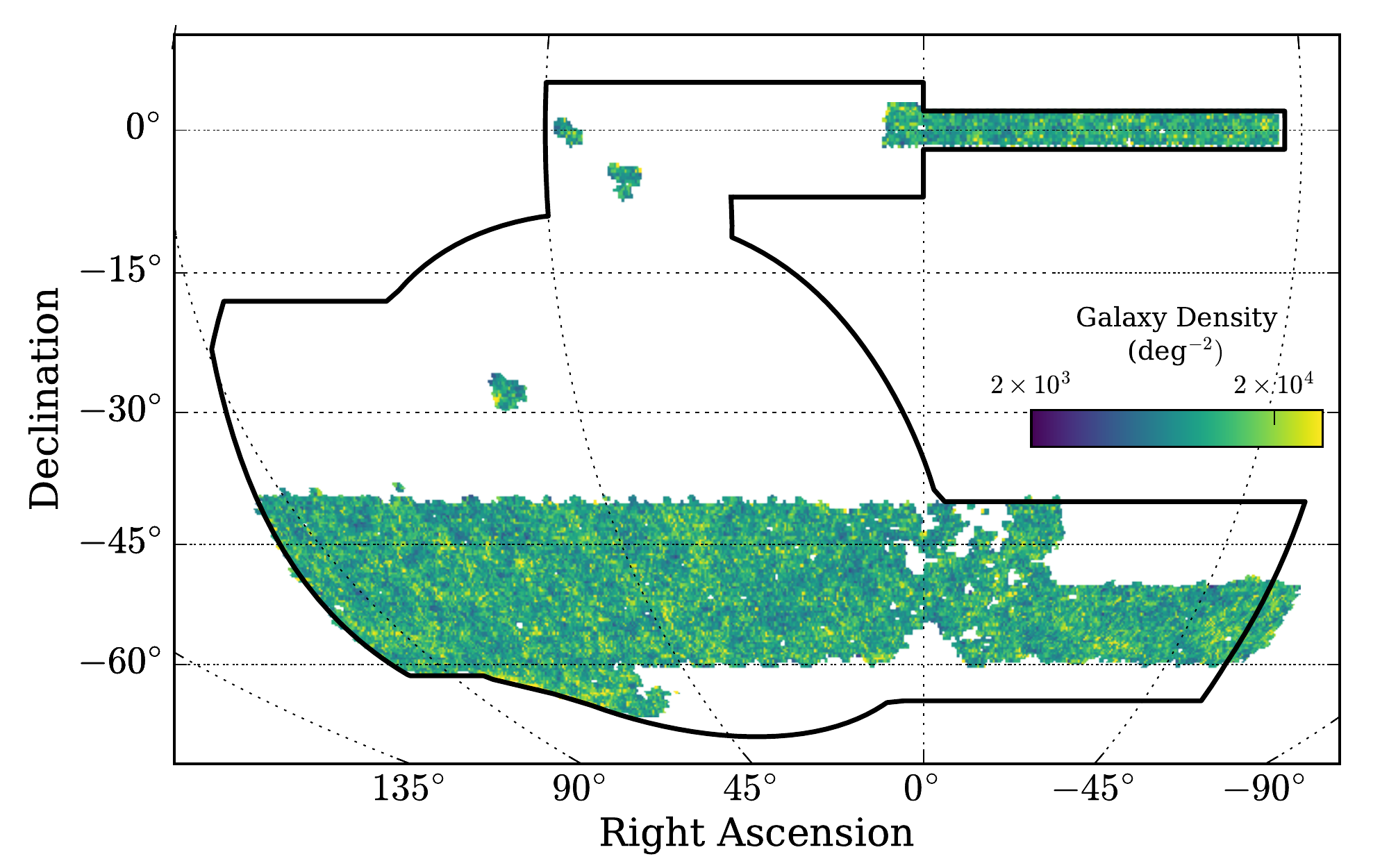}
\caption{\label{fig:stargal_map} Density of objects with $i < 22$ passing the high-completeness star (left) and high-purity galaxy (right) \modest selections.
The linear color scales represent the density of catalog objects and are the same for both panels.
The density of objects has been corrected for the coverage fraction of each pixel (\secref{depth}).
}
\end{figure*}

The \modest selection provides an initial baseline for object classification and is found to be sufficient for characterizing the distributions of stars and galaxies in \gold (Figures \ref{fig:modest_dndm} and \ref{fig:stargal_map}).
Multi-variate machine-learning techniques and template-fitting algorithms have the potential to provide much better object classification \citep[\eg][\etc]{Fadely:2012, Soumagnac:2015}. 
Several advanced object classification techniques are currently being explored within DES and will be detailed in future publications \citep{Sevilla:2017}.
We emphasize that \modest has been optimized for \emph{galaxy} selection. 
Several alternative selections have been suggested for more complete samples of stars \citep[\eg][]{Bechtol:2015,Drlica-Wagner:2015}.

\subsection{Photometric Redshift Estimation}
%(Hoyle, Bonnet, Lin, Rau, Hartley)
\label{sec:photoz}

In this section we briefly summarize the approach to \photoz estimation and validation for DES Y1 science analyses. %for the \gold data set. 
While \photoz estimates were provided as part of the initial \gold data set, it was realized that individual cosmology analyses benefit from \photoz estimation and validation customized to their distinct science samples.
Therefore, we present a general overview of the \photoz estimation and validation procedures, and we refer the reader to upcoming publications dedicated to \photoz estimation for distinct DES analyses \citep[\eg,][]{Hoyle:2017,Gatti:2017,Davis:2017,Cawthon:2017}.

\Photoz estimates were generated with two distinct algorithms: the machine-learning code \code{DNF} \citep{DeVicente:2016}, and a modified version of the template code \code{BPZ} \citep{Benitez:2000, Hoyle:2017}. 
These two codes are representative of common machine learning and template fitting \photoz estimation techniques.
Both algorithms utilized spectroscopic data for training, and a detailed discussion of the spectroscopic sample can be found in \citet{Gschwend:2017}.

For many cosmological analyses, we are interested in accurately characterizing the statistical distribution of galaxies in tomographic bins of redshift and less interested in predicting the redshift of any individual galaxy.
Thus, we applied two independent techniques targeted at validating the statistical properties of our predicted \photoz distributions \citep{Hoyle:2017,Davis:2017}.
\begin{enumerate}
\item We performed a direct validation of the color-redshift relationship by matching galaxies from DES science samples to galaxies with multi-band photometry obtained within the COSMOS field \citep{Laigle:2016}.
This choice of validation data mitigated the impact of redshift or galaxy-type dependent selection biases, which can affect spectroscopic surveys \citep[\eg,][]{Bonnett:2016,Hartley:2017}.
However, the 30-band \photoz estimates from COSMOS have a larger intrinsic uncertainty than spectroscopically determined redshifts.
In addition, validating performance on a $\roughly 2 \deg^2$ field leads to large uncertainty due to cosmic variance, which was estimated using the Buzzard suite of \LCDM simulations \citep{Sanchez:2017,Wechsler:2017,DeRose:2017}. 
\item A second, independent indirect validation technique relies on the clustering-redshift technique \citep{Newman:2008,Menard:2013,Schmidt:2013}.
We selected a luminous red galaxy sample \citep[redMaGiC;][]{Rozo:2016}, which has well-determined \photoz estimates, as a reference and divided this sample into redshift bins of width $\Delta z = 0.02$. 
We then divided the full sample of DES objects into tomographic redshift bins based on predicted \photoz and cross correlated the data in each tomographic bin with each of the more finely binned redMaGiC reference samples. 
We measured the excess angular cross-correlation signal, which is proportional to the redshift distribution. 
We calibrated a constant redshift offset in each tomographic bin between the \photoz predictions and the clustering signal. 
We estimated the errors arising from the evolution of galaxy-dark matter halo bias and discrepancies in the shape of the clustering reshift distribution by repeating the same analysis using the Buzzard simulations \citep{Gatti:2017,Cawthon:2017}. 
\end{enumerate}

Both validation techniques possess associated uncertainties.
The direct validation technique has comparable uncertainties from sample variance (since COSMOS covers a $\roughly 2 \deg^2$ region of the sky) and systematic uncertainty in matching the morphological and color-magnitude-error distribution of the galaxy sample. 
In contrast, we find that the dominant systematic uncertainties for the indirect validation technique come from the clustering bias evolution of the binned source galaxy samples and incorrectness in the shape of the \photoz distribution.
In addition, we are unable to perform indirect clustering validation for tomographic bins with $z \gtrsim 1$ owing to limited redMaGiC reference data at these redshifts.

The most important \photoz performance metric for cosmic shear analyses is the bias of the estimated mean of a redshift distribution in a tomographic bin with respect to the unknown true mean redshift in that bin \citep{Bonnett:2016}.
We characterized the \photoz accuracy from the \photoz bias distribution, defined as the difference between the average measured photometric redshift and the average true redshift distribution, $\Delta z = \langle z_{\rm true} \rangle - \langle z_{\rm phot} \rangle$. 
Since the true redshift distribution is unknown, we employed the direct and indirect validation techniques described above to estimate $\langle z_{\rm true} \rangle$ and $\Delta z$ in four tomographic bins with $0.2 < z < 1.3$.
We find that both techniques yield $|\Delta z| \lesssim 0.02$ with an uncertainty of comparable magnitude when applied to the BPZ estimates for the primary subsample of the \gold catalog used for cosmic shear analyses \citep{Hoyle:2017,Zuntz:2017}.

\begin{figure*}[th]
\center
\includegraphics[width=\columnwidth]{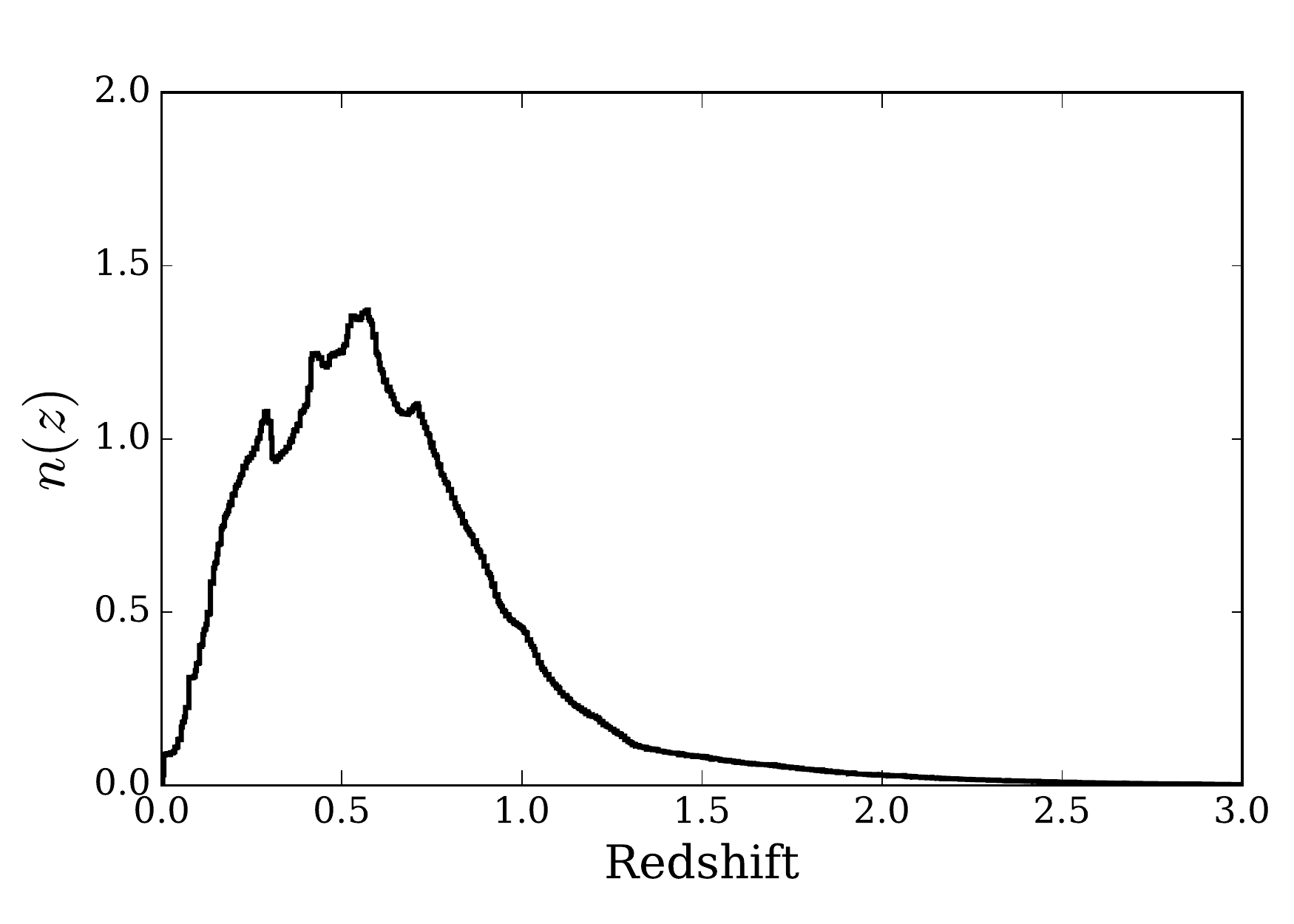}
\includegraphics[width=\columnwidth]{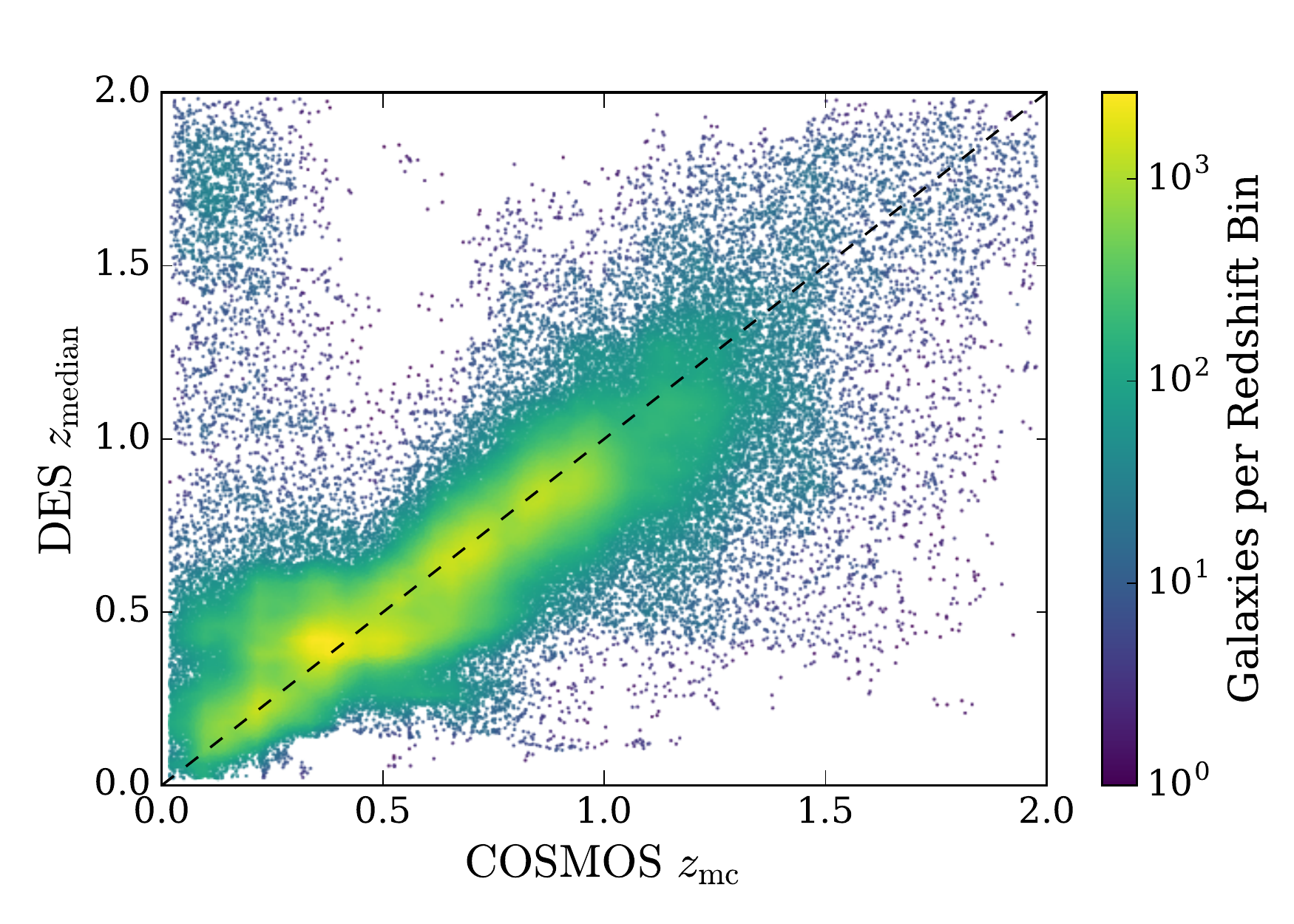}
\caption{\label{fig:photoz} Performance of the BPZ \photoz estimates for the weak-lensing subsample of \gold.
(Left): The $n(z)$ distribution generated from a random Monte Carlo sampling \citep[\ie,][]{Wittman:2009} of the BPZ $P(z)$ distribution for each galaxy.
(Right): Comparison between a random sampling of the 30-band COSMOS $P(z)$ \citep{Laigle:2016} and the median of the DES BPZ $P(z)$ for each galaxy in the overlapping sample. 
Points are colored by the local density of comparison galaxies in $0.04 \times 0.04$ redshift bins.
}
\end{figure*}

We present several other results from the validation of the BPZ template code optimized over the redshift range $0.2 < z < 1.3$ for the primary Y1 weak-lensing shear catalog \citep{Zuntz:2017}.
In \figref{photoz}, we show the $n(z)$ distribution for the weak-lensing shear catalog derived from \gold.
The $n(z)$ distribution is found to be in good agreement with the $n(z)$ predicted from COSMOS when cosmic variance and other associated systematic uncertainties are accounted for \citep{Hoyle:2017}.
We also show a comparison between the redshift estimate from a random sampling of the 30-band COSMOS $P(z)$ \citep{Laigle:2016} and the median \photoz derived from DES Y1 using BPZ.
Structure along the line of sight is visible in the higher-resolution COSMOS redshifts but is not resolved by DES.
For the full weak-lensing subsample, the normalized median absolute deviation (NMAD) of the quantity $(z_{\rm DES} - z_{\rm COSMOS})/(1+z_{\rm COSMOS})$ is $0.08 - 0.09$, depending on the point estimate used to determine the DES BPZ redshift.
When restricted to $i \leq 22$, the \photoz NMAD decreases to $0.06 - 0.07$.
Due to the strict selection requirements of the weak-lensing subsample, the NMAD for the full \gold galaxy sample is slightly larger ($\roughly 0.12$).
The photometric redshift accuracy for forthcoming DES Y1 cosmology analyses will be documented in more detail in \citet{Hoyle:2017,Cawthon:2017,Davis:2017,Gatti:2017}.

\section{Conclusion}
% (Drlica-Wagner, Sevilla, Rykoff)
\label{sec:discussion}
 
During its first year, DES imaged $\roughly 2000 \deg^2$ of the southern sky in each of the $g$, $r$, $i$, $z$ and $Y$ photometric filters.
These data have been processed, calibrated, coadded, cataloged, and characterized to form the DES \gold cosmology data sample, which covers $\roughly 1800 \deg^2$ with a depth of three to four tilings per band and a photometric calibration accuracy of $\lesssim 2\%$.
The photometric calibration uniformity of the \gold catalog was validated and adjusted via an SLR technique, which also corrects for the effects of interstellar extinction on the calibrated magnitudes of objects.
The development of \gold was driven by the goal of producing a maximal sample of Y1 data while minimizing the impact of systematic features.
Several ancillary maps characterizing the DES survey and its performance were produced as part of \gold.
In addition, a simple star-galaxy classifier and several photometric redshift estimates were also produced as necessary precursors to many DES science analyses.
The \gold data set is intended to be used as the nominal starting point for cosmological analyses with the DES Y1 data.
 
The next DES coadded data set will consist of exposures from the first three seasons of DES and will increase both the survey coverage ($\roughly 5000 \deg^2$) and depth (five to six tilings per band).
Improvements to the Blanco telescope infrastructure, data processing algorithms, and photometric calibration are expected to yield higher-quality data.
In addition, many of the improvements developed for \gold have been integrated into the core DESDM processing pipeline \citep[\eg,][]{Morganson:2017} and into automated tools for science catalog creation \citep[\eg,][]{FaustiNeto:2017}.
However, we anticipate that future data sets will still require the construction and validation of a high-quality data sample to serve as the basis for cosmological analyses.
On the longer term, we expect that a similar procedure for the assembling and validating cosmology data samples will be necessary for future surveys, such as LSST.
We hope that the production of \gold will help serve as a road map for assembling cosmology-ready data samples for future large photometric surveys.

\section*{Acknowledgments}

Funding for the DES Projects has been provided by the U.S. Department of Energy, the U.S. National Science Foundation, the Ministry of Science and Education of Spain, 
the Science and Technology Facilities Council of the United Kingdom, the Higher Education Funding Council for England, the National Center for Supercomputing 
Applications at the University of Illinois at Urbana-Champaign, the Kavli Institute of Cosmological Physics at the University of Chicago, 
the Center for Cosmology and Astro-Particle Physics at the Ohio State University,
the Mitchell Institute for Fundamental Physics and Astronomy at Texas A\&M University, Financiadora de Estudos e Projetos, 
Funda{\c c}{\~a}o Carlos Chagas Filho de Amparo {\`a} Pesquisa do Estado do Rio de Janeiro, Conselho Nacional de Desenvolvimento Cient{\'i}fico e Tecnol{\'o}gico and 
the Minist{\'e}rio da Ci{\^e}ncia, Tecnologia e Inova{\c c}{\~a}o, the Deutsche Forschungsgemeinschaft and the Collaborating Institutions in the Dark Energy Survey. 

The Collaborating Institutions are Argonne National Laboratory, the University of California at Santa Cruz, the University of Cambridge, Centro de Investigaciones Energ{\'e}ticas, 
Medioambientales y Tecnol{\'o}gicas-Madrid, the University of Chicago, University College London, the DES-Brazil Consortium, the University of Edinburgh, 
the Eidgen{\"o}ssische Technische Hochschule (ETH) Z{\"u}rich, 
Fermi National Accelerator Laboratory, the University of Illinois at Urbana-Champaign, the Institut de Ci{\`e}ncies de l'Espai (IEEC/CSIC), 
the Institut de F{\'i}sica d'Altes Energies, Lawrence Berkeley National Laboratory, the Ludwig-Maximilians Universit{\"a}t M{\"u}nchen and the associated Excellence Cluster Universe, 
the University of Michigan, the National Optical Astronomy Observatory, the University of Nottingham, The Ohio State University, the University of Pennsylvania, the University of Portsmouth, 
SLAC National Accelerator Laboratory, Stanford University, the University of Sussex, Texas A\&M University, and the OzDES Membership Consortium.

Based in part on observations at Cerro Tololo Inter-American Observatory, National Optical Astronomy Observatory, which is operated by the Association of Universities for Research in Astronomy (AURA) under a cooperative agreement with the National Science Foundation.

The DES data management system is supported by the National Science Foundation under Grant Numbers AST-1138766 and AST-1536171.
The DES participants from Spanish institutions are partially supported by MINECO under grants AYA2015-71825, ESP2015-88861, FPA2015-68048, SEV-2012-0234, SEV-2016-0597, and MDM-2015-0509, 
some of which include ERDF funds from the European Union. IFAE is partially funded by the CERCA program of the Generalitat de Catalunya.
Research leading to these results has received funding from the European Research
Council under the European Union's Seventh Framework Program (FP7/2007-2013) including ERC grant agreements 240672, 291329, and 306478.
We  acknowledge support from the Australian Research Council Centre of Excellence for All-sky Astrophysics (CAASTRO), through project number CE110001020.

This manuscript has been authored by Fermi Research Alliance, LLC under Contract No. DE-AC02-07CH11359 with the U.S. Department of Energy, Office of Science, Office of High Energy Physics. The United States Government retains and the publisher, by accepting the article for publication, acknowledges that the United States Government retains a non-exclusive, paid-up, irrevocable, world-wide license to publish or reproduce the published form of this manuscript, or allow others to do so, for United States Government purposes.

Support for DG was provided by NASA through Einstein Postdoctoral Fellowship grant number PF5-160138 awarded by the Chandra X-ray Center, which is operated by the Smithsonian Astrophysical Observatory for NASA under contract NAS8-03060.
ACR is supported by CNPq process 157684/2015-6.

This work is based on observations obtained with MegaPrime/MegaCam, a joint project of CFHT and CEA/IRFU, at the Canada-France-Hawaii Telescope (CFHT) which is operated by the National Research Council (NRC) of Canada, the Institut National des Sciences de l'Univers of the Centre National de la Recherche Scientifique (CNRS) of France, and the University of Hawaii. This research used the facilities of the Canadian Astronomy Data Centre operated by the National Research Council of Canada with the support of the Canadian Space Agency. CFHTLenS data processing was made possible thanks to significant computing support from the NSERC Research Tools and Instruments grant program.

This publication makes use of data products from the Two Micron All Sky Survey, which is a joint project of the University of Massachusetts and the Infrared Processing and Analysis Center/California Institute of Technology, funded by the National Aeronautics and Space Administration and the National Science Foundation.

This research was made possible through the use of the AAVSO Photometric All-Sky Survey (APASS), funded by the Robert Martin Ayers Sciences Fund.

The UKIDSS project is defined in \citet{Lawrence:2007}. UKIDSS uses the UKIRT Wide Field Camera \citep[WFCAM;][]{Casali:2007}. The photometric system is described in \citet{Hewett:2006}, and the calibration is described in \citet{Hodgkin:2009}. The pipeline processing and science archive are described in Irwin et al (2009, in prep) and \citet{Hambly:2008}. \nocite{Irwin:2009}

Funding for SDSS-III has been provided by the Alfred P. Sloan Foundation, the Participating Institutions, the National Science Foundation, and the U.S. Department of Energy Office of Science. The SDSS-III web site is http://www.sdss3.org/.

SDSS-III is managed by the Astrophysical Research Consortium for the Participating Institutions of the SDSS-III Collaboration including the University of Arizona, the Brazilian Participation Group, Brookhaven National Laboratory, Carnegie Mellon University, University of Florida, the French Participation Group, the German Participation Group, Harvard University, the Instituto de Astrofisica de Canarias, the Michigan State/Notre Dame/JINA Participation Group, Johns Hopkins University, Lawrence Berkeley National Laboratory, Max Planck Institute for Astrophysics, Max Planck Institute for Extraterrestrial Physics, New Mexico State University, New York University, Ohio State University, Pennsylvania State University, University of Portsmouth, Princeton University, the Spanish Participation Group, University of Tokyo, University of Utah, Vanderbilt University, University of Virginia, University of Washington, and Yale University.

\facility{Blanco (DECam)} 
\software{\SExtractor \citep{Bertin:1996}, \PSFEx \citep{Bertin:2011}, \scamp \citep{Bertin:2006}, \swarp \citep{Bertin:2002,Bertin:2010}, \mangle \citep{Hamilton:2004,Swanson:2008}, \healpix \citep{Gorski:2005},\footnote{\url{http://healpix.sourceforge.net}} \code{astropy} \citep{Astropy:2013}, \code{matplotlib} \citep{Hunter:2007}, \code{numpy} \citep{numpy:2011}, \code{scipy} \citep{scipy:2001}, \code{healpy},\footnote{\url{https://github.com/healpy/healpy}} \code{fitsio}\footnote{\url{https://github.com/esheldon/fitsio}}, \ngmix \citep{Sheldon:2014}.\footnote{\url{https://github.com/esheldon/ngmix}}}

\appendix
\numberwithin{figure}{section}
\numberwithin{table}{section}

\section{Photometric Calibration}
% (Tucker, Drlica-Wagner, Rykoff, ...)
\label{app:calibration}
 
In this Appendix, we provide more details on the photometric calibration of Y1A1, including nightly calibration (\appref{psm}), global calibration (\appref{gcm}), and a SLR adjustment (\appref{slr}). 
We note that the nightly and global calibration steps followed on the procedure of \citet{Tucker:2007} and were performed on the single-epoch catalog data {\em before} coaddition. 
In contrast, the SLR adjustment was performed on the weighted-average magnitudes of multiple single-epoch catalogs and is applied directly to the coadded object catalogs.
A collection of transformation equations between DES and several other surveys is provided in \appref{transform}.
 
\subsection{Nightly Photometric Calibration}
\label{app:psm}
 
\begin{figure*}[th]
\center
\includegraphics[width=0.8\textwidth]{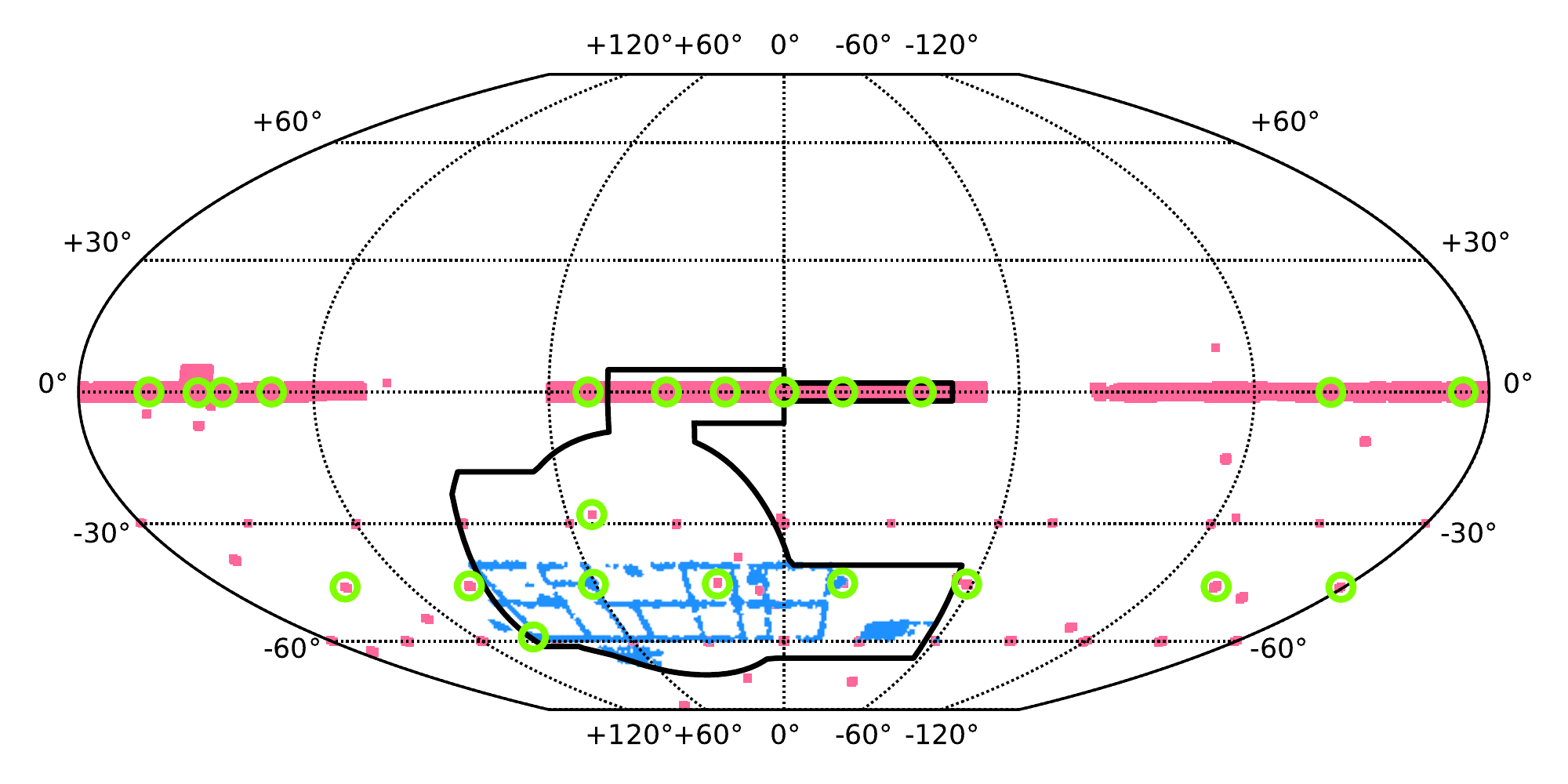}
\caption{\label{fig:standards} 
Standard stars used for the photometric calibration of DES Y1.
Nightly standard stars fields (\tabref{nightly_std_fields}) are marked with green circles while other secondary standards from SDSS and \citet{Smith:2017} are shown in pink.
The grid work of tertiary standards is shown in blue.
}
\end{figure*}

The first step in DES Y1 photometric calibration used observations of standard-star fields to derive a set of calibration coefficients for each photometric night.
A subset of the standard-star fields listed in \tabref{nightly_std_fields} were observed at different airmasses at the beginning and end of each DES night or half night (\secref{observations}).
The DES nightly standard-star fields are predominantly located in the equatorial fields of SDSS Data Release 9 \citep[DR9;][]{Ahn:2012}, with the addition of several fields from the Southern $u'g'r'i'z'$ Standard Network \citep{Smith:2017}.\footnote{\url{http://www-star.fnal.gov}} 
Devoted observations of these standard-star fields were supplemented by DES survey observations that overlapped the standard-star fields.
For $Y$ band, we used stars from the equatorial fields of the UKIRT Infrared Deep Sky Survey Data Release 6 \citep[UKIDSS DR6; ][]{Lawrence:2007}  matched against SDSS stars. 
All nightly standard stars were transformed to an initial DES AB photometric system via matching to objects in SDSS and UKIDSS (\appref{transform}). 
The spatial distribution of the DES standard-star fields is shown in \figref{standards}.

Due to its provenance, primarily from SDSS DR9, we refer to the set of DES nightly standards as {\em secondary} standards. 
The {\em fundamental} standard for SDSS was the F subdwarf star BD$+$17$^{\circ}$~4708, which was used for calibrating the set of SDSS {\em primary} standards \citep{Smith:2002}. 
These primary standards were in turn used (indirectly) in the ubercalibration of SDSS \citep{Padmanabhan:2008}.  
Thus, the DES secondary standards tie the absolute flux calibration of DES to the SDSS primary standards, to BD$+$17$^{\circ}$~4708, and ultimately to the AB magnitude system.
 
\begin{deluxetable}{lllrrrrr}[t!]
\tabletypesize{\tablesize}
\tablecaption{DES Nightly Standard-star Fields\label{tab:nightly_std_fields}}
\tablewidth{0pt}
\tablehead{
  \colhead{Field Name} & 
  \colhead{RA} & 
  \colhead{Dec} & 
  \multicolumn{5}{c}{Exposure Time (sec)} \\
  \colhead {} &
  \colhead {J2000} &
  \colhead {J2000} &
  \colhead{$g$} & 
  \colhead{$r$} & 
  \colhead{$i$} & 
  \colhead{$z$} & 
  \colhead{$Y$}
} 
\startdata
\multicolumn{8}{c}{{\em Preferred Fields\tablenotemark{a}}} \\ 
SDSS\,J2140-0000  & 21:40:00 & $+$00:00:00 & 15 & 15 & 15 & 15 & 20 \\
SDSS\,J2300-0000  & 23:00:00 & $+$00:00:00 & 15 & 15 & 15 & 15 & 20 \\
SDSS\,J0000-0000  & 00:00:00 & $+$00:00:00 & 15 & 15 & 15 & 15 & 20 \\
SDSS\,J0100-0000  & 01:00:00 & $+$00:00:00 & 15 & 15 & 15 & 15 & 20 \\
SDSS\,J0200-0000  & 02:00:00 & $+$00:00:00 & 15 & 15 & 15 & 15 & 20 \\
SDSS\,J0320-0000  & 03:20:00 & $+$00:00:00 & 15 & 15 & 15 & 15 & 20 \\
SDSS\,J0843-0000  & 08:43:00 & $+$00:00:00 & 15 & 15 & 15 & 15 & 20 \\
SDSS\,J0933-0005  & 09:33:00 & $-$00:05:00 & 15 & 15 & 15 & 15 & 20 \\
SDSS\,J0958-0010  & 09:58:00 & $-$00:10:00 & 15 & 15 & 15 & 15 & 20 \\
SDSS\,J1048-0000  & 10:48:00 & $+$00:00:00 & 15 & 15 & 15 & 15 & 20 \\
SDSS\,J1227-0000  & 12:27:00 & $+$00:00:00 & 15 & 15 & 15 & 15 & 20 \\
SDSS\,J1442-0005  & 14:42:00 & $-$00:05:00 & 15 & 15 & 15 & 15 & 20 \\
C26202/HST      & 03:32:30 & $-$27:46:05 & 15 & 15 & 15 & 15 & 20 \\
MaxVis          & 06:30:00 & $-$58:45:00 & 15 & 15 & 15 & 15 & 20 \\ 
\tableline 
\multicolumn{8}{c}{{\em Supplemental Fields\tablenotemark{b}}} \\ 
SA~E1-A         & 01:24:50 & $-$44:33:40 &  3 &  3 &  3 &  3 &  5 \\
SA~E2-A         & 04:03:00 & $-$44:41:45 &  3 &  3 &  3 &  3 &  5 \\
SA~E3-A         & 06:42:54 & $-$45:05:06 &  3 &  3 &  3 &  3 &  5 \\
SA~E4-A         & 09:23:44 & $-$45:21:02 &  3 &  3 &  3 &  3 &  5 \\
SA~E5-A         & 12:04:11 & $-$45:24:03 &  3 &  3 &  3 &  3 &  5 \\
SA~E6-A         & 14:45:33 & $-$45:15:34 &  3 &  3 &  3 &  3 &  5 \\
SA~E8-A         & 20:07:22 & $-$44:37:01 &  3 &  3 &  3 &  3 &  5 \\
SA~E9-A         & 22:45:37 & $-$44:22:47 &  3 &  3 &  3 &  3 &  5 \\
\enddata
\tablenotetext{a} {The preferred fields (with the exception of C26202/HST and MaxVis) have photometric standard stars covering the entire DECam focal plane. This permits photometric zeropoints to be determined for every CCD using a single exposure.}
%C26202/HST\tablenotemark{b}      & 03:32:30 & -27:46:05 & 30 & 15 & 15 & 15 & 20 \\
%\tablenotetext{b} {We offset these fields by $5\arcmin$ to the North to place a particular star or set of stars at the center of the DECam CCD35.}
\tablenotetext{b} {The supplemental fields have photometric standard stars covering a $10 \arcmin \times 10 \arcmin$ region and are typically used for expanding the range of airmasses when no suitable primary field is observable.  These particular supplemental fields come from the Southern $u’g’r’i’z’$ Standard Stars project \citep{Smith:2017}.}

\end{deluxetable}

Nightly observations of the secondary standards were used to fit a set of photometric equations.  
These photometric equations, which are based on those used by SDSS \citep{Tucker:2006}, have the form
\begin{align}
g_0 &= -2.5\log_{10}(F_g) - a_g - b_g \times ( (g-r)_0 - (g-r)_{\rm fid} ) - k_g \times X \label{eqn:psm_g} \\ 
r_0 &= -2.5\log_{10}(F_r) - a_r - b_r \times ( (g-r)_0 - (g-r)_{\rm fid} ) - k_r \times X \label{eqn:psm_r} \\ 
i_0 &= -2.5\log_{10}(F_i) - a_i - b_i \times ( (i-z)_0 - (i-z)_{\rm fid} ) - k_i \times X \label{eqn:psm_i} \\ 
z_0 &= -2.5\log_{10}(F_z) - a_z - b_z \times ( (i-z)_0 - (i-z)_{\rm fid} ) - k_z \times X \label{eqn:psm_z} \\ 
Y_0 &= -2.5\log_{10}(F_Y) - a_Y - b_Y \times ( (z-Y)_0 - (z-Y)_{\rm fid} ) - k_Y \times X \label{eqn:psm_Y}
\end{align}
\noindent where $\lambda_0$ ($\lambda={g,r,i,z,Y})$ is the calibrated standard-star magnitude in the DES system,
$F_\lambda$ is the observed PSF flux (counts/sec),
$a$ is the photometric zeropoint for the night,
$b$ is the instrumental color term coefficient,
$(g-r)_0$, $(i-z)_0$, $(z-Y)_0$ are the calibrated standard-star colors,
$(g-r)_{\rm fid}$, $(i-z)_{\rm fid}$, and $(z-Y)_{\rm fid}$ are fiducial reference colors (chosen so that the effects of $b$ are relatively small for a star of typical color within the DES footprint),
$k$ is the first-order extinction coefficient, 
and $X$ is the airmass of the observation.  
The values for the fiducial colors are $(g-r)_{\rm fid} =  0.53$, $(i-z)_{\rm fid}= 0.09$, and $(z-Y)_{\rm fid} = 0.05$.
 
\begin{figure*}[t]
\centering
\includegraphics[width=\textwidth]{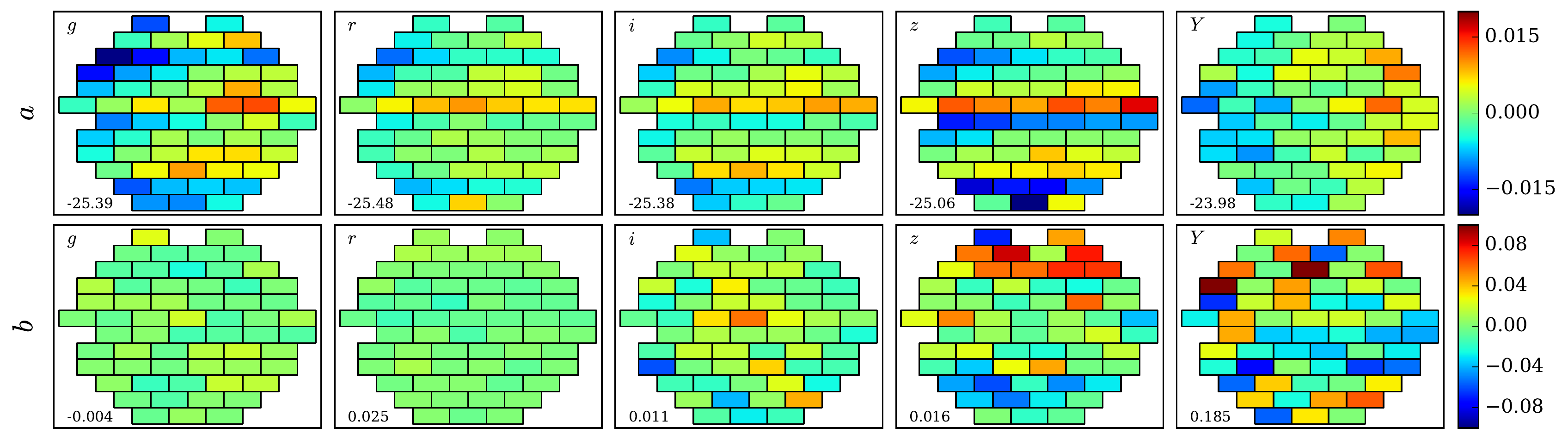}
\caption{\label{fig:psm_ab_vs_xy} Median fit values from DES Y1 for the $a$ and $b$ coefficients as a function of position on the DECam focal plane. 
The color scale represents the offset in the median fit value for each CCD with respect to the median for the focal plane as listed in the bottom left of each panel.
The spatial variations in photometric zeropoints across the DECam focal plane are typically less than 0.02--0.03 mag for Y1A1.}
\end{figure*}

\begin{figure}[h]
\centering
\includegraphics[width=0.5\columnwidth]{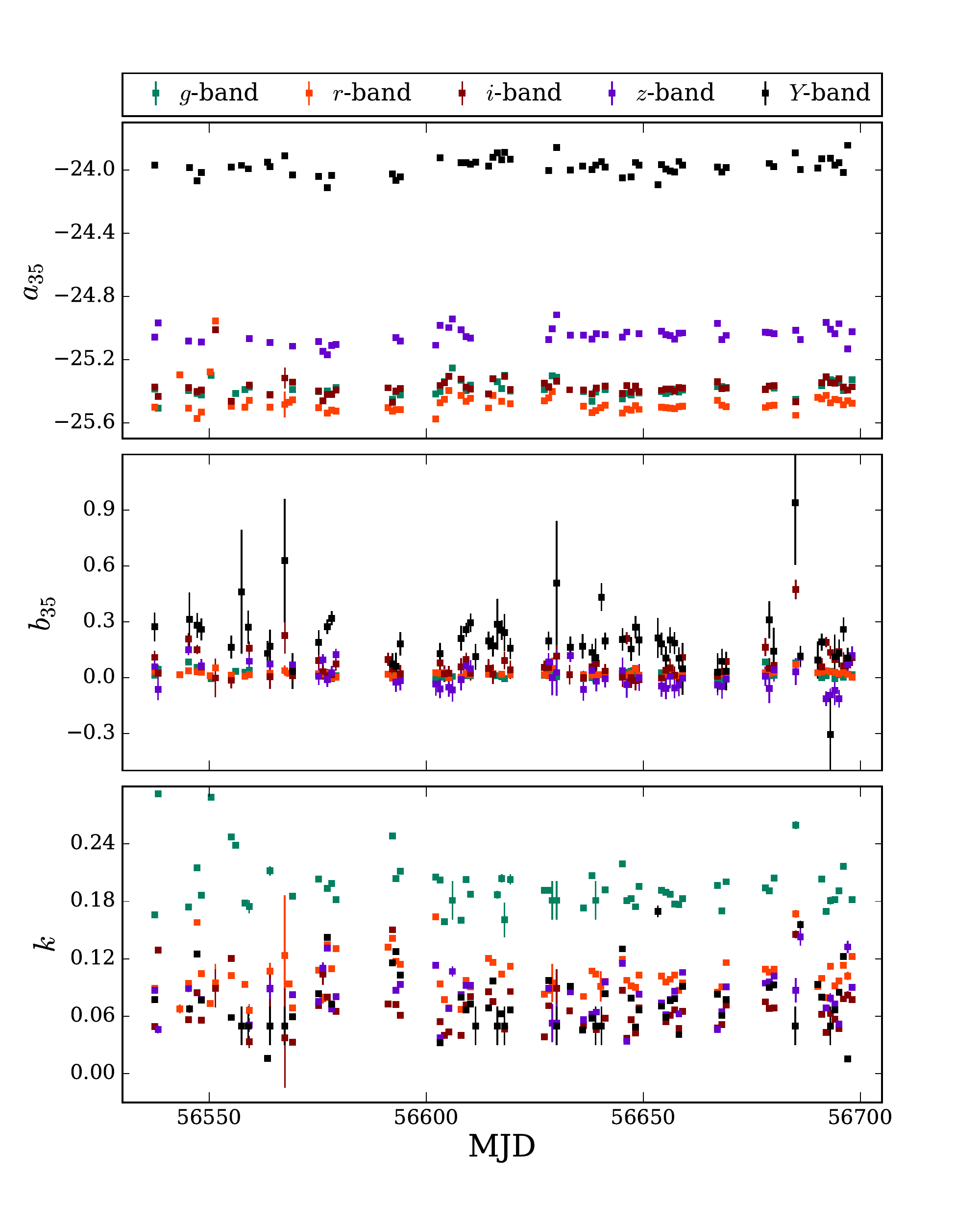}
\caption{\label{fig:psm_vs_time} Nightly fit values for the $a$ coefficient of CCD35 (top), the $b$ coefficient of CCD35 (middle), and the $k$ coefficient  of the DECam focal plane (bottom) for nights in DES Y1. 
For the $a$ and $b$ coefficients, the trends in CCD35 are found to be representative of the trends in the other CCDs.
}
\end{figure}

Separate $a$ and $b$ coefficients were determined for each functioning science CCD, while a single value of $k$ was assumed for the full focal plane.
The nightly values of $a$ track the overall throughput of the DECam instrument at the location of each CCD, and variations in $a$ mostly track the gradual accumulation of dust on the Blanco primary mirror.
The nightly values of $b$ track variations in the {\em shape} of the total filter response curve at the location of each CCD (including atmospheric transmission). 
Under photometric conditions, the value of $k$ should not vary across the focal plane, and a single value of $k$ was fit for the full focal plane.
Variations in the nightly values of $k$ track the relative throughput of the atmosphere at CTIO.
The median values of $a$ and $b$ are shown for each science CCD in \figref{psm_ab_vs_xy}, and the nightly variations of $a$, $b$, and $k$ are shown in \figref{psm_vs_time}.
The site-average values for the $a$, $b$, and $k$ coefficients are tabulated in \tabref{psm_site_values}.

We note that, despite the use of star flats and pupil corrections, there are still minor variations in the zeropoints across the focal plane in \figref{psm_ab_vs_xy}.  
These variations can be attributed to two main sources: 
(1) the DES starflat procedure is subject to small flat/planar gradients across the focal plane, with the understanding that the photometric calibration procedure will remove such gradients; and 
(2) CCD-to-CCD variations in quantum efficiency have not been fully accounted for in the Y1A1 image processing, and this is also reflected in the smaller
scale between-CCD variations in the zeropoints.
The DES Y3 processing has largely corrected for variations in quantum efficiency while \citet{Y3FGCM} show that gradients in the star flats can be successfully removed by the photometric calibration.
 
The code used to perform the nightly fits is called the Photometric Standards Module (\PSM).\footnote{\url{https://github.com/DarkEnergySurvey/PSM}}
We note that \PSM not only fits Equations (\ref{eqn:psm_g})--(\ref{eqn:psm_Y}), but also performs an automated initial culling of non-photometric data using the outputs of RASICAM \citep{Lewis:2010,Reil:2014}.  
\PSM also culls dome-occulted exposures (identified by a strong gradient in $a$ across the focal plane) and performs iterative sigma-clipping to achieve a good solution for a night.  
For the Y1 data set, we also culled nights with rms fit residuals $> 0.025\magn$ or an atypical ($\roughly 2$--$3\sigma$ outlier) fit value for the first-order extinction. 
The typical relative calibration scatter from the \PSM solution is $\roughly 0.02 \magn$ rms.
This scatter includes the contribution from stellar shot noise in the standard-star observations.

\begin{deluxetable}{ccrrrrr} 
\tabletypesize{\tablesize}
\tablecaption{\label{tab:psm_site_values} DES Y1 Average \PSM Fit Values}
\tablewidth{0pt}
\tablehead{
  \colhead{Coeff.} &
  \colhead{Band} & 
  \colhead{Median} & 
  \colhead{Mean} & 
  \colhead{$\sigma$} & 
  \colhead{Mean error} & 
  \colhead{Number} 
}
\startdata
\multirow{5}{*}{$a$} & $g$ & $-$25.396 & $-$25.389 & 0.044 & 0.001 & 3717  \\    
                     & $r$ & $-$25.501 & $-$25.485 & 0.079 & 0.001 & 4072  \\    
                     & $i$ & $-$25.386 & $-$25.380 & 0.059 & 0.001 & 3806  \\    
                     & $z$ & $-$25.056 & $-$25.056 & 0.049 & 0.001 & 3080  \\    
                     & $Y$ & $-$23.976 & $-$23.979 & 0.051 & 0.001 & 3627  \\ 
\tableline
\multirow{5}{*}{$b$} & $g$ &  $-$0.004 & $-$0.004 & 0.017 & $<$0.001 & 3717  \\
                     & $r$ &   0.024   &  0.024   & 0.014 & $<$0.001 & 4072  \\
                     & $i$ &   0.012   &  0.012   & 0.057 & 0.001 & 3806  \\
                     & $z$ &   0.020   &  0.020   & 0.071 & 0.001 & 3080  \\
                     & $Y$ &   0.187   &  0.187   & 0.123 & 0.002 & 3627  \\ 
\tableline
\multirow{5}{*}{$k$} & $g$ &   0.191 & 0.196 & 0.027 & 0.003 &   63 \\ 
                     & $r$ &   0.099 & 0.102 & 0.021 & 0.003 &   69 \\ 
                     & $i$ &   0.065 & 0.068 & 0.024 & 0.003 &   64 \\ 
                     & $z$ &   0.083 & 0.081 & 0.023 & 0.003 &   52 \\ 
                     & $Y$ &   0.070 & 0.075 & 0.031 & 0.004 &   61 \\ 
\enddata
\tablecomments{Statistics were calculated for nights in Y1 with a good \PSM fit. 
The $a$ and $b$ values were calculated individually for each CCD, while the $k$ values were calculated for the full focal plane. 
}
\end{deluxetable}

\subsection{Global Calibration}
% (Tucker, Drlica-Wagner)
\label{app:gcm}
 
%\COMMENT{DLT: More details in DES-doc\#6584, 8375, 8456, and 8708}

In addition to deriving the nightly calibration coefficients for each photometric night in DES Y1, we would like to calibrate exposures taken under cloudy conditions and exposures where the nightly solution failed (\eg, due to contrails).
We would also like to improve on the $\roughly2\%$ rms relative calibration uncertainty achieved by the \PSM solution. 
To achieve both of these goals, we applied a global calibration to simultaneously calibrate all overlapping CCD images in the Y1 data set.
In addition to calibrating images that lacked a \PSM solution, the global calibration can achieve a relative calibration between overlapping images at the level of $0.003\magn\ (0.3\%)$ rms, even if the exposures were taken under cloudy conditions.

The global calibration was implemented as a Global Calibrations Module (\GCM).\footnote{\url{https://github.com/DarkEnergySurvey/GCM}} 
The \GCM generalizes the procedure of \citet{Glazebrook:1994} by replacing overlapping image ``frames'' with arbitrarily shaped overlapping catalog data sets (these are still conventionally referred to as ``images'').
The procedure is summarized briefly as follows.
\begin{enumerate}
\item For each filter, consider $n$ data sets for which $(1, \ldots, m)$ are uncalibrated and $(m+1,\ldots,n)$ are calibrated. In most cases these data sets represent object catalogs from individual exposures or CCD images. However, the calibrated data set consists of standard stars spanning the entire Y1A1 footprint (\figref{standards}).
\item Compile a list of all unique pairs of observations of a common star on two data sets. 
\item For a given pair of images, $i,j$, let 
\begin{equation}
\Delta_{ij} = \underset{\rm pairs}{\rm median}(m_i - m_j),
\end{equation}
\noindent where $m_i$ is the magnitude of a star in image $i$, $m_j$ is the magnitude of the same star in image $j$, and the median is calculated over matched pairs of stars. Note that $\Delta_{ij} = -\Delta_{ji}$.  
\item Let $ZP_i$ be a floating zero-point that can be applied to the data set from image $i$ to produce calibrated magnitudes. For images that are already calibrated ($i > m$), we fix $ZP_i = 0$.
\item Let $\theta_{ij}$ define a function that selects overlapping image pairs. We define $\theta_{ij} = 1$ if $i=j$ or if $i$ and $j$ overlap; otherwise $\theta_{ij} = 0$.  
\item To find calibrated zeropoints for each image, we minimize the sum of squares,
\begin{equation}
S = \sum_{i,j} \theta_{ij} (\Delta_{ij} + ZP_i - ZP_j )^{2}. \label{eqn:gcm_minimize}
\end{equation}
\item We derive a calibrated magnitude for each object detected on image $i$ (where $i < m$) by adding $ZP_i$ to the raw instrumental magnitudes.
\end{enumerate}
 
% gcm_mag_psf = desdm_mag_psf + 2.5*log10(exptime) - 25
% desdm_mag_psf = -2.5*log10(flux_psf) + 25
% gcm_mag_psf = -2.5*log10(cts_psf) + 2.5*log10(exptime) 
% sextractor flux_psf is really cts_psf
% gcm_raw_mag_psf = -2.5*log10(flux_psf)
% gmc_cal_mag_psf = -2.5*log10(flux_psf) + 25 + ZPi
 
\begin{figure}[t]
\center
\begin{minipage}{0.49\textwidth}
\includegraphics[width=\textwidth]{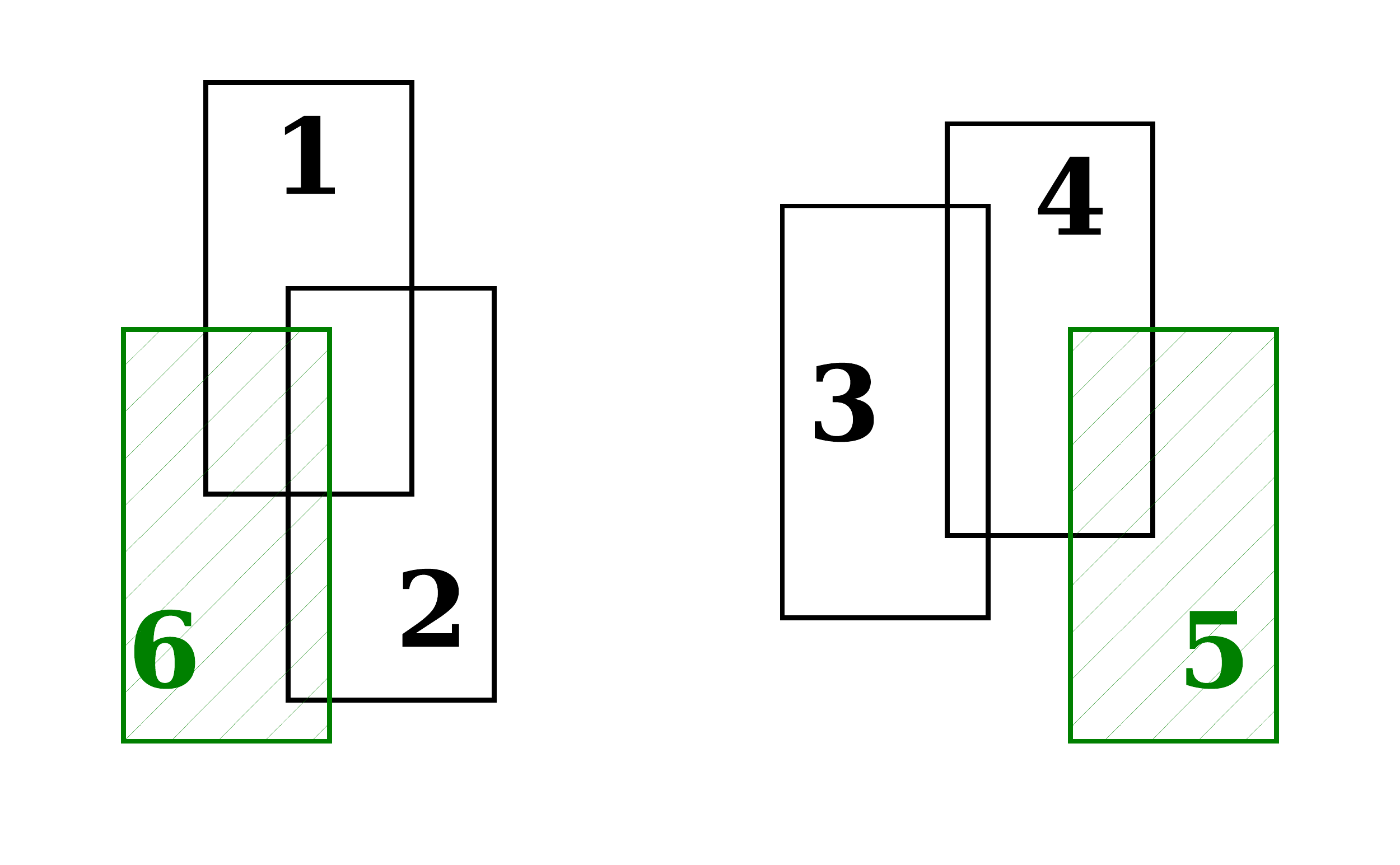}
\end{minipage}
\begin{minipage}{0.49\textwidth}
\input{gcm_example.tex}
\end{minipage}
\caption{\label{fig:GCM_example} A schematic of the \GCM algorithm based on Fig.~1 of \cite{Glazebrook:1994}.  {\em Left:}  The stars in images 5 and 6 have been previously calibrated while the stars in the other images are uncalibrated.  The algorithm minimizes the zeropoint offsets from all the overlapping images.   Images that have a connected path via overlapping images to a reference image can be calibrated to that reference image. {\em Right:} The corresponding matrix equation for this set of images.
}
\end{figure}

In \figref{GCM_example}, we show a simple example of the \GCM algorithm on two disconnected groups of three overlapping data sets (\ie, images).  
In each group, one of the overlapping images has been previously calibrated and serves as the reference against which the other images in its grouping are calibrated.  
To be calibrated, an uncalibrated image needs either to overlap a calibrated image (\eg, the left group in \figref{GCM_example}) or to have an unbroken path of overlapping images to a calibrated image (\eg, image 3 in the right group of \figref{GCM_example}).  
In the right panel of \figref{GCM_example} we show the matrix equation that minimizes Equation~(\ref{eqn:gcm_minimize}) for this particular set of images \citep{Glazebrook:1994}.  
Note that, via this matrix equation, the zeropoints for the two calibrated images (images 5 and 6) have been fixed to a value of zero ($1 \times ZP_{5} = 0$ and $1 \times ZP_{6} = 0$), since no offset is applied to these previously calibrated images.

Following the prescription of \citet{Glazebrook:1994}, we estimate the rms magnitude residual for each CCD image, $i$, from overlap with other CCD images, $j$, as
\begin{equation}
\rm{rms}_i = \sqrt{ \frac{\sum_{j} \theta_{ij} (\Delta_{ij} + ZP_i - ZP_j )^2}{\sum_{j} \theta_{ij} } }.
\end{equation}
The rms distribution over all CCD images is a measure of the internal (reproducibility) errors on small scales (the scales of overlapping CCD images) and is a measure of the precision of the overall \GCM solution.

In principle, the \GCM method is very precise, but carries the caveat that any small systematic gradients in the flat fielding of individual images can cause low-amplitude gradients over large scales.  
We used the set of secondary standard stars and a sparse gridwork of tertiary standard stars (\figref{standards}) as an ``anchor'' to keep the GCM fit from drifting due to any small systematic gradients in the \finalcut exposures.  
We note that the sparse gridwork of tertiary standards in the SPT region was extracted from stars in photometric exposures, calibrated by the nightly \PSM results.  
From the full set of \PSM-calibrated exposures in the SPT region, we selected a sparse gridwork of thin (1-degree-wide) ``struts'' of constant right ascension and constant declination.
This morphology was chosen so that the tertiaries would anchor the \GCM solution on large scales ($>10$--$15\deg$), but on smaller scales the calibration would be dominated by the \GCM solution for overlapping uncalibrated exposures.
 
The \GCM algorithm relies on having at least one calibrated image or data set to anchor each isolated image group. To identify isolated groups of exposures, we employed a group-finding algorithm developed for studies of galaxy clusters and large-scale structure \citep{Huchra:1982}.
For \finalcut, there were several disconnected image groups -- in particular, the SPT region, the S82 region, the four SN fields (SN-E was treated as an isolated group even though it overlaps the SPT area), the COSMOS field, and the VVDS-14h field (see \figref{footprint}).  
We therefore ran \GCM separately on each of these eight regions.  
The S82, COSMOS, and VVDS-14h fields overlap with the equatorial region of SDSS and were anchored by the secondary standards (mostly derived from SDSS).  
SPT was anchored by the aforementioned gridwork of tertiaries, supplemented with individual fields from DES SV. 
The SN fields also used individual standard-star fields from SV for their calibrators.  As with the gridwork of tertiary standards, the individual SV fields had been previously calibrated using nightly results from the \PSM code \citep{Wyatt:2014}. 
 
The S82 region and the smaller individual fields (COSMOS, VVDS-14h, and the SN fields) were each calibrated with a single pass of the \GCM.
This run treated the catalog from each individual CCD image as the unit to be calibrated and yielded zeropoint offsets for each CCD directly.  
Due to its large area, the SPT region was calibrated from multiple iterations of the \GCM.  
The first pass treated the full catalog from each exposure (59 or 60 functioning science CCDs) as the unit to be calibrated.  
This could be done because, due to the star flat procedure, all the CCDs on a given exposure have very nearly the same zeropoint (at least for exposures taken under photometric conditions).  
In this pass, small (2--3\%) variations in the relative zeropoint across the focal plane were temporarily removed using the median $a$ coefficients for each CCD (\figref{psm_ab_vs_xy}).
In this manner, each exposure was temporarily flat-fielded across the focal plane to reduce exposure-scale photometric gradients.  
For the first pass, only exposures that were classified as having been observed under photometric conditions -- as determined by RASICAM -- were allowed in the \GCM fit.  
The first pass yielded a set of zeropoint offsets -- one per exposure -- for all the (apparently) photometric exposures in the SPT region.  
The second run of \GCM was essentially identical to the first, but it removed outlier exposures -- ones with particularly ``noisy'' or discrepant zeropoints.
For both the first and second runs, the sparse gridwork of tertiaries and the handful of individual calibrated SV fields (\figref{standards}) were used as the calibrated data set for the \citet{Glazebrook:1994} algorithm. 
Again, this yielded a set of zeropoint offsets -- one per exposure -- for all the photometric exposures in the SPT region.  
These individual CCD zeropoint offsets were applied to all the CCD images in the set of photometric exposures included in the second-pass run of \GCM, creating a set of ``quaternary'' standard stars covering nearly all of the SPT region.  
In the third and final run of the \GCM for the SPT region, the catalog from each individual CCD image was treated -- as in the case of \GCM runs for S82, COSMOS, VVDS-14h, and the SN fields -- as the unit to be calibrated. 
Furthermore, {\em all} CCD images from the SPT region -- those from photometric exposures and those from non-photometric exposures -- were included in the \GCM fit. 
For this third pass of the \GCM, the newly created quaternary standard stars were used as the calibrated data set. 
This third pass of the \GCM for the SPT region yielded a set of zeropoint offsets for each CCD image, which was used to calibrate the \gold single-epoch CCD images in advance of the image coaddition process.

\subsection{Photometric Calibration Adjustment}
\label{app:slr}

\begin{figure*}[t]
\center
\includegraphics[width=0.85\textwidth]{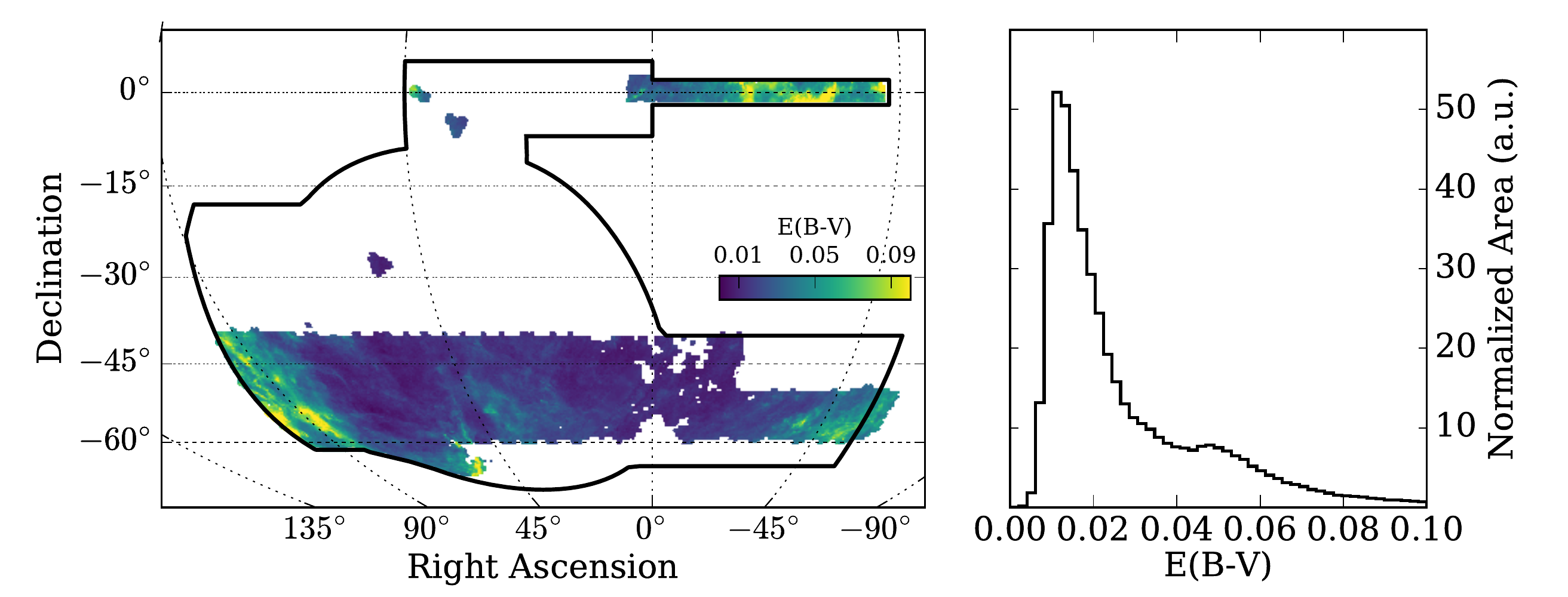}
\caption{\label{fig:sfd_ebv} Interstellar extinction, $E(B-V)$, over the Y1A1 GOLD footprint taken from \citet{Schlegel:1998}. The DES footprint was explicitly chosen to occupy a low-extinction region at high Galactic latitude.
}
\end{figure*}

To correct for residual color non-uniformity in the photometric calibration and to account for Galactic reddening (i.e., \figref{sfd_ebv}), the GCM calibration was adjusted at the catalog level using SLR (\secref{slr}). 
A reference stellar locus was empirically derived from the globally calibrated DES Y1A1 stellar objects in the region of the Y1A1 footprint with the smallest $E(B-V)$ value from \citet{Schlegel:1998}.  
Corrections were computed for the \var{WAVGCALIB\_MAG\_PSF} magnitudes described in \secref{catalog}.  
Our stellar selection was based on the weighted average of the \var{spread\_model} quantity for the matched objects ($|\wavgspreadmodel[r]| < 0.003$).  
We selected coadd objects with $S/N > 10$ in $i$ band and $S/N > 5$ in at least two other bands ($grzY$).  
We segmented the sky into equal-area pixels using the \HEALPix scheme~\citep{Gorski:2005}, starting with a relatively fine grid, $\nside=512$ ($\sim0.01\,\mathrm{deg}^2$).  
If there were fewer than 200 stars in a pixel, then we appended neighboring pixels using the \code{get\_all\_neighbors} function from \healpy, enlarging the pixel chunks until they contained at least 200 stars.  
To reduce computation time in high-density regions near the LMC, when there were more than 2000 stars per pixel we randomly down-sampled.
Approximately $97\%$ of the wide-area survey footprint was fit in chunks of 9 pixels containing a median of $\sim400$ stars and yielding an effective resolution of $\sim0.1\,\mathrm{deg}^2$.
We applied a modified version of the \code{BigMACS} SLR code \citep{Kelly:2014a}\footnote{\url{https://code.google.com/p/big-macs-calibrate/}} to calibrate each star from the reference exposure with respect to the empirical stellar locus.
 The absolute calibration was set against the $i$-band magnitude derived from the GCM solution, which was dereddened using the SFD map with a reddening law of $A_I = 1.947 \times E(B-V)_{\rm SFD}$. 
This extinction correction was derived following the prescription of \citet{Cardelli:1989} with $R_V = 3.1$, but updated for the DES $i$-band throughput using optical-NIR coefficients from \citet{O'Donnell:1994} assuming a source spectrum that is constant in spectral flux density per unit wavelength, $f_{\lambda}$ ($\erg \cm^{-2} \second^{-1} \AA^{-1}$).
The flat SED was chosen to represent the wide range of stellar SEDs from the Pickles ATLAS \citep{Pickles:1998} and SEDs of galaxies over the range of redshifts probed by DES \citep{Arnouts:2011}.

Variations in the average metallicity of the stellar populations used for the SLR will introduce systematic shifts that are not due to photometric variation or Galactic reddening \citep[e.g.,][]{High:2009a}.  
For DES Y1A1, these shifts are largest for the $g$ band, where they can have a 1-2\% effect on the calibration.
A larger effect can be found in the vicinity of the LMC, which we avoid for extragalactic science.
The effect of metallicity variations can be much worse at lower Galactic latitudes and in bluer filters (i.e., $u$ band).

The final product was an SLR correction map at a resolution of $\nside=512$ that we implemented with a bi-linear interpolation to obtain magnitude and flux corrections for the full \gold catalog.
The resulting SLR-adjusted magnitudes used in the \gold catalog are thus already corrected for Galactic reddening.  

\subsection{Photometric Transformation Equations}
\label{app:transform}

We have derived transformation equations between various surveys and the DES system. 
We document these transformation equations here for reference.

We define a transformation from SDSS/UKIDSS to the DES system to place the nightly standard star exposures on an initial DES AB photometric system (\secref{psm}):
\begin{align}
g_{\des} & =  g_{\sdss} - 0.104 \times (g-r)_{\rm sdss} + 0.01 \label{eqn:des_g}\\ 
r_{\des} & =  r_{\sdss} - 0.102 \times (g-r)_{\rm sdss} + 0.02 \label{eqn:des_r}\\
i_{\des} & =  i_{\sdss} - 0.256 \times (i-z)_{\rm sdss} + 0.02 \label{eqn:des_i}\\
z_{\des} & =  z_{\sdss} - 0.086 \times (i-z)_{\rm sdss} + 0.01 \label{eqn:des_z}\\
Y_{\des} & =  Y_{\ukidss} + 0.238 \times (z_{\rm sdss} - Y_{\rm ukidss}) + 0.634 . \label{eqn:des_Y}
\end{align}
\noindent These transformation equations were derived in a hybrid manner:  the color coefficients were determined by matching data from the DES SV data set with data from SDSS DR9 (or, in the case of the $Y$ band, with a combination of UKIDSS DR6 $Y$ band and SDSS DR9) and fitting the result.  
The zeropoint for each relation was determined from synthetic AB photometry.
We applied the DES, SDSS, and UKIDSS filter curves to the \cite{Pickles:1998} stellar library and measured the offset between the two synthetic magnitudes at zero color for each filter band.  
We note that the large zeropoint offset for the $Y$-band transformation  is due to the fact that the UKIDSS data are in the Vega magnitude system, while the $Y_{\des}$ is set to the AB magnitude system.  
These transformation equations are valid for stars with $(g-r)_{\sdss} < 1.2$.
For an individual object, the transformation from SDSS/UKIDSS to DES will depend on interstellar extinction.
The DES footprint occupies a region of low extinction, and we estimate that the median correction due to reddening in the $g$ band is $0.8 \mmag$ (90\% of Y1A1 GOLD has a $g$-band correction of $< 2\mmag$).
Median extinction corrections for the other bands are a factor of $\gtrsim4$ lower than $g$-band.

We validate the relative calibration accuracy of \gold by comparing the calibrated magnitudes of stars in the \gold catalog against those derived from a combination of APASS \citep{Henden:2014} and 2MASS \citep{Skrutskie:2006}.
We selected stellar objects from the \gold catalog using \modest (\secref{sgsep}) and perform a $2\arcsec$ match to the APASS and 2MASS catalogs. 
We then fit a set of transformation equations to map from $g_\apass$, $r_\apass$, and $J_\twomass$ to a predicted magnitude in each of the DES filters:
\begin{align}
g_\des & = g_\apass - 0.0642 \times (g - r)_\apass - 0.0239 \label{eqn:g_apass} \\
r_\des & = r_\apass - 0.1264 \times (r - i)_\apass - 0.0098 \label{eqn:r_apass} \\
i_\des & = r_\apass - 0.4145 \times (r_\apass - J_\twomass - 0.81) - 0.391 \label{eqn:i_apass} \\
z_\des & = (J_\twomass+0.81) + 0.3866 \times (r_\apass - J_\twomass - 0.81) - 0.0414 \label{eqn:z_apass} \\
Y_\des & = (J_\twomass+0.81) + 0.2938 \times (r_\apass - J_\twomass - 0.81) - 0.0443 \label{eqn:Y_apass}.
\end{align}
 
\noindent Equations \ref{eqn:i_apass}-\ref{eqn:Y_apass} are derived from a global fit of the \gold data set and are valid for stars for which $r_\apass - J_\twomass < 1.81$.
We find a cleaner and tighter relation using the hybrid APASS/2MASS $(r_\apass - J_\twomass)$ color rather than a purely APASS $(r_\apass - i_\apass)$ color for these transformation equations.
These transformation equations explicitly remove any absolute calibration offset between the two data sets and can be used to test for spatial non-uniformity between the GCM calibration and these external catalogs (\figref{app_apass2mass}).
We note that the residual structure seen in the S82 region of \figref{app_apass2mass} does not appear in comparisons with SDSS DR10 or DES Y3, suggesting that this structure is a feature introduced by APASS.

To validate the completeness and contamination of the \gold catalog, we perform a comparison with the CFHTLenS data in the W4 field.
In this case, we are interested in the transformed magnitude of \emph{all} objects, so we perform no stellar selection.
We use matched objects to derive a set of transformation equations from the CFHTLenS $g', r', i', z'$ filters to the DES $g, r, i, z$ system:
\begin{align}
g_\des &= g_\cfht + 0.062 (g_\cfht - r_\cfht) + 0.058 \\ 
r_\des &= r_\cfht - 0.078 (g_\cfht - r_\cfht) + 0.021 \\ 
i_\des &= i_\cfht - 0.179 (i_\cfht - z_\cfht) + 0.062 \\ 
z_\des &= z_\cfht - 0.139 (i_\cfht - z_\cfht) + 0.053.
\end{align}
\noindent We find that these equations should be valid for objects with $g - r < 1.2$ and $i - z < 1.0$.

\subsection{Calibration Validation}
\label{app:cal_acc}

\begin{figure}[h]
\center
\includegraphics[width=0.65\columnwidth]{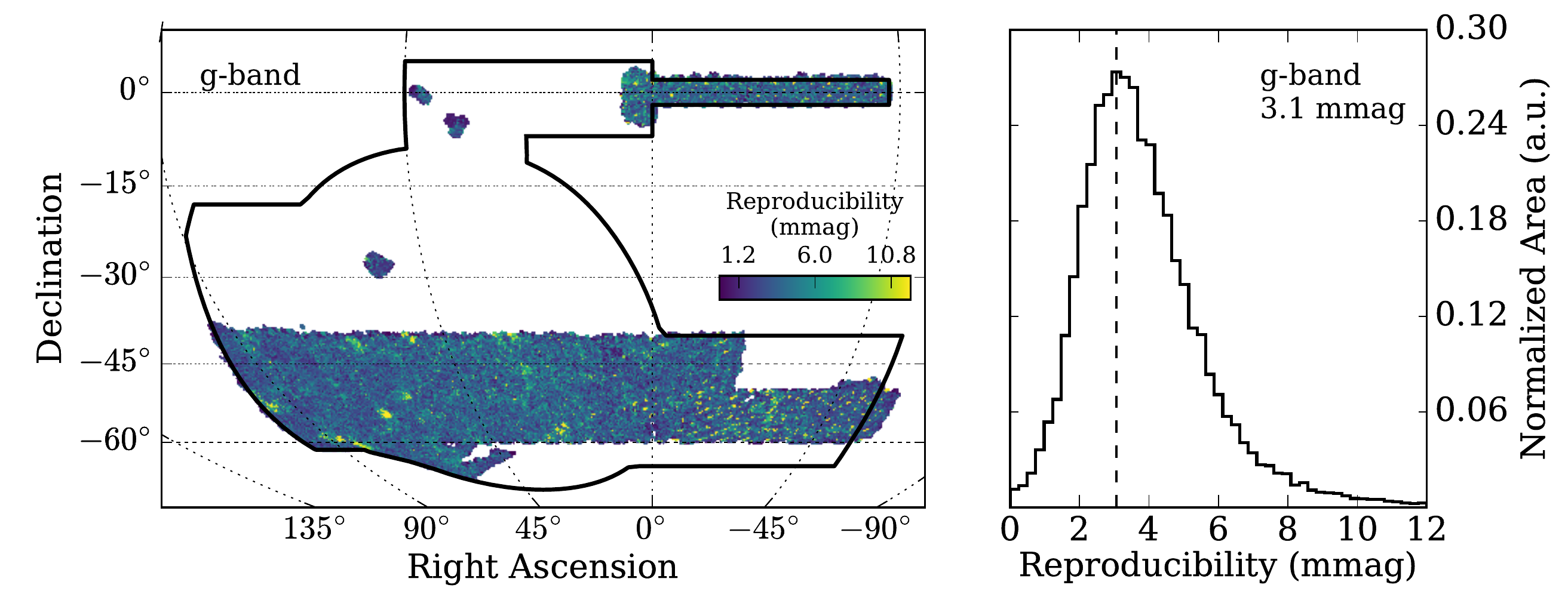}
\includegraphics[width=0.65\columnwidth]{y1a1_gcm_zprms_r.pdf}
\includegraphics[width=0.65\columnwidth]{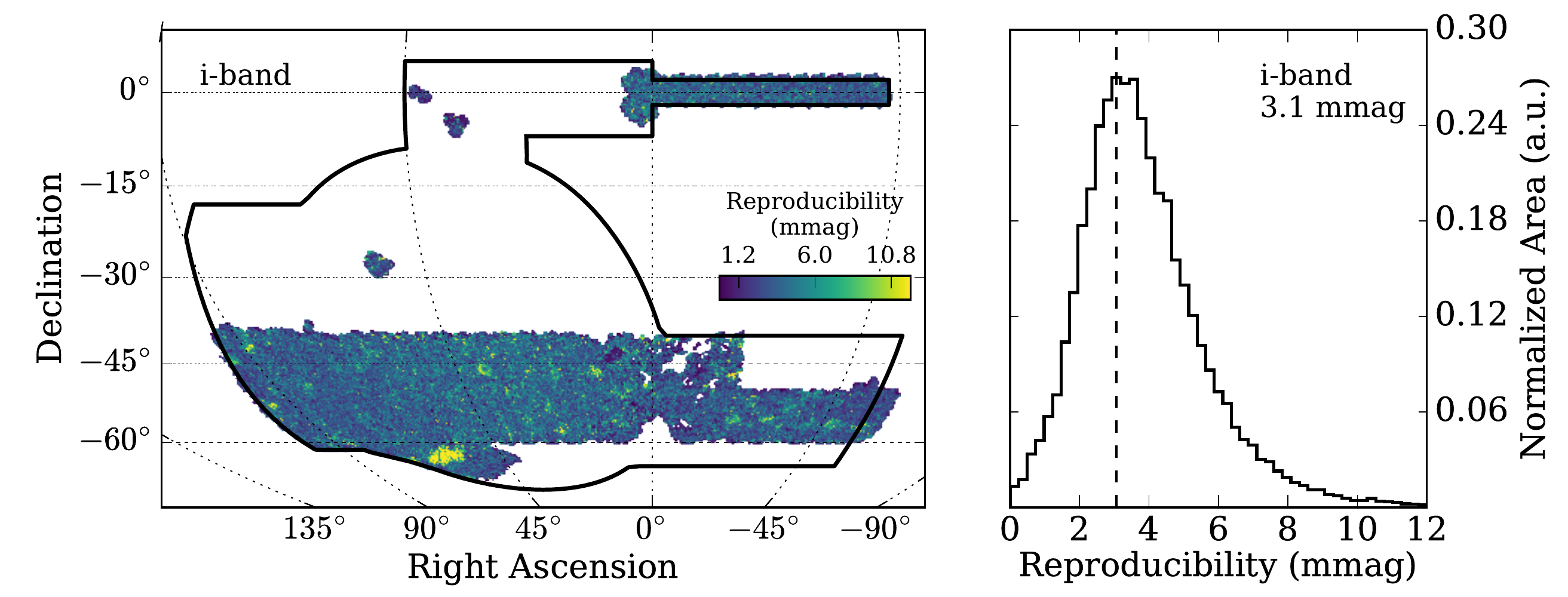}
\includegraphics[width=0.65\columnwidth]{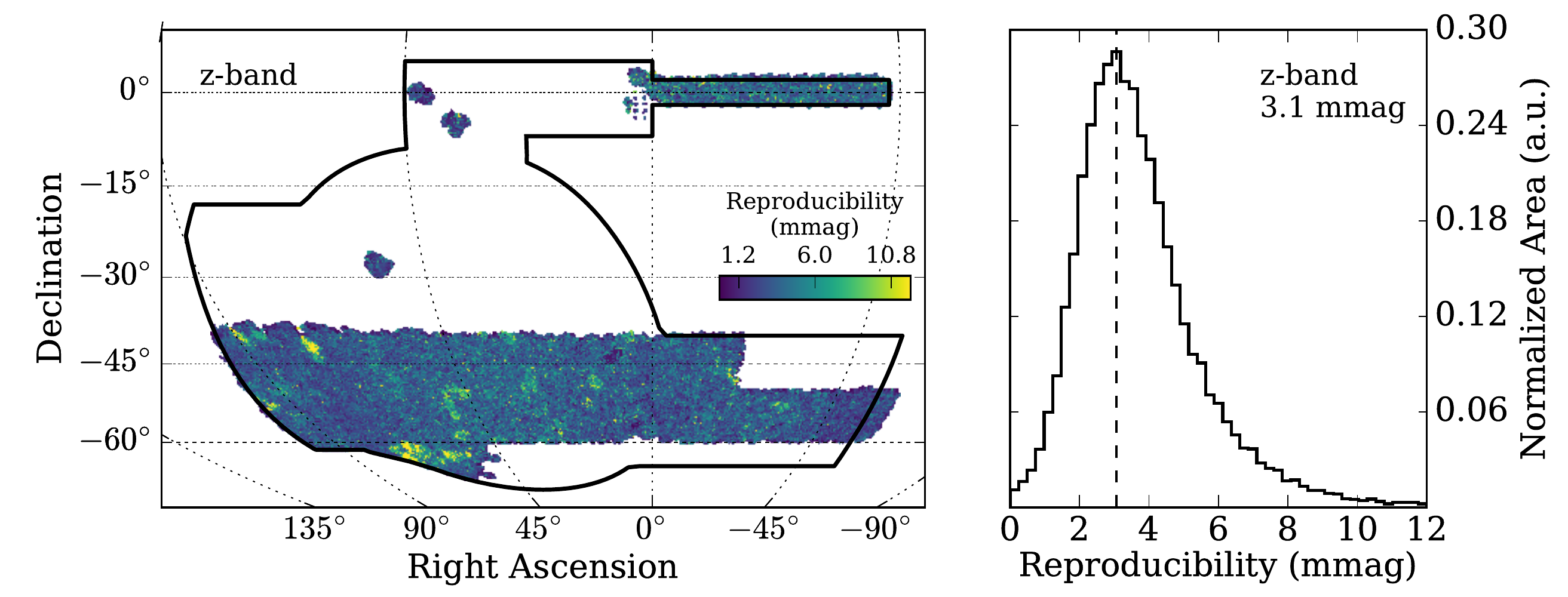}
\includegraphics[width=0.65\columnwidth]{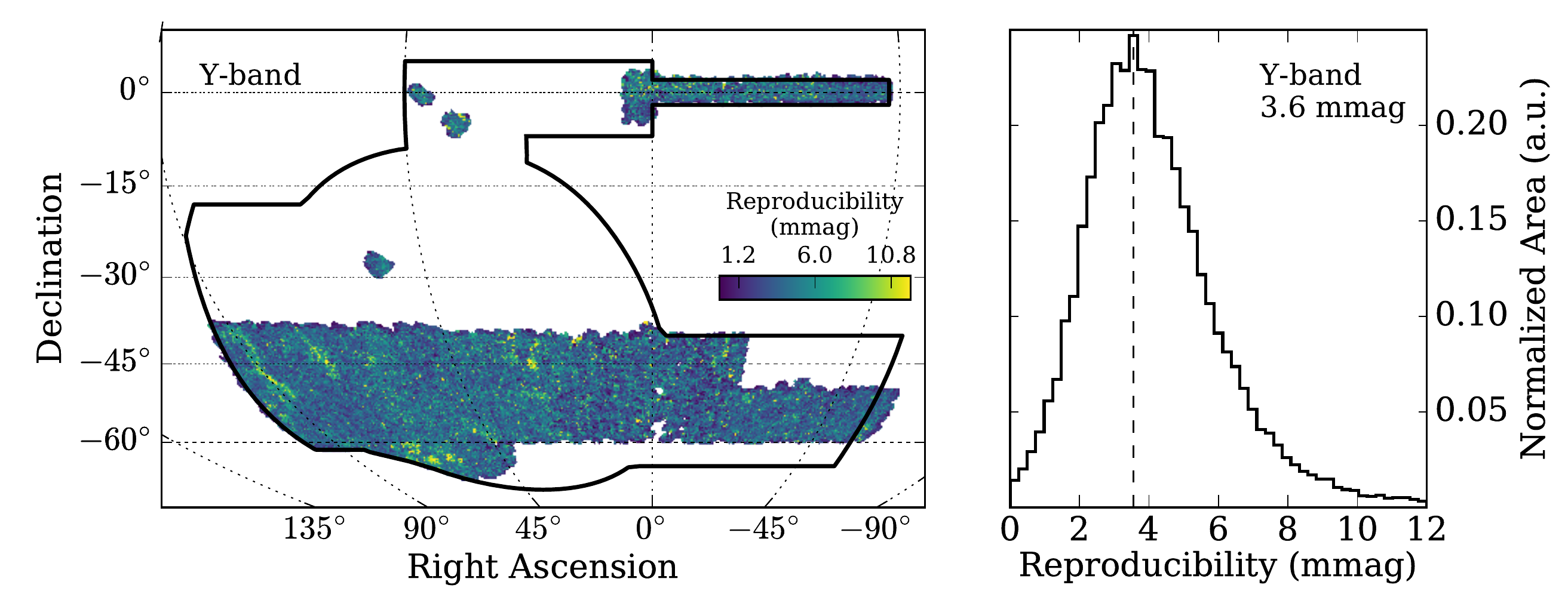}
\caption{\label{fig:app_gcm_zprms} Internal rms errors in the photometric zeropoint reproducibility per CCD for DES Y1A1. 
The zeropoint rms is calculated by comparing the calibrated magnitudes of stars in overlapping CCDs. 
Note that these data include observations taken in both clear and cloudy conditions. 
Typical internal reproducibility errors are $\roughly 3~\mmag$ ($\roughly 0.3\%$).
The color scale in the left panels represents the rms internal calibration uncertainty in mmag.}
\end{figure}

\begin{figure}[!h]
\center
\includegraphics[width=0.65\columnwidth]{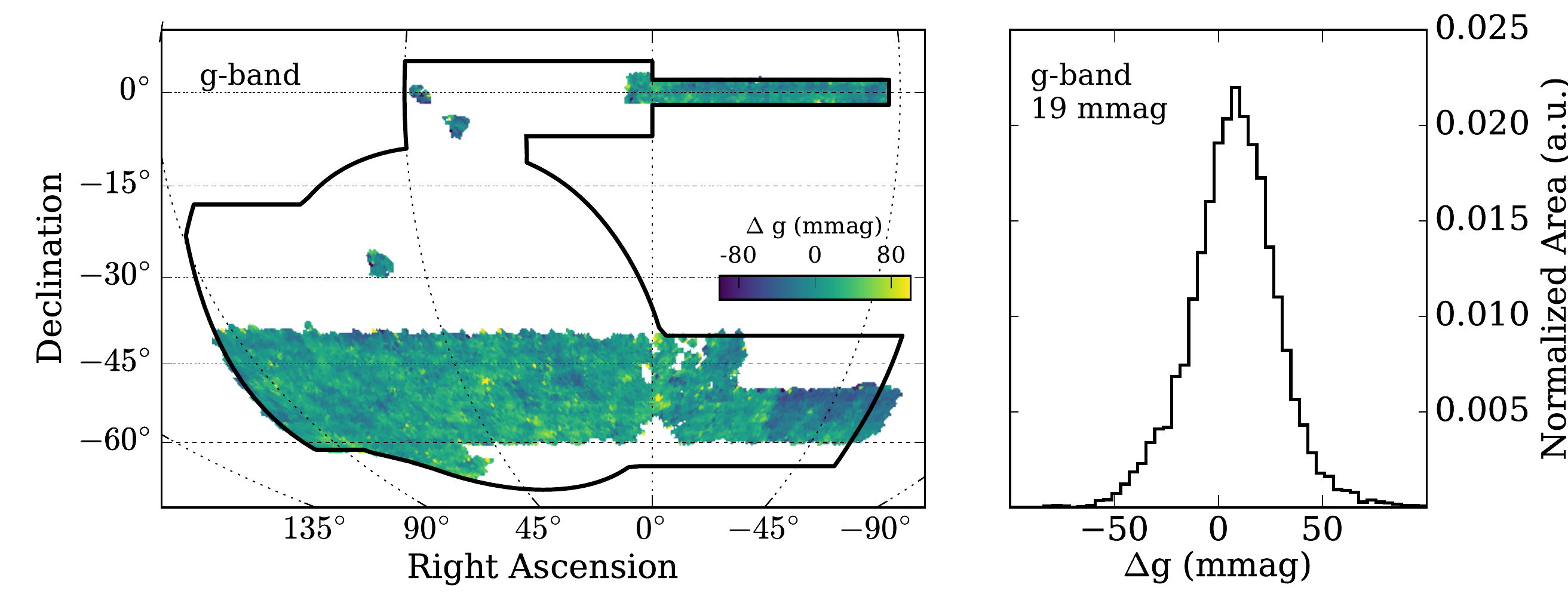}
\includegraphics[width=0.65\columnwidth]{y1a1_apass2mass_gcm_n128_r.pdf}
\includegraphics[width=0.65\columnwidth]{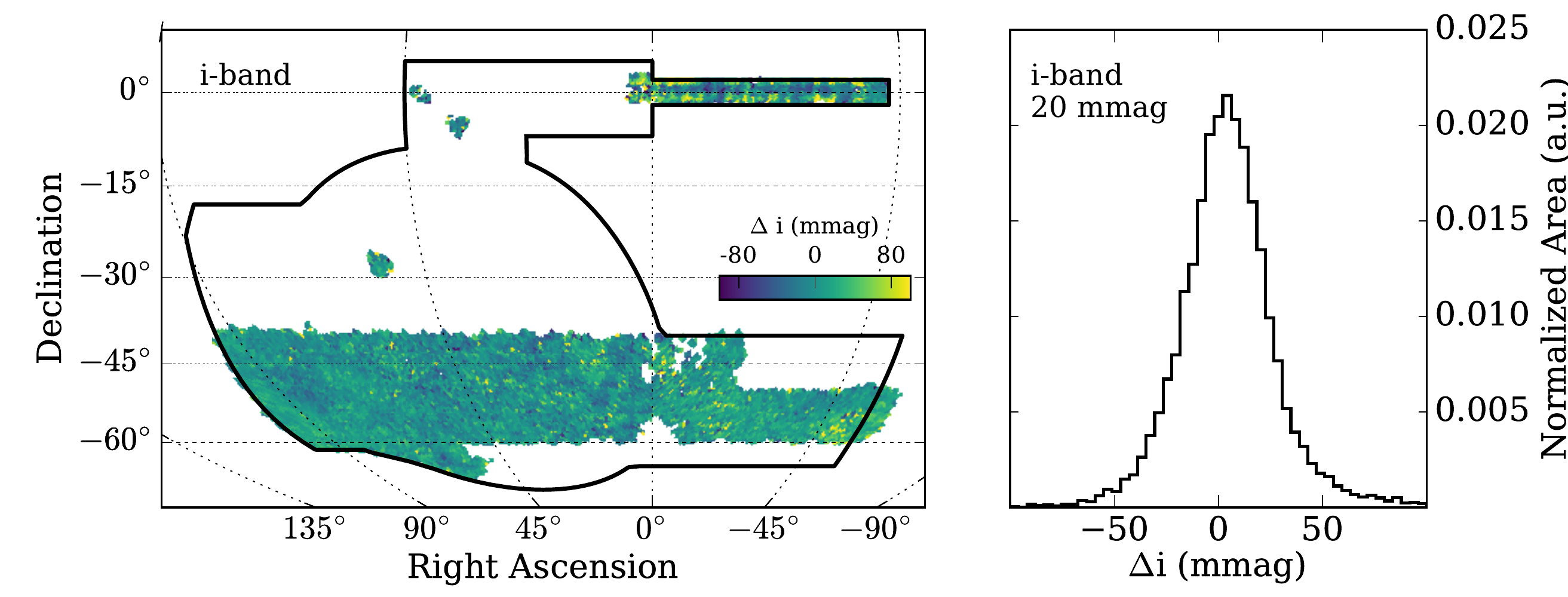} 
\includegraphics[width=0.65\columnwidth]{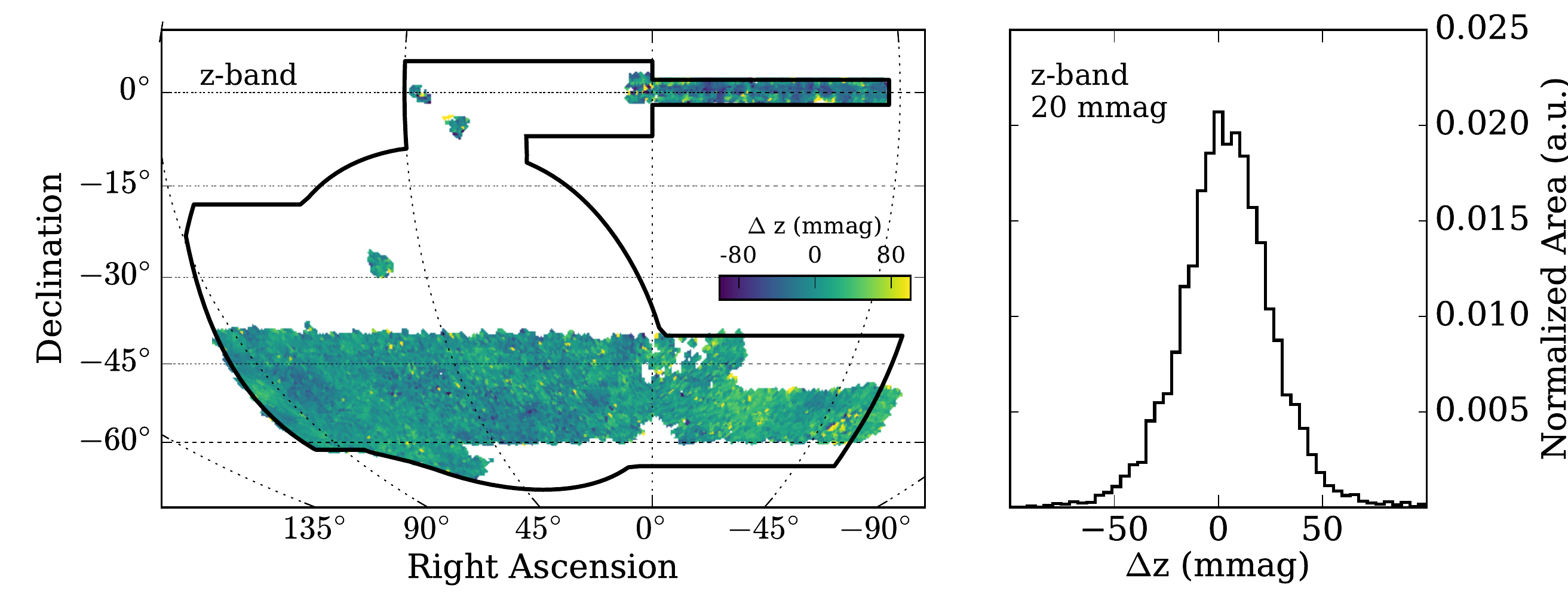} 
\includegraphics[width=0.65\columnwidth]{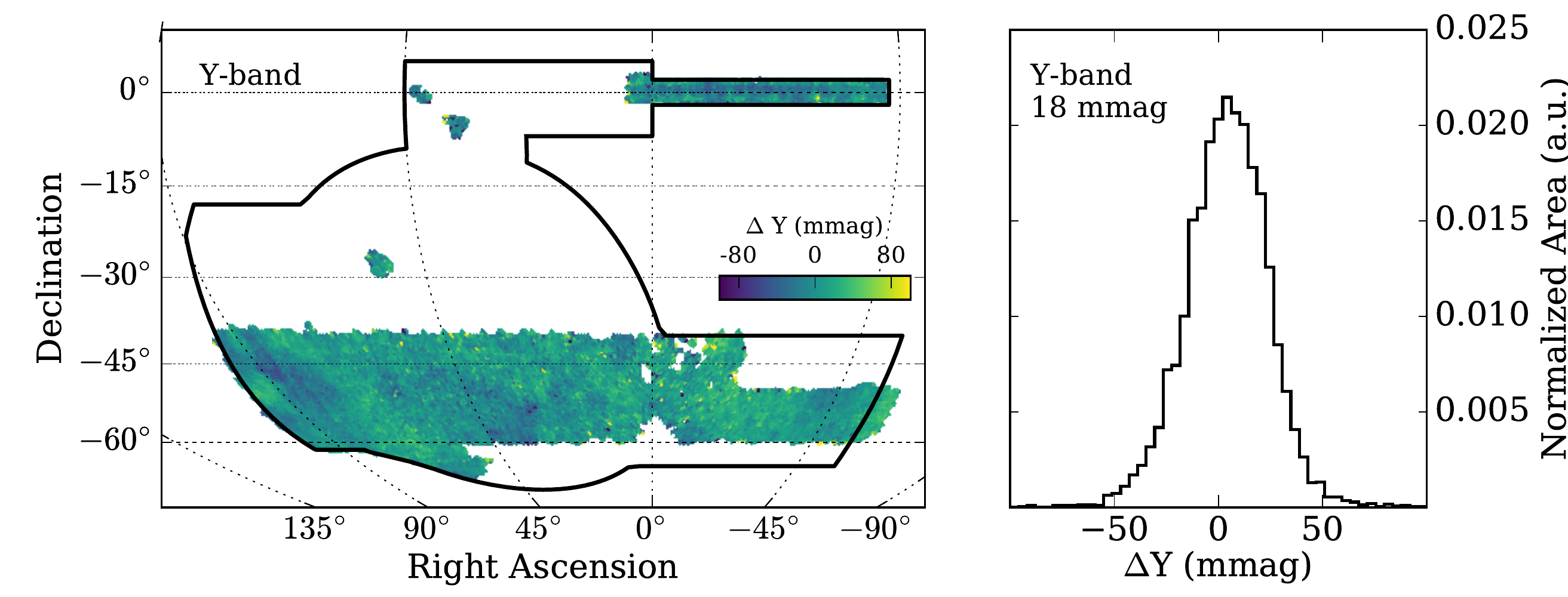}
\caption{\label{fig:app_apass2mass} Comparison of stellar magnitudes from the DES Y1A1 \GCM and those estimated from APASS/2MASS transformed into the DES filter system (Equations \ref{eqn:g_apass}-\ref{eqn:Y_apass}). 
The sky plots (left) show the median magnitude offset for stars binned into $\roughly 0.2 \deg^2$ \healpix pixels.
The \GCM calibrated magnitudes are consistent with the transformed values from APASS/2MASS with $\sigma_{68} \sim 20 \mmag$ (calculated between the 16th and 84th percentiles). 
Note that the GCM $g$-band calibration disagrees with APASS/2MASS by $\roughly 4\%$ in the eastern portion of the SPT region (${\rm RA}<-20$), motivating the SLR adjustment described in \secref{slr}.
}
\end{figure}

\begin{figure}[h]
\center
\includegraphics[width=0.65\columnwidth]{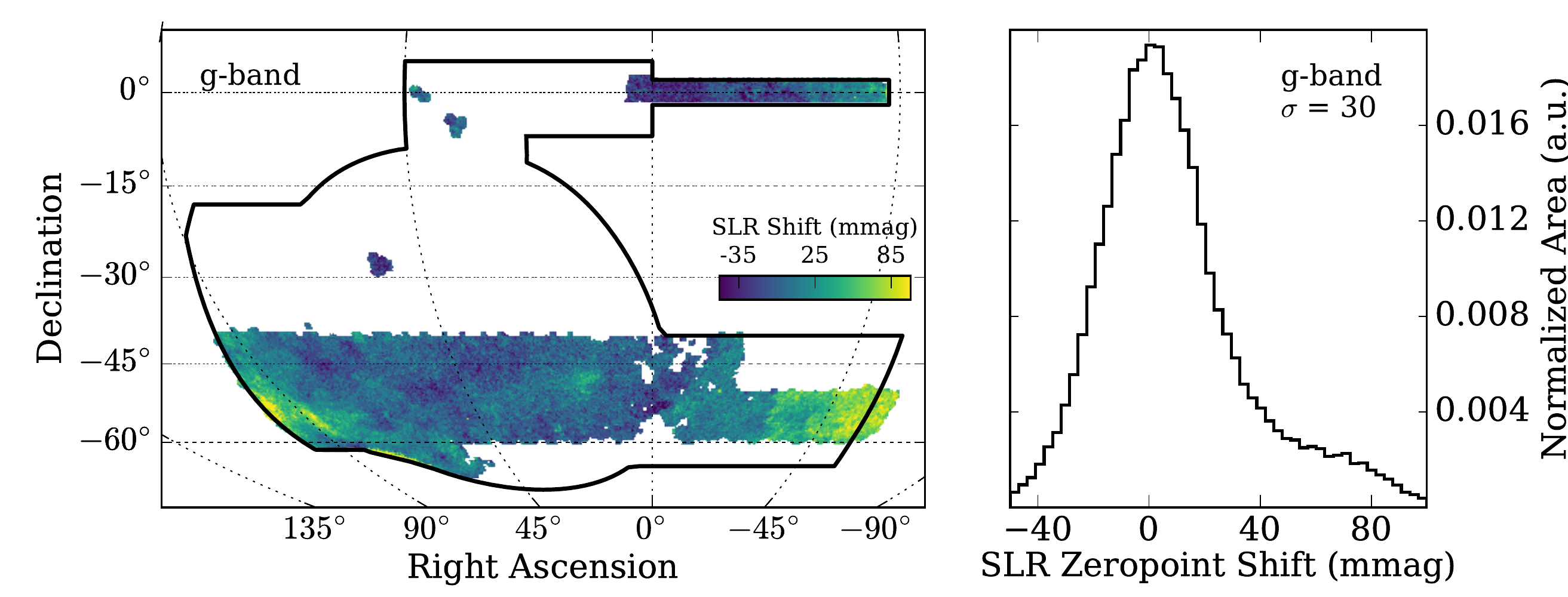}
\includegraphics[width=0.65\columnwidth]{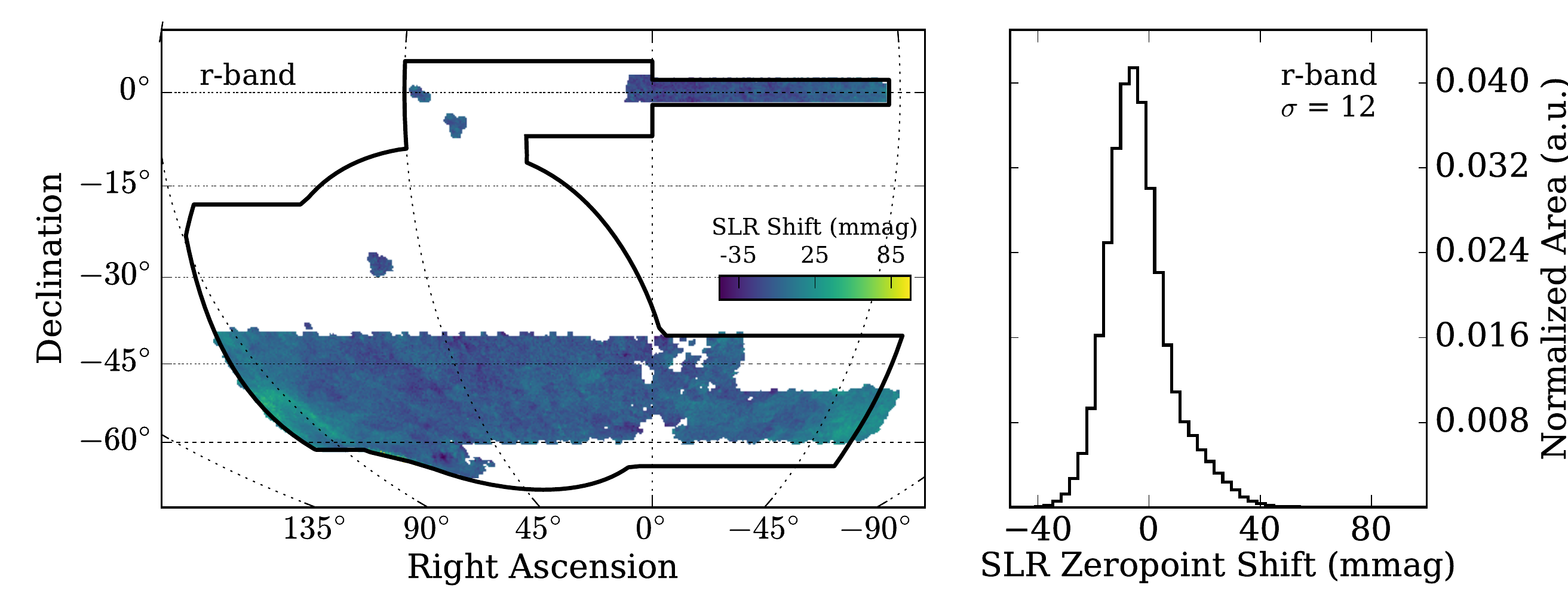}
\includegraphics[width=0.65\columnwidth]{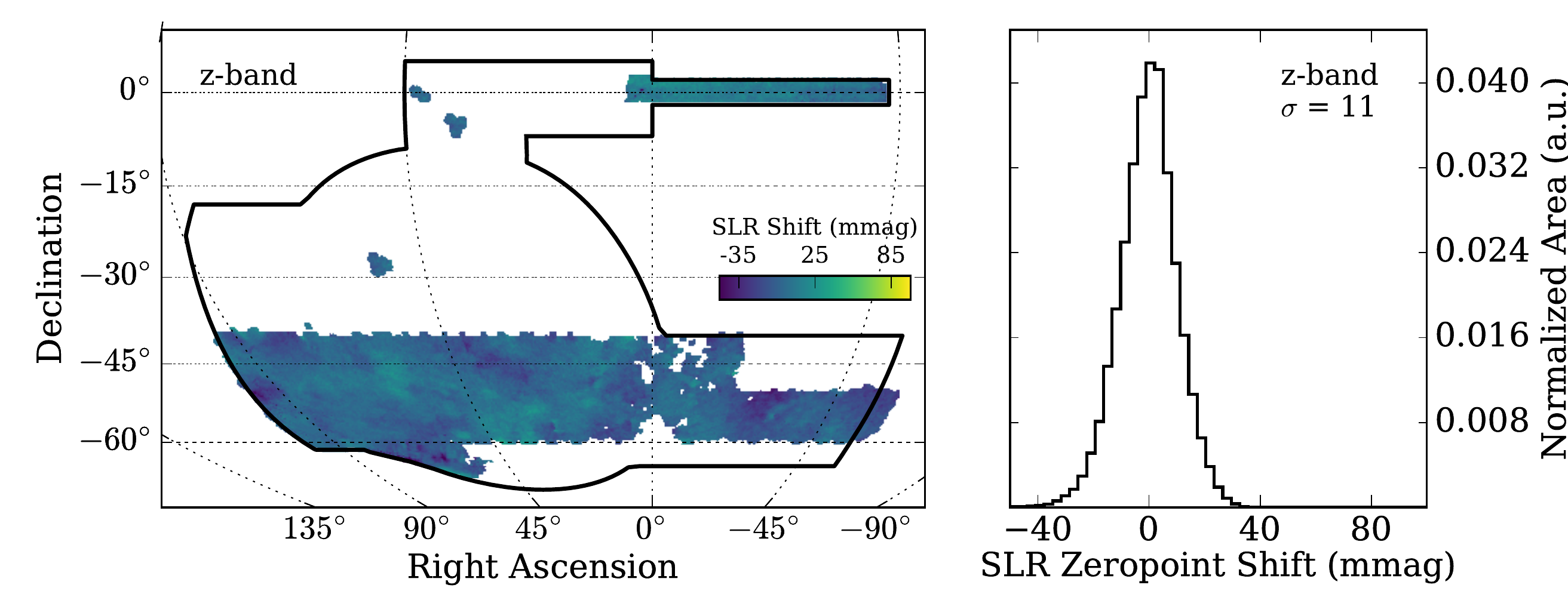}
\includegraphics[width=0.65\columnwidth]{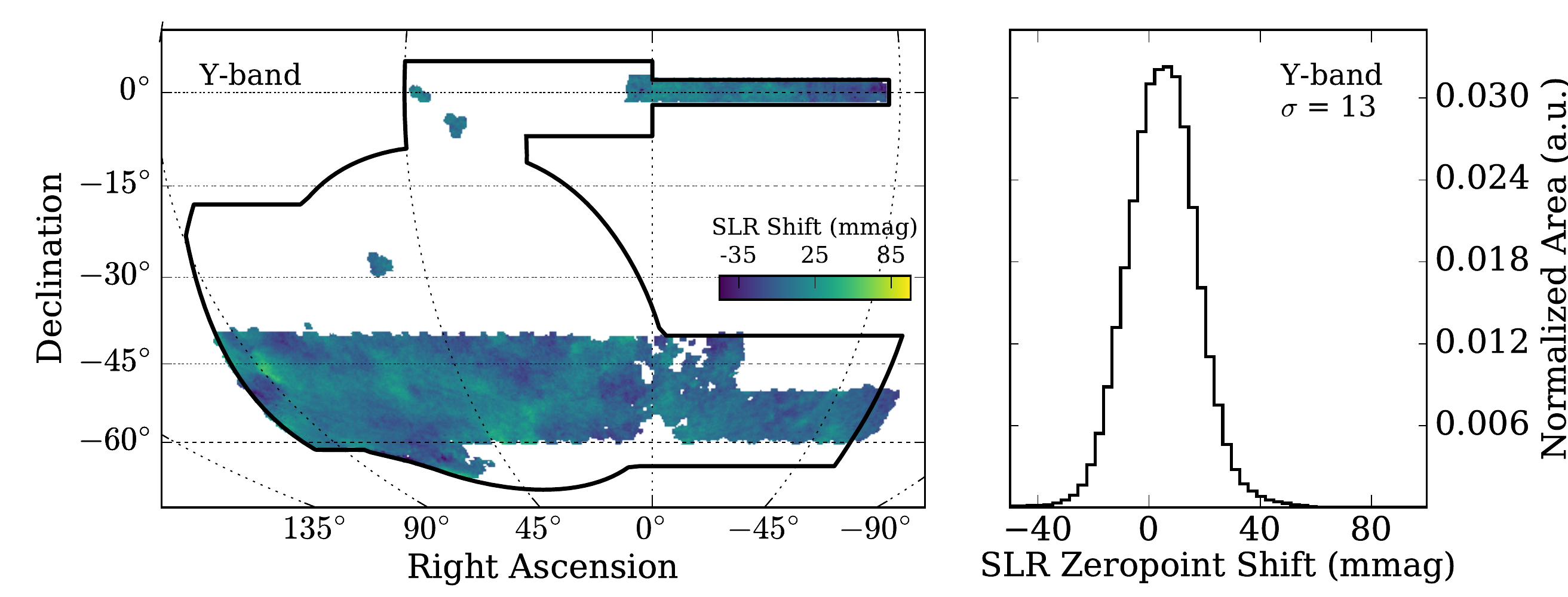}
\caption{\label{fig:app_gcm_vs_slr}
Adjustment to the \GCM photometric zeropoints derived from the SLR fit, after removing the contribution from interstellar extinction using the SFD maps and reddening from \citet{O'Donnell:1994}. 
The width of these distributions represents adjustments to the calibration uniformity and differences between the interstellar extinction derived from the stellar locus and interstellar dust maps.
The SLR adjustment is generally $\roughly 10 \mmag$ (rms) over most of the area.
A larger adjustment is made in the $g$ band, which reflects the larger impact of reddening in the blue filters and a region of non-uniformity in the west of the footprint.
There is no adjustment to the \GCM $i$ band because the SLR fit is tied to the dereddened magnitudes of stars in that band.}
\end{figure}

\begin{figure}[h]
\center
\includegraphics[width=0.65\columnwidth]{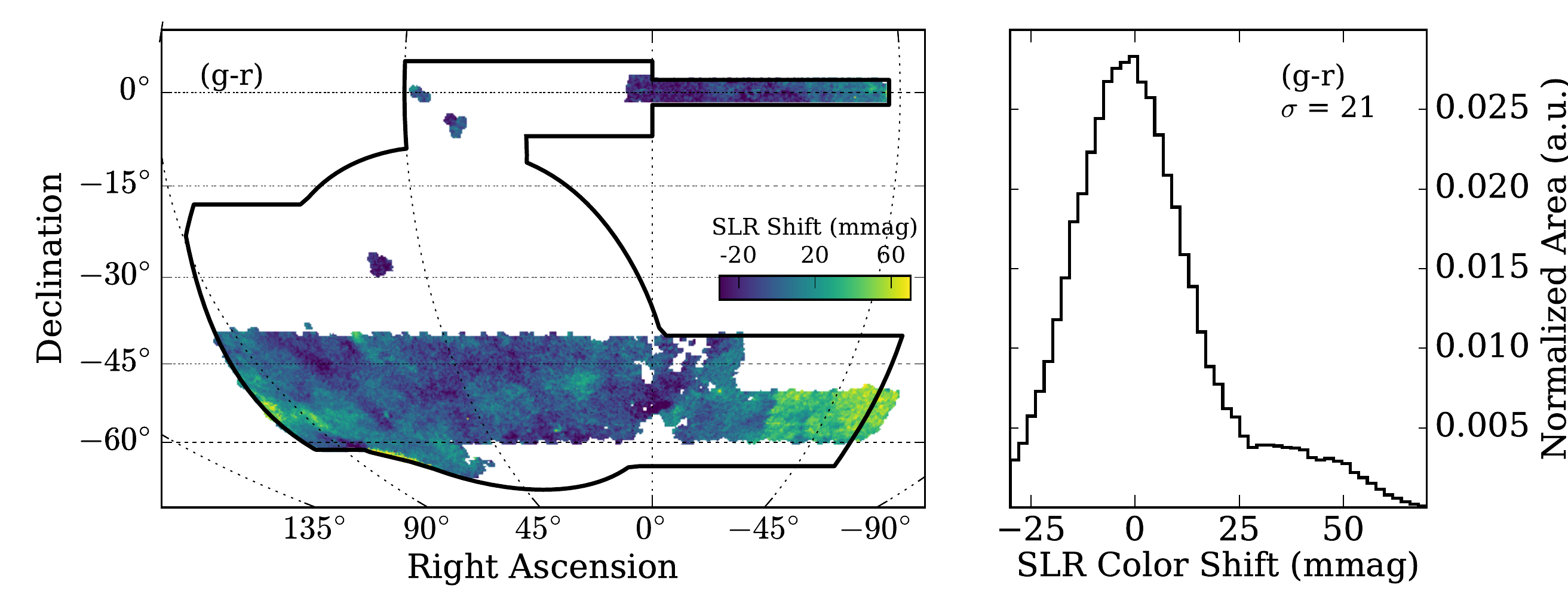}
\includegraphics[width=0.65\columnwidth]{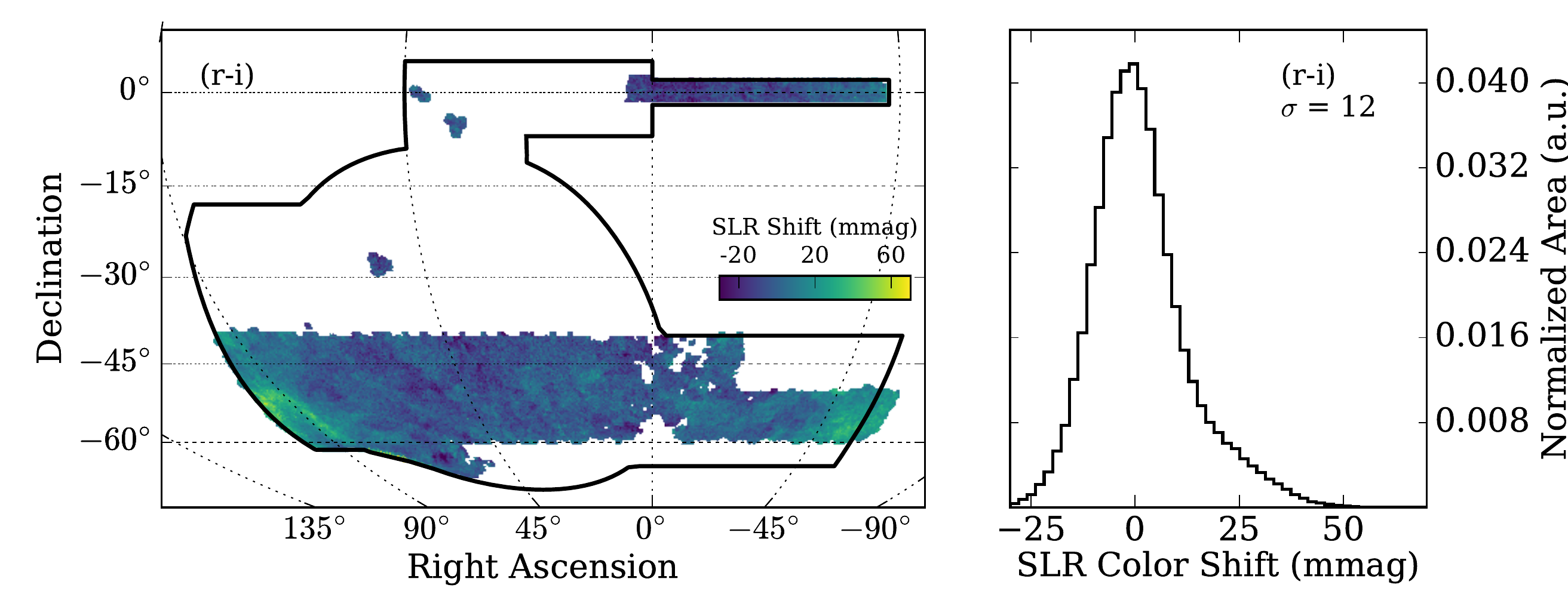}
\includegraphics[width=0.65\columnwidth]{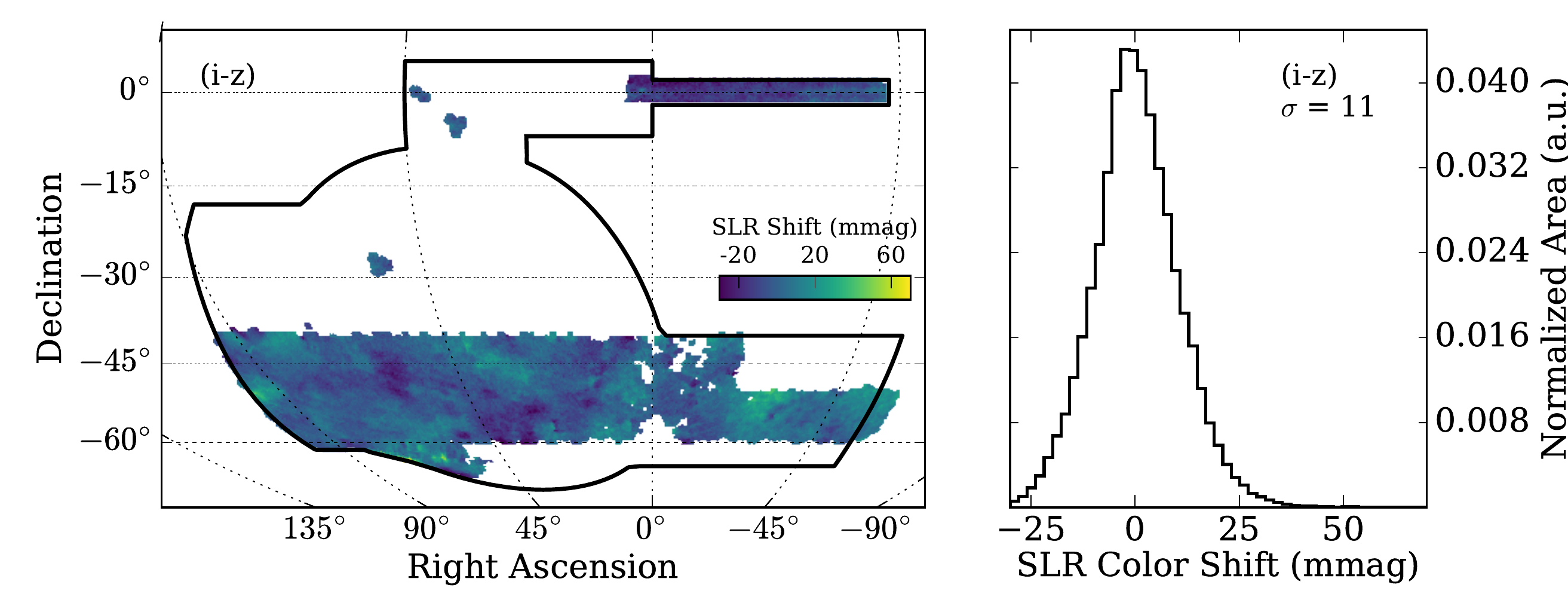}
\caption{\label{fig:app_slr_zp} Color uniformity of the SLR adjustment applied to the GCM zeropoints. 
The adjustment was largest for the $(g - r)$ color in the eastern portion of the SPT region.
The color non-uniformity in this region was one of the specific motivations for the SLR calibration adjustment.
After the SLR adjustment was applied, the color of stars was found to be uniform at the 1--2\% level across the footprint.
}
\end{figure}

In this section we show ancillary plots of the performance and validation of the Y1A1 photometric calibration (Figures \ref{fig:app_gcm_zprms} -- \ref{fig:app_slr_zp}).

\section{Co-add Source Detection}
\label{app:chimean}

The \coadd source detection was performed on a normalized ``detection image'' formed from a nonlinear combination of the $r$, $i$, and $z$ coadded images. 
The original \swarp combination formula for computing the value of a pixel of the detection image is \citep{Bertin:2010}:
\begin{equation}
\label{eq:chi}
\chi = \sqrt{\frac{\sum_{c\le n} w_c f_c^2}{n}},
\end{equation}
where $f_c$ is the background-subtracted pixel value, $w_c$ is the weight
of the pixel in channel $c$, and $n$ is the number of valid inputs.
Compared to the standard $\chi^2$ combination proposed by
\cite{Szalay:1999}, $\chi$ leads to a less skewed noise distribution (if
one assumes that input noise follows a Gaussian distribution), while
maintaining identical detection capabilities. However, both estimators
have a bias that depends on $n$, which leads to visible seams between
regions with a different number of input images. This motivated the
implementation of two new normalized image combination schemes in
\swarp, with a variable offset applied to the original (still assuming
that the inputs are normally and independently distributed).
\code{CHI-MEAN} is recentered on the mean \citep[e.g.,][]{Evans:2000}:
\begin{equation}
\code{CHI-MEAN} =  \frac{\sqrt{\sum_{c\le n} w_c f_c^2} - \mu}{\sqrt{n -
\mu^2}},
\end{equation}
with %\citep{}
\begin{equation}
\mu = \sqrt{2}\frac{\Gamma((n+1)/2)}{\Gamma(n/2)},
\end{equation}
while \code{CHI-MODE} is recentered on the mode of the distribution:
\begin{equation}
\code{CHI-MODE} = \frac{\sqrt{\sum_{c\le n} w_c f_c^2} - \sqrt{n -
1}}{\sqrt{n - \mu^2}}.
\end{equation}

\begin{figure}[th]
\center
\includegraphics[width=0.49\columnwidth]{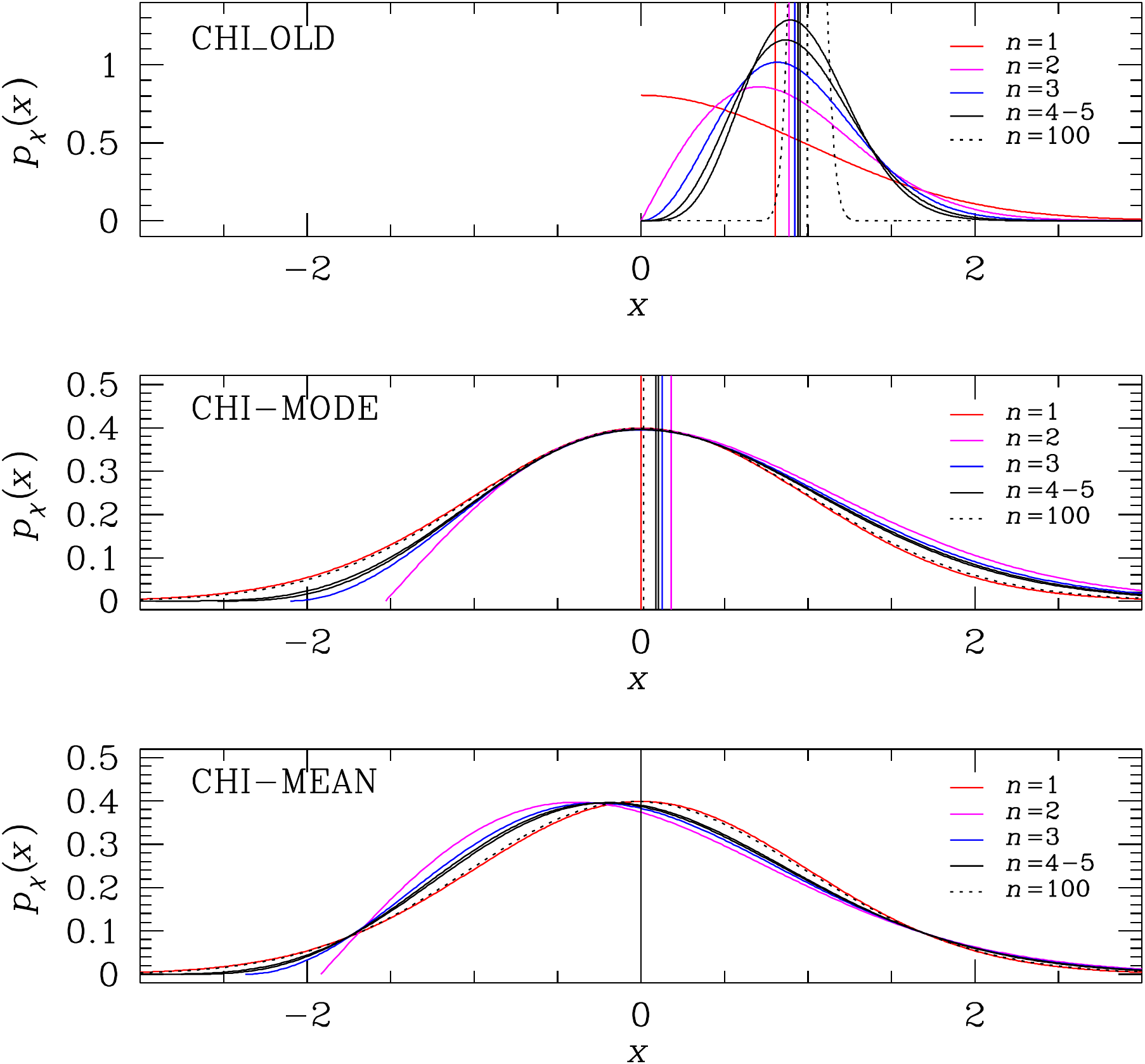}
\raisebox{0.03\height}{\includegraphics[width=0.49\columnwidth]{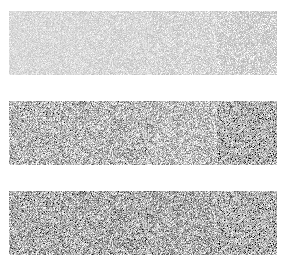}}
\label{fig:chidist}
\caption{
(Left) Normalized distribution of the value $x$ of a detection image pixel for
the original $\chi$ (\code{OLD\_CHI}, top), \code{CHI-MODE} (middle) and
\code{CHI-MEAN} (bottom) estimators when the inputs to the co-add are
normally and independently distributed. $n$ is the number of input
images. The means of the distributions are shown as vertical lines.
(Right) Gamma-corrected close-up of a $\chi$ (\code{OLD\_CHI}, top),
\code{CHI-MODE} (middle) and \code{CHI-MEAN} (bottom) detection image
computed from a set of 8 input images with zero-mean Gaussian white
noise. The number of inputs decreases by steps of 64 pixels from left (eight
inputs) to right (one input). \code{CHI-MEAN} detection images are
virtually seamless, even at the transition between one and two input images.
}
\end{figure}

The left panel of \figref{chidist} shows a comparison of the distributions obtained
from the original $\chi$, \code{CHI-MODE} and \code{CHI-MEAN} estimators
for Gaussian input noise. 
The right panel of \figref{chidist} shows that the \code{CHI-MEAN} estimator generates the most seamless stacking results, and it was used to produce the \coadd detection images.

\section{Catalog Depth Maps}
\label{app:depth}

\begin{figure}[!ht]
\centering
\includegraphics[width=0.65\columnwidth]{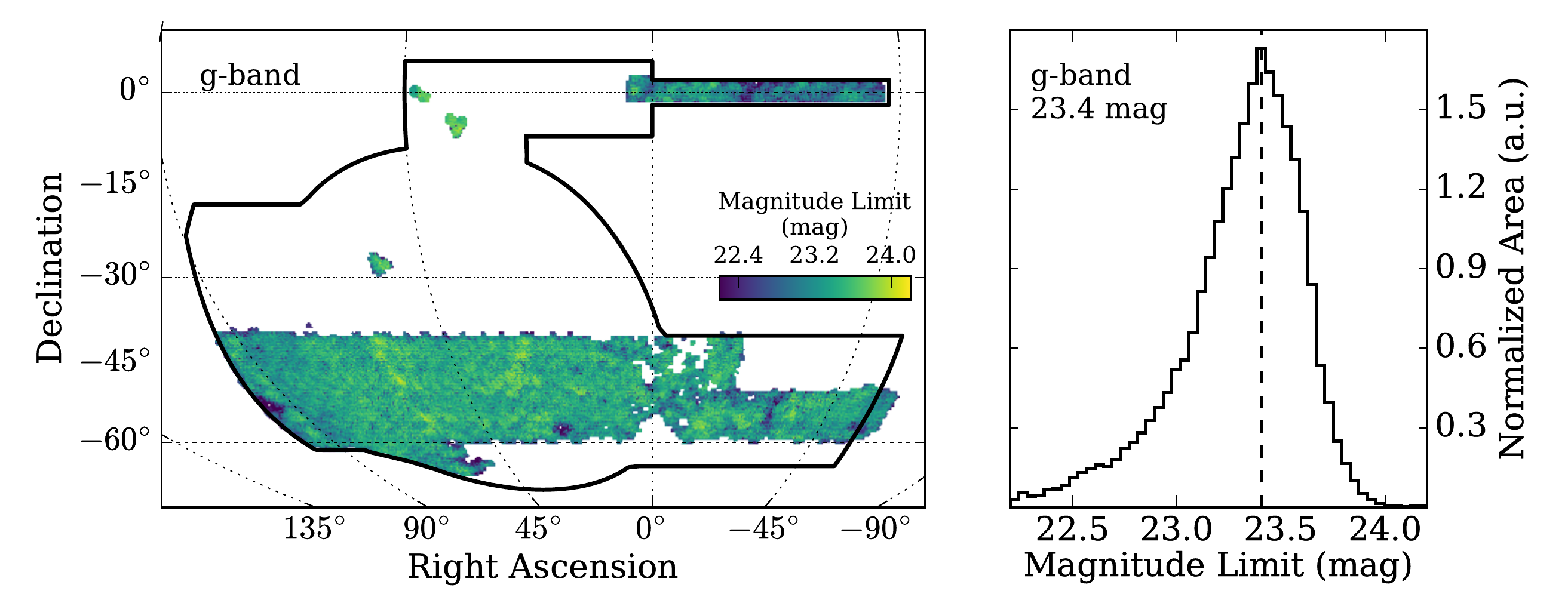}
\includegraphics[width=0.65\columnwidth]{y1a1_gold_1_0_2_wide+d04_auto_nside4096_r_10sigma.pdf}
\includegraphics[width=0.65\columnwidth]{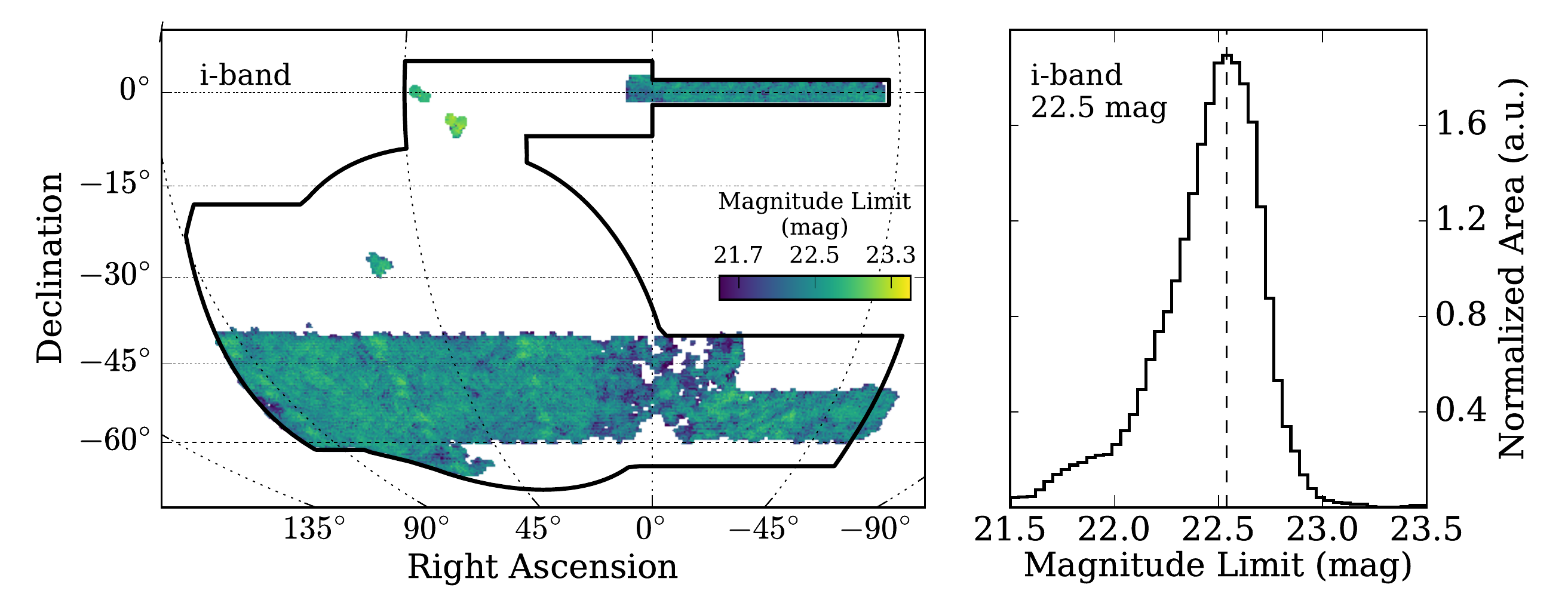}
\includegraphics[width=0.65\columnwidth]{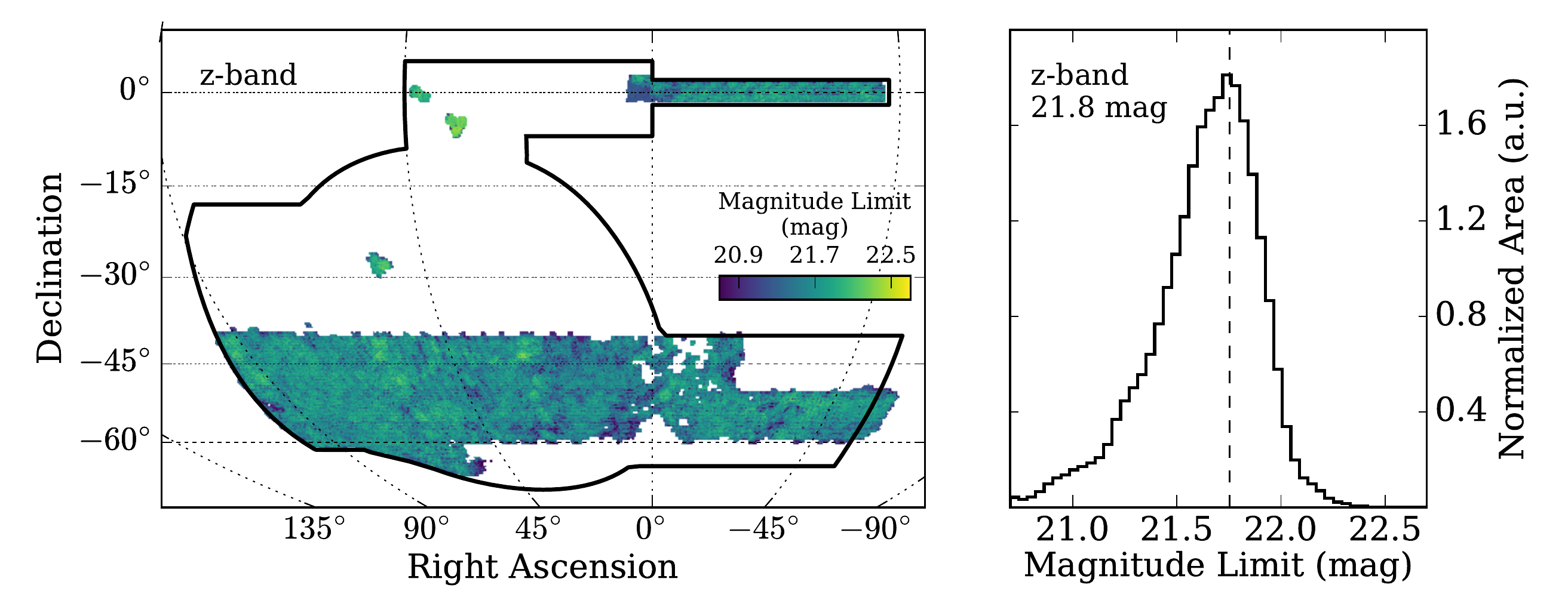}
\includegraphics[width=0.65\columnwidth]{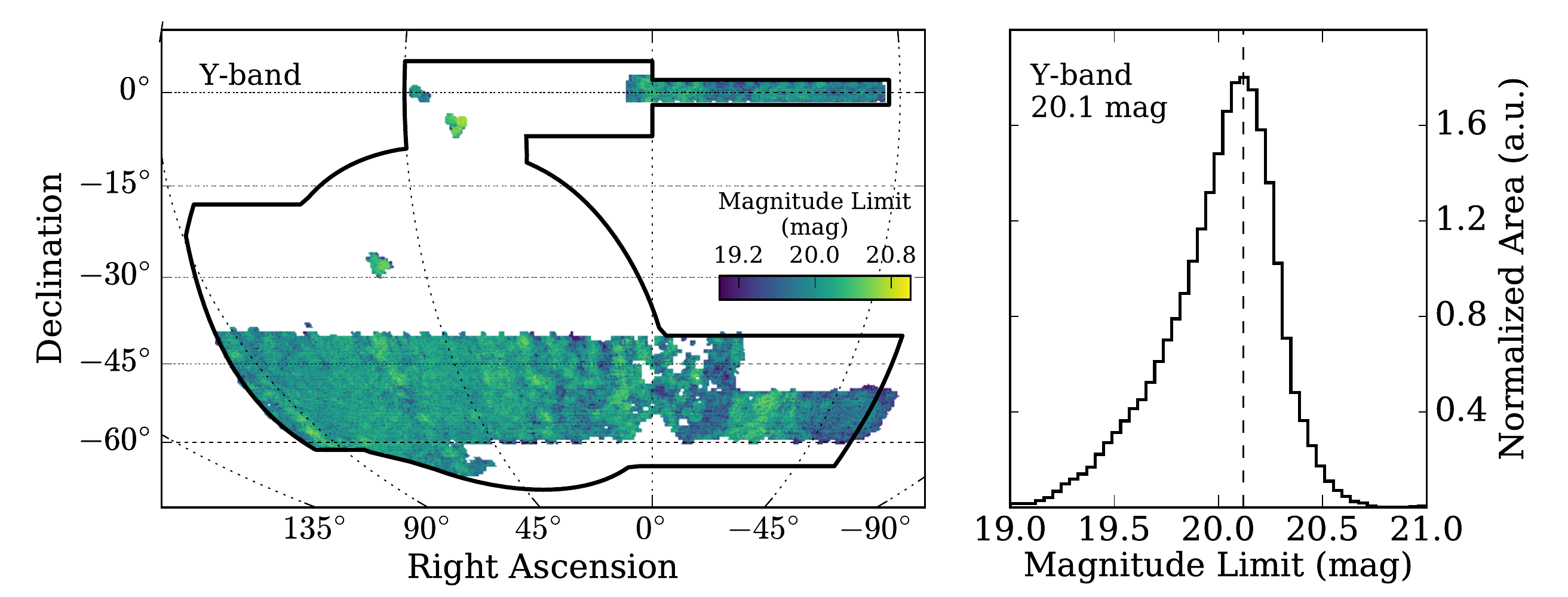}
\caption{\label{fig:app_depth} 
Sky maps and normalized histograms of the $10\sigma$ limiting magnitude for galaxies fit with \var{MAG\_AUTO}.
The mode of the limiting magnitude distribution is shown in the right panel of each row.
The derivation of the limiting magnitude is described in more detail in \secref{depth}.
}
\end{figure}

\begin{figure}[!ht]
\centering
\includegraphics[width=0.65\columnwidth]{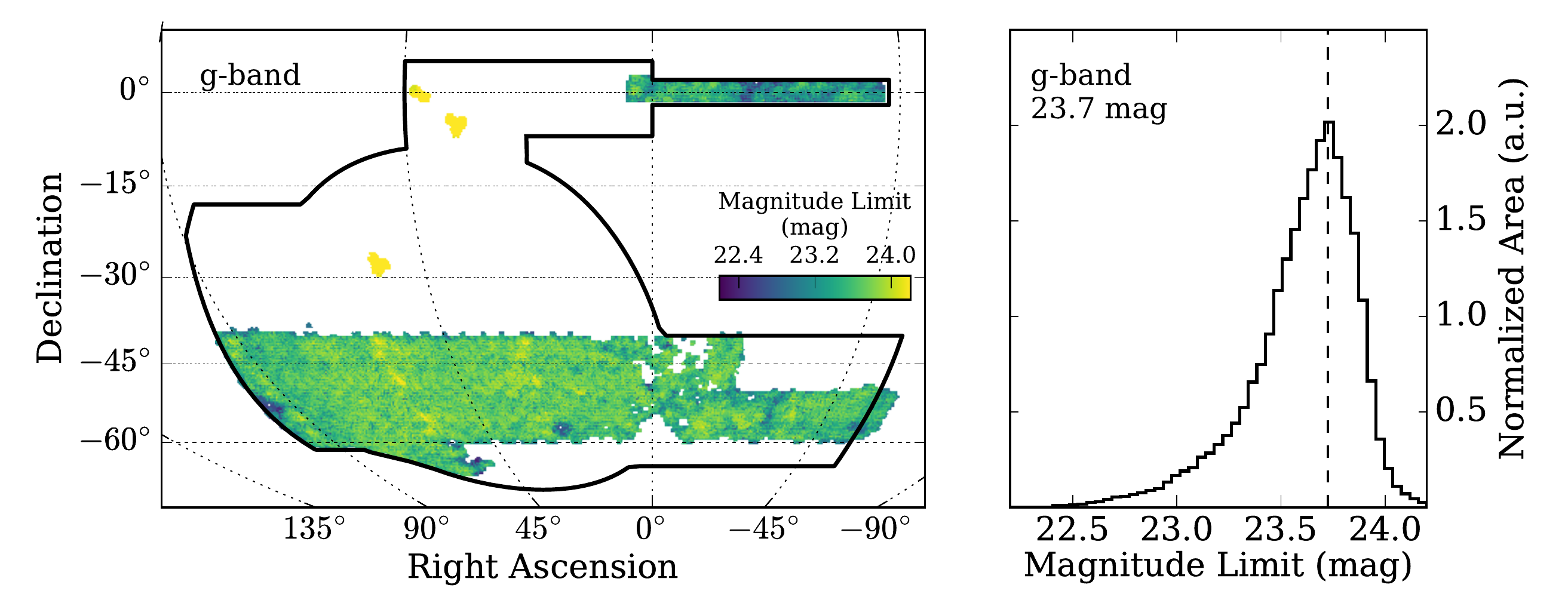}
\includegraphics[width=0.65\columnwidth]{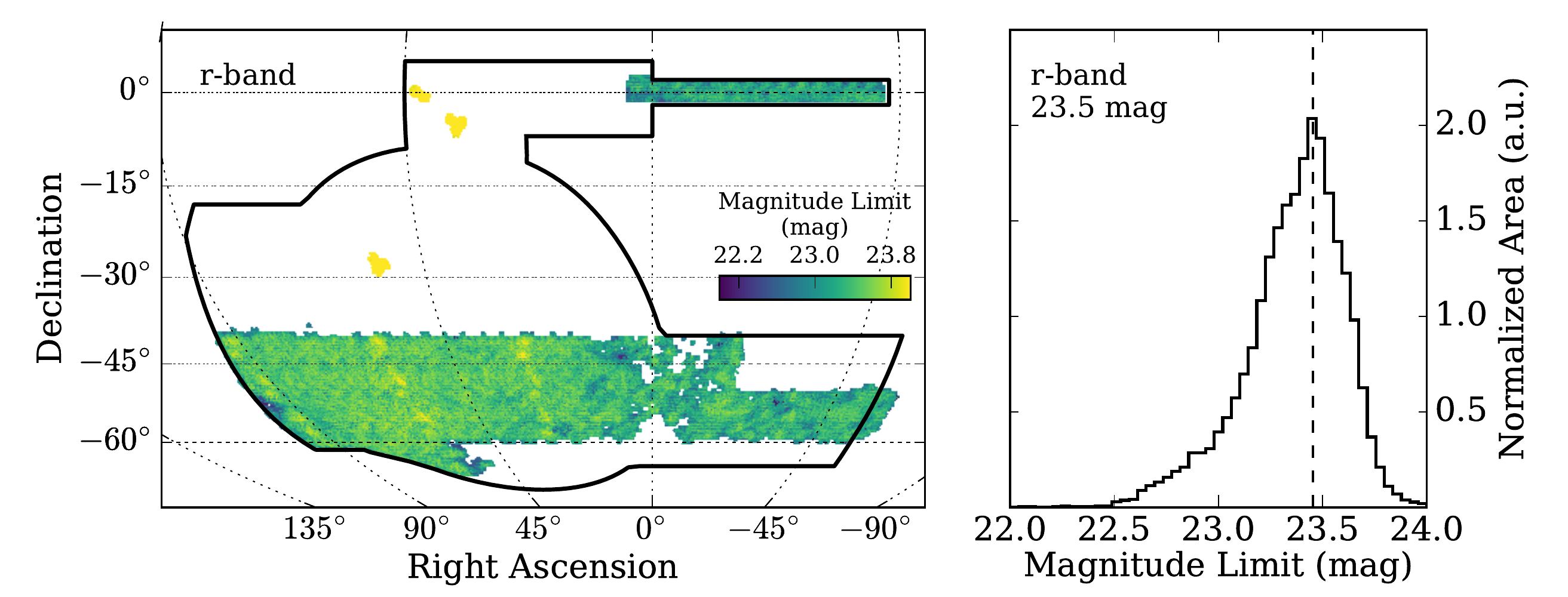}
\includegraphics[width=0.65\columnwidth]{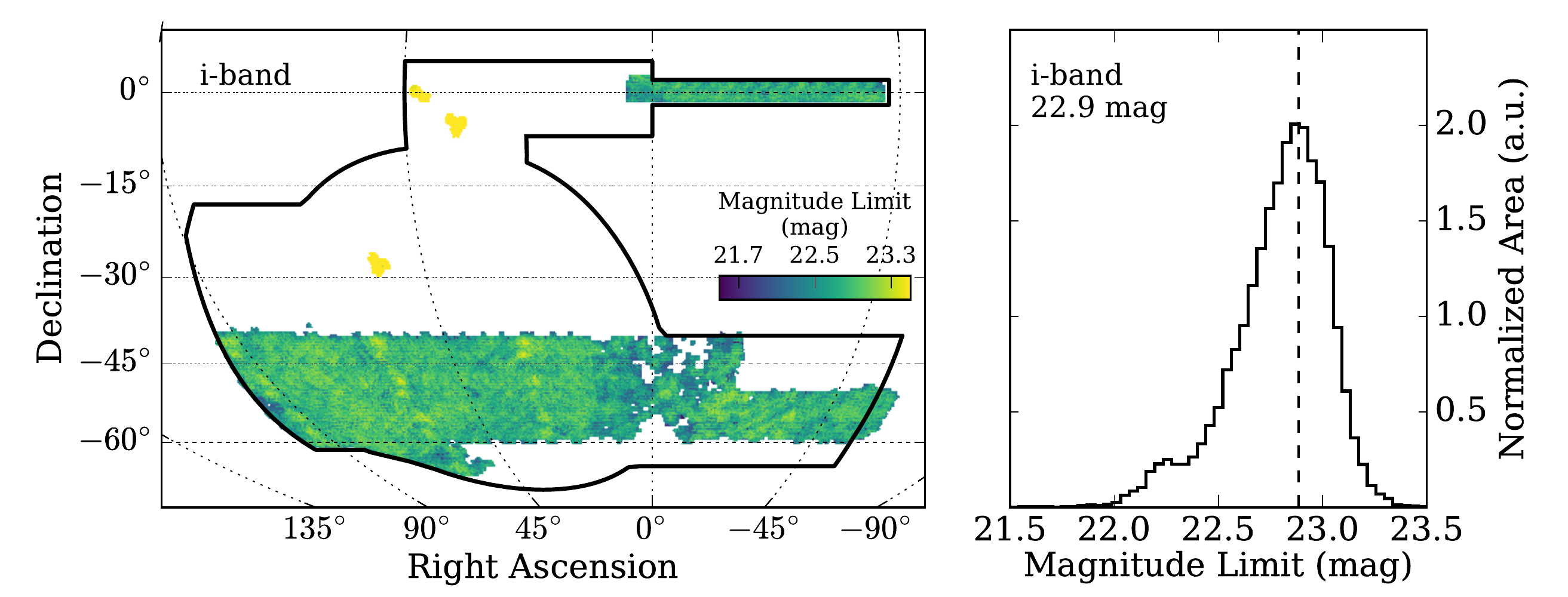}
\includegraphics[width=0.65\columnwidth]{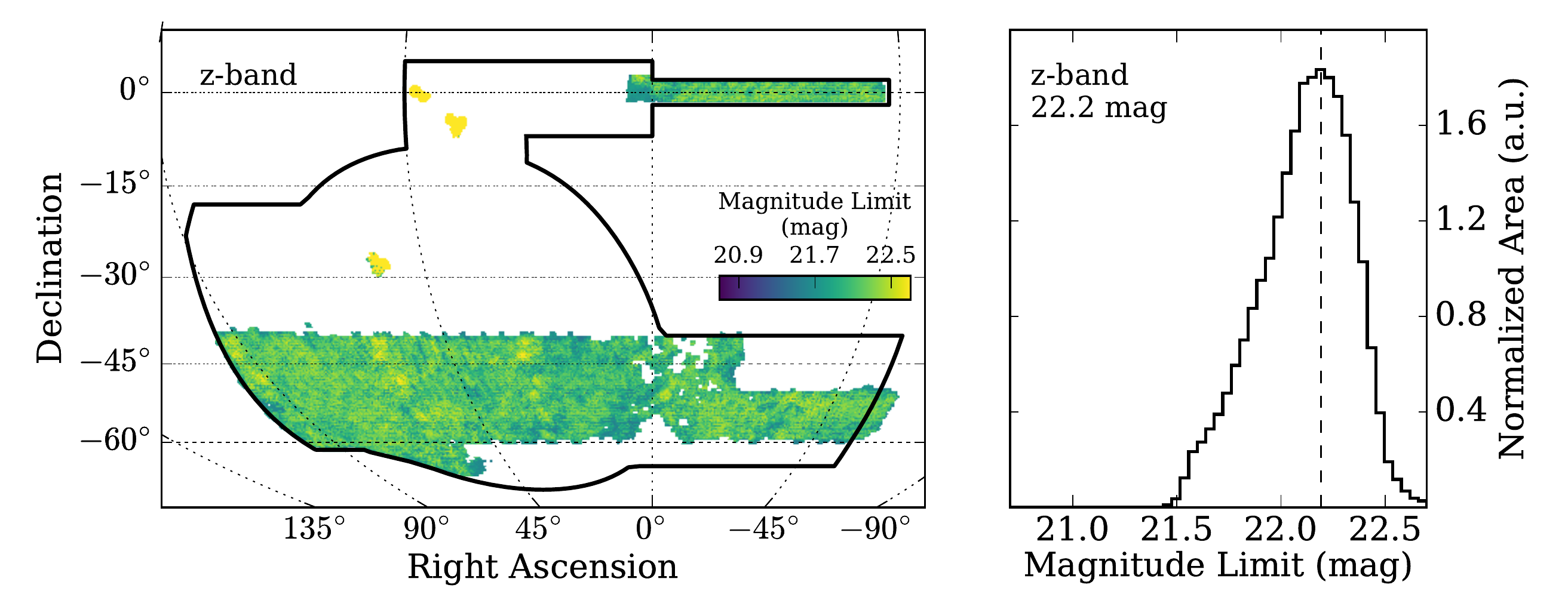}
\caption{\label{fig:app_mof_depth} 
Sky maps and normalized histograms of the $10\sigma$ limiting magnitude for galaxies fit with the MOF \var{CM\_MAG}.
The magnitude range for these figures is the same as the \var{MAG\_AUTO} magnitude limits shown in \figref{app_depth}.
The mode of the limiting magnitude distribution is shown in the right panel of each row.
Note that these magnitude limits include the deeper D10 coadds of the SN fields.
The derivation of the limiting magnitude is described in more detail in \secref{depth}.
}
\end{figure}

In this appendix we collect a set of figures documenting the $10\sigma$ limiting magnitude of the \gold catalog as described in \secref{depth}.
We include depth maps both for the \var{MAG\_AUTO} values derived from the coadded images (\figref{app_depth}) and for the \var{CM\_MAG} values derived from multi-epoch, multi-object fitting (\figref{app_mof_depth}).

\clearpage

\bibliographystyle{\bibsty}
\bibliography{main}

\end{document}